\documentclass[12pt,a4paper]{article}
\usepackage[utf8]{inputenc}

\usepackage{jheppub}

\usepackage{float}

\newcommand{\bq}{{q}}
\newcommand{\bk}{{k}}
\newcommand{\bl}{{\ell}}
\newcommand{\dd}{\mathrm{d}}
\newcommand{\hd}{\hat{\dd}}
\newcommand{\hdelta}{\hat{\delta}}

\usepackage{amsmath}
\usepackage{amssymb}
\usepackage{braket}

\usepackage{xcolor}
\usepackage{graphicx}

\definecolor{emerald}{rgb}{0.31, 0.78, 0.47}

\usepackage{tikz}
\def\lexp{\biggl\langle\!\!\!\biggl\langle}
\def\rexp{\biggr\rangle\!\!\!\biggr\rangle}
\newcommand{\KMOCav}[1]{\lexp #1 \rexp}

\usepackage[compat=1.1.0]{tikz-feynman}
\definecolor{airforceblue}{rgb}{0.36, 0.54, 0.66}
\definecolor{blue(ncs)}{rgb}{0.0, 0.53, 0.74}
\definecolor{caribbeangreen}{rgb}{0.0, 0.8, 0.6}

\usepackage{relsize}

\renewcommand{\[}{\begin{equation}\begin{aligned}}
\renewcommand{\]}{\end{aligned}\end{equation}}
\newcommand{\Aop}{\mathbb{A}}
\newcommand{\Fop}{\mathbb{F}}
\newcommand{\Rop}{\mathbb{R}}
\newcommand{\waveshape}{\alpha_\eta(k)}
\renewcommand{\Im}{\operatorname{Im}}
\renewcommand{\Re}{\operatorname{Re}}

\newcommand{\Principal}{\operatorname{PV}}
\renewcommand{\v}[1]{\mathbf{#1}}

\newcommand{\E}{\mathcal{E}}
\newcommand{\SCut}{\operatorname{SCut}}
\newcommand{\CCut}{\operatorname{CCut}}
\newcommand{\ICut}{\operatorname{ICut}}

\title{Radiation and Reaction at One Loop}

\author[1]{Asaad Elkhidir}
\author[1,2]{Donal O'Connell}
\author[1,2]{Matteo Sergola}
\author[2,3,4]{Ingrid A. Vazquez-Holm}

\affiliation[1]{Higgs Centre, School of Physics and Astronomy, University of Edinburgh, EH9 3FD, Scotland}
\affiliation[2]{Kavli Institute for Theoretical Physics, University of California, Santa Barbara, CA 93106-4030, USA}
\affiliation[3]{Department of Physics and Astronomy, Uppsala University, Box 516, 75120 Uppsala, Sweden}
\affiliation[4]{Nordita, Stockholm University and KTH Royal Institute of Technology, Hannes Alfv\'ens v\"ag 12, 10691 Stockholm, Sweden}

\emailAdd{A.E.H.Elkhidir@sms.ed.ac.uk}
\emailAdd{donal@ed.ac.uk}
\emailAdd{matteo.sergola@ed.ac.uk}
\emailAdd{ingrid.holm@physics.uu.se}

\abstract{
We study classical radiation fields at next-to-leading order using the methods of scattering amplitudes.
The fields of interest to us are sourced when two massive, point-like objects scatter inelastically, and can be computed from one-loop amplitudes.
The real and imaginary parts of the amplitudes play important but physically distinct roles in the radiation field.
We argue that the imaginary part captures the effects of radiation reaction.
This aspect of radiation reaction is directly linked to cuts of one-loop amplitudes which expose Compton trees.
We also discuss the fascinating interplay between renormalisation, radiation reaction and classical field theory from this perspective.
}

\begin{document}

\maketitle

\section{Introduction}

Gravitational waveforms sourced by compact binary coalescence events are now the basic physical observable in precision studies of General Relativity (GR).
Future gravitational wave observatories will work at higher signal-to-noise ratios, and therefore will be sensitive to more subtle aspects of the waveform.
This presents an exciting challenge for the theoretical physics community: to develop tools allowing for efficient determination of gravitational waveforms at new levels of precision~\cite{Antonelli:2019ytb,Kalogera:2021bya, Khalil:2022ylj,Buonanno:2022pgc}.

Precision computations in GR are a challenge because of the non-linearity of its perturbative structure. 
One hope for simplifying this non-linearity comes from the study of scattering amplitudes in quantum field theory.
Scattering amplitudes have a remarkable property known as the ``double copy'', which allows us to obtain scattering amplitudes in gravitational theories given amplitudes in much simpler Yang-Mills theories \cite{Kawai:1985xq,Bern:2008qj,Bern:2010ue,Bern:2010yg}.
Furthermore, elegant and powerful tools have been developed to compute scattering amplitudes with remarkable ease.
In this article, we will make heavy use of generalised unitarity~\cite{Bern:1994zx,Bern:1994cg}.
This method allows us to construct loop-level scattering amplitudes from tree amplitudes.
We can further combine the double copy and generalised unitarity, effectively building the dynamical information necessary for gravitational waveforms from tree amplitudes in Yang-Mills theory.
The union of generalised unitarity and the double copy has already proven very fruitful in the study of General Relativity, and provides
a fresh perspective on the relativistic two-body problem~\cite{Bjerrum-Bohr:14, Bjerrum-Bohr:16,76,Guevara:2018wpp, Bern:2020gjj, Moynihan:2020gxj, Cristofoli:2020uzm, Bern:2020buy,Parra-Martinez:2020dzs, Haddad:2020tvs, AccettulliHuber:2020oou, Moynihan:2020ejh, Manu:2020zxl, Sahoo:2020ryf,delaCruz:2020bbn, Bonocore:2020xuj,Mogull:2020sak,Emond:2020lwi,Cheung:2020gbf,Mougiakakos:2020laz,Carrasco:2020ywq,Kim:2020cvf,Bjerrum-Bohr:2020syg, Gonzo:2020xza,delaCruz:2020cpc,Herrmann:2021lqe,Bern:2021dqo, DiVecchia:2021bdo,Herrmann:2021tct,  Bjerrum-Bohr:2021vuf, Brandhuber:2021kpo,  Bjerrum-Bohr:2021din, Bautista:2021llr, Cristofoli:2021vyo, Bautista:2021wfy,  Aoude:2021oqj, Brandhuber:2021eyq,   yutinspin, Brandhuber:2021bsf, Cristofoli:2021jas,  Cho:2022syn, Bern:22, Alessio:2022kwv,FebresCordero:2022jts, Menezes:2022tcs}.

To date, much of the work on amplitudes and classical gravity has focused on understanding the interaction potential between gravitating masses. 
In contrast, our interest is directly in the radiation emitted during a dynamical process. We make use of a method (known as the KMOC formalism) for constructing radiation fields from amplitudes which has been developed in recent years \cite{ Luna:2016due,Goldberger:2016iau, Shen:2018ebu, Kosower:2018adc, Cristofoli:2021vyo, Jakobsen:2021smu, Herrmann:2021lqe, Adamo:2022qci}.
The basic idea is to use a quantum-mechanical language to describe the event; the physical observable to be computed becomes the expectation value of the Riemann curvature.
In the classical limit, this expectation will equal the classical curvature, up to quantum corrections which can be systematically dropped.
The formalism is general and can be applied to field strengths in a variety of theories: 
electromagnetism, Yang-Mills (YM) theory, and gravity.

As their name suggests, the methods of scattering amplitudes are most directly applicable to events where two objects scatter, generating radiation, rather than to the more physically relevant bound binaries. 
Nevertheless methods exist to connect scattering and bound physics.
In certain examples, we can simply analytically continue observables from the scattering to the bound case~\cite{Cho:2018upo, Kalin:2019rwq, Kalin:2019inp, Cho:2021arx}.
More generally, it is possible to build effective field theories (EFTs) describing general binary dynamics ~\cite{Goldberger:2004jt,Goldberger:2005cd,Goldberger:2009qd,Levi:2015msa,Levi:2015uxa,Levi:2015ixa,Levi:2016ofk,Goldberger:2016iau,Foffa:2016rgu,Goldberger:2017frp,Goldberger:2017vcg,Cheung:2018wkq,Foffa:2019hrb,Foffa:2019rdf,Foffa:2019yfl,Foffa:2019eeb,Levi:2019kgk,Blanchet:2019rjs,Goldberger:2019sya,Aoude:2020onz,Levi:2020kvb,Blumlein:2020znm,Levi:2020uwu,Levi:2020lfn,Blumlein:2020pyo,Foffa:2020nqe,Goldberger:2020fot,Blumlein:2021txj,Foffa:2021pkg,Brandhuber:2021kpo,Brandhuber:2021eyq,Almeida:2021xwn,Kim:2021rfj,Edison:2022cdu,Kim:2022pou,Mandal:2022nty,Kim:2022bwv,Almeida:2022jrv,Mandal:2022ufb}, see ~\cite{Porto:2016pyg,Levi:2018nxp} for reviews. 
These EFTs can be matched to scattering data and then applied to the bound case, effectively forming a bridge between the two.  

Effective field theory is a powerful tool, and has been used very successfully in classical gravitational wave physics for many years now~\cite{Goldberger:2004jt}. 
Most closely connected to this article are the ``post-Minkowskian''  effective theories~\cite{Kalin:2020mvi,Kalin:2020fhe,Kalin:2020lmz,Dlapa:2021npj,Dlapa:2021vgp,Kalin:2022hph,Dlapa:2022lmu} 
and  worldline quantum field theories \cite{Schubert:1996jj,Ahmadiniaz:2016vai,Mogull:2020sak,Jakobsen:2021smu,Jakobsen:2021lvp,Edwards:2021elz,Jakobsen:2021zvh,Jakobsen:2022fcj,Jakobsen:2022psy,Jakobsen:2022zsx,Comberiati:2022ldk}. 
These methods were applied to describe conservative binary dynamics at $\mathcal{O}(G)$ and $\mathcal{O}(G^2)$~\cite{Cheung:2018wkq}, at $\mathcal{O}(G^3)$~\cite{Bern:2019nnu, Bern:2019crd} and at $\mathcal{O}(G^4)$ in references~\cite{Bern:2021dqo, Bern:2021yeh}.
Recently, the ``HEFT'' approach \cite{Damgaard:2019lfh,Aoude:2020onz,Brandhuber:2021kpo, Brandhuber:2021eyq} has been introduced to  the array of EFT based methods. Inspired by the success of heavy-quark effective theory ~\cite{Isgur:1989vq,Georgi:1990um,Luke:1992cs,Neubert:1993mb,Manohar:2000dt}, the HEFT approach implements the classical limit as a large mass limit. 
The resulting decoupling between heavy and light degrees of freedom exposes certain simplifications associated with the classical limit and combines nicely with the double copy \cite{ Damgaard:2019lfh,  Aoude:2020onz, Haddad:2020tvs, Brandhuber:2021kpo, Brandhuber:2021eyq,  Brandhuber:2021bsf}. 

In this article, we build on previous work which studied scattering encounters between classical, point-like objects at leading order (LO)~\cite{Goldberger:2016iau,Luna:2017dtq,Shen:2018ebu,Kosower:2018adc,Cristofoli:2021vyo}.
We describe the structure of field strength observables at next-to-leading order (NLO) in terms of scattering amplitudes.
As we will see, the structure is remarkably simple. 
The waveform, as determined by amplitudes, naturally has two pieces.
These are associated with the real and imaginary parts (defined more carefully in section~\ref{sec:fs}) of a one-loop five-point amplitude.
As usual in scattering amplitudes, the imaginary part captures interesting dispersive physics and is closely related to radiation reaction.
Because of this connection, we include a detailed discussion of the imaginary part, and connect it to classical approaches to radiation reaction.
We show that the imaginary part captures the physics of the Abraham-Lorentz-Dirac force in electrodynamics, and argue that radiation reaction in Yang-Mills theory and in gravity can also be understood through the imaginary part.
The relevant diagrams involve cutting one massive and one massless line, and involve Compton amplitudes.
We refer to these diagrams as ``Compton cuts''.
Our treatment makes it clear that this aspect of radiation reaction double-copies in a straightforward manner at NLO.

We begin in section~\ref{sec:fs} with a discussion of the general structure of field-strength observables at NLO before describing some technical simplifications we can take advantage of at this order in section~\ref{sec:simp}, culminating in a detailed algorithm for computing the classical waveform.
We then turn to explicit examples.
First, in section~\ref{radiation}, we determine the radiation at one loop which is associated with the real part of the scattering amplitude. 
We argue that this part of the radiation field is classically associated with essentially conservative forces (eg the Lorentz force in electromagnetism).
We focus on examples in electrodynamics and Yang-Mills theory.
In section~\ref{reactionnn}, we turn to the imaginary part of the amplitude.
We discuss Compton cuts in detail in electrodynamics, Yang-Mills theory and in gravity, arguing that these cuts capture the physics of radiation reaction.
As radiation reaction is intimately associated with renormalisation, we present a discussion of the renormalisation of the one-loop five-point amplitude in QED which determines the radiation field in section ~\ref{rennn2}. 
In this section, we justify the omission of certain cuts which one might naively think could contribute to the classical radiation field. 
We show that these cuts are fully quantum after renormalisation in the on-shell scheme.
Along the way we discuss infrared divergences.
In section~\ref{sec:confirm} we turn to a detailed classical verification of our results in the context of electrodynamics.
We explicitly match the contribution of the Abraham-Lorentz-Dirac force to the Compton cuts arising in the imaginary part.
We conclude with a discussion of our results in section~\ref{sec:concl}.
In appendix \ref{cutextra} we remark on some classically relevant cuts~\cite{Caron-Huot:2023vxl} which contribute to the imaginary part although they are unrelated to radiation reaction.  
Appendix~\ref{coleman} contains a classical perspective on the Abraham-Lorentz-Dirac self-field inspired by an old article of Coleman.
This perspective is intended to clarify how the quantum-mechanical approach coincides with the classical approach to radiation reaction.
Finally, in appendix~\ref{sec:fullqed} we collect our results for the integrand of the waveform in electrodynamics.

\subsubsection*{Note added}
While finalising this paper we learned about parallel research presented in references~\cite{wave1} and \cite{wave2}
which contain some overlap with our work.
Further related work by one of the present authors and collaborators will appear in the forthcoming article~\cite{wave3}.
We thank the authors for cooperating with us in the submission of our work, and for sharing advance copies of their drafts with us. 
After completing the first version of our preprint we learned of the important work~\cite{Caron-Huot:2023vxl} and have updated our manuscript to incorporate some nomenclature and especially by adding the cuts described in appendix \ref{cutextra}.

\section{Field strengths from amplitudes}\label{sec:fs}

Our goal is to compute the radiation field generated by a scattering event involving two point-like classical objects using the methods of scattering amplitudes.
The basic observable of interest is the field strength (in electrodynamics and YM theory) or the Riemann curvature (in gravity), both of which are very similar in structure; we will refer to both generically as ``field strengths''.
The waveform, as measured by gravitational wave observatories, is closely connected to the curvature. 
For us the difference will be immaterial, so we will also refer to these field strengths as waveforms.
In this section, we explain how to determine field strengths from scattering amplitudes at next-to-leading order (NLO) accuracy.
We begin with a short review of the connection between amplitudes and observables.

\subsection{States and observables}

Field strengths, as observables in themselves, were first discussed from the perspective of amplitudes in references~\cite{Cristofoli:2021vyo, Monteiro:2020plf} and were recently reviewed in~\cite{Travaglini:2022uwo,Kosower:2022yvp}.
Amplitudes are quantum-mechanical objects, so we must start by specifying an initial quantum state which happens to be in the domain of validity of the classical approximation.
If we also arrange initial conditions so that we may rely on the classical approximation throughout the scattering event,
the correspondence principle guarantees that the quantum treatment will agree with a classical treatment up to small quantum corrections which we systematically drop.
This is the basic philosophy of the KMOC approach~\cite{Kosower:2018adc} to extracting classical physics from scattering amplitudes.
We will follow the notation of KMOC closely below.

We choose our initial state to be
\[\label{eq:initialState}
\ket{\psi} = \int \dd \Phi(p_1, p_2) \, \phi_b(p_1, p_2) \ket{p_1, p_2} \,,
\]
where, following~\cite{Kosower:2018adc}, we write\footnote{Note that working in $D$ spacetime dimensions calls for an energy scale, required by dimensional consistency: $\dd ^D p\to \mu^{4-D} \dd ^Dp$. In our expressions we will however set $\mu=1$ for notational simplicity; this scale can always be reintroduced by dimensional analysis whenever necessary. We sometimes take $D=4$ when dealing with manifestly finite integrals.}
\[
\dd \Phi(p) \equiv \frac{\dd^D p}{(2\pi)^D} (2\pi) \delta(p^2 - m^2) \, \Theta(p^0) \equiv \hd^D p \, \hdelta(p^2-m^2) \, \Theta(p^0)
\]
for the on-shell phase-space measure of a single particle with mass $m$. Hatted derivatives and delta-functions are defined to absorb factors of $2\pi$.
Meanwhile the ket $\ket{p_1,p_2}$ involves two different particles --- quanta of entirely different quantum fields --- with masses $m_1$ and $m_2$ and momenta $p_1$ and $p_2$.

Since a plane-wave state of a massive particle has no classical interpretation, we have placed our particles in a wavepacket $\phi_b(p_1, p_2)$. 
This wavepacket should individually localise each particle with an uncertainty which is very small compared to any relevant classical scale in our process (for example, the impact parameter).
As an example, we could choose
\[
\label{eq:wavepacketChoice}
\phi_b(p_1, p_2) &\equiv e^{i b_1 \cdot p_1} e^{i b_2 \cdot p_2} \phi_1(p_1) \phi_2(p_2)\,, 
\]
where $\phi_i(p_i)$ are sharply-peaked functions of the momenta.
The two-particle wavefunction in equation~\eqref{eq:wavepacketChoice} displaces particle $i$ by a distance $b_i$ relative to an origin; then the impact parameter is $b_{12} = b_1 - b_2$.

We will soon find it very convenient to extend our notation for phase-space measures by writing the measure for several particles as
\[
\dd \Phi(p_1, p_2, \ldots) = \dd \Phi(p_1) \dd \Phi(p_2) \cdots \,.
\]
We also define the appropriate delta-function $\delta_\Phi(p)$ with respect to this measure such that
\[
\int \dd \Phi(p) \, \delta_\Phi(p-p') \, f(p) = f(p') \,,
\]
for any smooth function $f(p)$.

Our basic task is to compute the future expectation value of a field strength operator. 
In Yang-Mills theory, for example, the relevant operator is the field strength tensor
\[
\Fop_{\mu\nu}^a = \partial_\mu \Aop_\nu^a - \partial_\nu \Aop_\mu^a + g f^{abc} \Aop^b_\mu \Aop^c_\nu \,.
\]
There is one immediate simplification from working in the far-field limit. 
In the far field, the expectation value of the Yang-Mills potential $\Aop(x)$ is inversely proportional to the large radius $r$ between the observer and the scattering event \cite{Buonanno:2022pgc, Cristofoli:2021vyo}. 
We will only be interested in this leading $1/r$ behaviour.
As a result we may replace the full non-Abelian field strength with its abelianised version:
\[
\Fop_{\mu\nu}^a \simeq \partial_\mu \Aop_\nu^a - \partial_\nu \Aop_\mu^a  \,.
\]
In gravity, we are only interested in the expectation value of the linearised Riemann tensor for the same reason.
It may be worth emphasising that there is still non-linear (non-Abelian) dynamics in the core of spacetime.

The state in the far future is $S\ket{\psi}$ since the $S$ matrix is the all-time evolution operator. 
Placing our detector at a position $x$ near lightlike future infinity, the observable of interest to us in Yang-Mills theory is
\[
\label{eq:observableDef}
F_{\mu\nu}^a(x) \equiv \braket{\psi | S^\dagger \Fop_{\mu\nu}^a(x) S | \psi} \,.
\]
To connect with scattering amplitudes, we use the mode expansion for the quantum field $\Aop_\mu^a$:
\[
\Aop_\mu^a(x) = \sum_\eta \int \dd \Phi(k) \left [ \varepsilon_\mu^\eta(k)  a_\eta^a(k) e^{-i k \cdot x} + \textrm{h.c.} \right] \,,
\]
so that\footnote{We define antisymmetrisation brackets with no factor of 2: $A_{[\mu} B_{\nu]} = A_\mu B_\nu - A_\nu B_\mu$.}
\[
\label{eq:genFieldStrength}
F_{\mu\nu}^a(x) = 2 \Re \sum_\eta \int \dd \Phi(k) \left [ -i k_{[\mu} \varepsilon_{\nu]}^\eta(k) \braket{ \psi | S^\dagger a_\eta^a(k) S | \psi } e^{-i k \cdot x} 
\right] \,.
\]

In gravity, defining the curvature expectation
\[
R_{\mu\nu\rho\sigma}(x) \equiv \braket{\psi | S^\dagger \Rop_{\mu\nu\rho\sigma}(x) S | \psi} \,,
\]
it similarly follows that
\[
\label{eq:gengrav}
R_{\mu\nu\rho\sigma}(x) = 
\kappa \Re \sum_\eta \int \dd \Phi(k) \left [ k_{[\mu} \varepsilon_{\nu]}^\eta(k) k_{[\rho} \varepsilon_{\sigma]}^\eta(k) \braket{ \psi | S^\dagger a_\eta(k) S | \psi } e^{-i k \cdot x} \right] \,.
\]
We have introduced the constant $\kappa$, defined in terms of Newton's constant by $\kappa = \sqrt{32\pi G}$, and the annihilation operator\footnote{We hope that context will suffice to distinguish annihilation operators of different messengers.} $a_\eta(k)$ of a graviton state with helicity $\eta$ and momentum~$k$. 
Since the physics is so similar in gravity and gauge theory it is useful to have a word referring to any of the force carriers we are interested in (photons, gluons or gravitons). 
We will refer to them as ``messengers''.

The field strength of equation~\eqref{eq:genFieldStrength} and the curvature~\eqref{eq:gengrav} both involve an integration over the phase space of a massless particle.
At large distances, this integral can be reduced to a one-dimensional Fourier transform using standard methods (see~\cite{Travaglini:2022uwo,Kosower:2022yvp} for a recent review).
Writing the observation coordinate as $x = (x^0, \v x)$ and introducing the retarded time $u = x^0 - |\v x|$, the results are 
\[
\label{eq:laterUse}
F_{\mu\nu}^a(x) = \frac{-1}{4\pi |\v x|} 2 \Re \int_0^\infty \hd \omega \, e^{-i \omega u} \sum_\eta k_{[\mu} \varepsilon^\eta_{\nu]}(k) \, \braket{ \psi | S^\dagger a_\eta^a(k) S | \psi }  \, ,
\]
in Yang-Mills theory, and 
\[\label{riemmm}
R_{\mu\nu\rho\sigma}(x) = \frac{\kappa}{4\pi |\v x|} \Re \int_0^\infty \hd \omega \, e^{-i \omega u} \sum_\eta \, i k_{[\mu} \varepsilon^\eta_{\nu]}(k)k_{[\rho} \varepsilon^\eta_{\sigma]}(k) \, \braket{ \psi | S^\dagger a_\eta(k) S | \psi } \, ,
\]
in gravity. Note that in both integrals  \eqref{eq:laterUse} and \eqref{riemmm} the momentum $k$ is set to be $k^\mu=\omega(1, \hat{\mathbf{x}}).$

Referring to the two equations above, it is clear that the key dynamical quantity to be determined is 
\[
\label{eq:waveshapeDef1}
\waveshape \equiv \braket{\psi | S^\dagger a_\eta(k) S | \psi}\,,
\]
where $a_\eta(k)$ is an annihilation operator for the relevant field.
We will refer to this quantity as the ``waveshape'' because once the waveshape is known, the relevant field strength (in frequency space) can be immediately determined using equations~\eqref{eq:laterUse} and~\eqref{riemmm}. Much of the body of this paper will be devoted to studying this waveshape and how it may be computed.

It may be worth adding that classical objects which are particularly closely related to the waveshape are the Newman-Penrose scalars. In fact, Newman-Penrose scalars in the frequency domain are proportional to the waveshape for a given choice of helicity~\cite{Cristofoli:2021vyo}. In this article we
found it to be simplest to work directly with the waveshape itself.
Waveshapes are also particularly important in the study of coherent states and their connection with classical fields.
The connection between amplitudes, coherent states and radiation was discussed in detail in reference~\cite{Cristofoli:2021jas}.

Having discussed the general connection between amplitudes and field strengths, let us now understand how to construct the waveshape from perturbative scattering amplitudes.
One obvious way to proceed is simply to extend the KMOC framework of \cite{Kosower:2018adc} to the computation of matrix element \label{eq:waveshapeDef} by expanding  $S = 1 + i T$.
This approach immediately leads to the leading order expression
\[\label{wslo}
\waveshape =  \int \dd\Phi&(p'_1, p'_2, p_1, p_2) \, \phi_b^*(p_1', p_2') \phi_b(p_1, p_2)   \hdelta^D(p_{\text{tot}}) \, i \mathcal{A}_{5,0}(p_1, p_2 \rightarrow p'_1, p'_2, k_\eta ) \,,
\]
where we are adopting the notation that $\mathcal{A}_{n,L}$ is an $n$ point,  $L$ loop amplitude. 
The waveshape is slightly more involved at one loop, where we encounter two terms
\[\label{ws1l}
\waveshape=  \int \dd\Phi&(p'_1, p'_2, p_1, p_2) \, \phi_b^*(p_1', p_2') \phi_b(p_1, p_2) \hdelta^D(p_{\text{tot}}) \left( \, i \mathcal{A}_{5,1}(p_1, p_2 \rightarrow p'_1, p'_2, k_\eta )
\right.\\&\left.+\int \dd\Phi(\tilde p_1, \tilde p_2) \hdelta^D(\tilde p_{\text{tot}})   \mathcal{A}_{5,0}(p_1, p_2 \rightarrow \tilde p_1, \tilde p_2, k_\eta) \mathcal{A}_{4,0}^*(\tilde p_1, \tilde p_2 \rightarrow p'_1, p'_2)
\right) \,,
\] 
with the delta functions imposing conservation of energy and momentum
\begin{equation}
    p_{\text{tot}}=p_1+p_2-p_1'-p_2'-k=0,\,\,\,\,\,\tilde p_{\text{tot}}=\tilde p_1+\tilde p_2-p_1'-p_2'=0,
\end{equation}
for external states and across the cut. 

It is easy to see that the structure of the one-loop waveshape  \eqref{ws1l} is indeed very similar to the impulse described in~\cite{Kosower:2018adc}: one  sums ($i$ times) the one-loop amplitude and the specific cut shown in equation \eqref{ws1l}.  However, in this article, we find it to be useful to rearrange the observable in a form which clarifies the physics while also simplifying aspects of the computation. 

\subsection{Real and imaginary parts}\label{impart}

One clue that there is another way of constructing the observable is the fact that the two terms in equation \eqref{ws1l} instruct us to sum $i$ times the amplitude and the cut of the amplitude. 
Since cuts arise from the imaginary parts of the amplitude it is clear that the combination $i \mathcal{A}_1 + \Im \mathcal{A}_1$ \emph{removes} an imaginary part of the amplitude.
However it is important to realise that the cut in equation~\eqref{ws1l} is not the complete imaginary part of the amplitude: the whole imaginary part is the sum of several distinct cuts.
The usefulness of real and imaginary parts of amplitudes in the construction of KMOC-style classical observables was first emphasised in reference~\cite{Herrmann:2021tct} which studied the impulse in classical scattering. 
Real and imaginary parts also play a crucial role in eikonal methods, see for instance references~\cite{Bern:2020gjj,DiVecchia:2022nna,DiVecchia:2022piu, Cristofoli:2021jas,Damgaard:2021ipf}.

We begin with the waveshape~\eqref{eq:waveshapeDef1} in the form
\[
\waveshape &= \int \dd\Phi(p'_1, p'_2, p_1, p_2) \, \phi_b^*(p_1', p_2') \phi_b(p_1, p_2) \braket{p_1' p_2' | S^\dagger a_\eta(k) S | p_1 p_2} \, .
\]
At the level of quantum field theory, the key quantity we need to compute is the in-in expectation (we will simply write ``expectation'' for brevity) $\braket{p_1' p_2' | S^\dagger a_\eta(k) S | p_1 p_2}$,  
so inspired by reference~\cite{Caron-Huot:2023vxl} we define
\[
\label{eq:expectationDef}
\E(p_1 p_2 \rightarrow p_1' p_2' k_\eta) \delta^D(p_1 + p_2 - p_1' -p_2' - k) \equiv -i \braket{p_1' p_2' | S^\dagger a_\eta(k) S | p_1 p_2} \,.
\]
Unitarity of the $S = 1 + iT$ matrix leads to the equivalent form
\[
\label{eq:expecationFromT}
\E(p_1 p_2 \rightarrow  p_1' &p_2' k_\eta) \delta^D(p_1 + p_2 - p_1' - p_2'-k) \\
&= \braket{p_1' p_2'| a_\eta(k) \Re T + \frac{i}{2} \left( [a_\eta(k), T^\dagger]T - T^\dagger [a_\eta(k), T]\right) | p_1,p_2} \,.
\]
To begin understanding $\E$, consider the lowest order contribution. 
We may then neglect terms quadratic in $T$, and find
\[\label{eq:treeExpect}
\E_{(5,0)}(p_1 p_2 \rightarrow  p_1' p_2' k_\eta) = \mathcal{A}_{(5,0)}(p_1 p_2 \rightarrow  p_1' p_2' k_\eta) \,.
\]
So at lowest order, the only contribution is the five-point tree amplitude.
At higher orders, $\E$ is still analogous to an amplitude but it differs in the details of the cuts which contribute; see reference~\cite{Caron-Huot:2023vxl} for a discussion.
Because of the close connection between the expectation $\E$ and amplitudes, we will use closely analogous notation for the expectation; for example $\E_{(n,L)}$ is an $n$-point $L$ loop expectation.

This paper is concerned with the next-to-leading order waveshape.
At this level the overall structure of the expectation~\eqref{eq:expecationFromT} is very simple. 
There is a ``real'' part $\braket{p_1' p_2' k_\eta | \Re T | p_1 p_2}$: 
an element of the real part of the $T$ matrix. 
We hasten to emphasise that this element of $\Re T$ need not be real, for example because the messenger helicity may be flipped on conjugation.
We could choose a real (linear) basis of messenger polarisations; then we find (see section~\ref{sec:PV}) that $\braket{p_1' p_2' k_\eta | \Re T | p_1 p_2}$ is actually real at our order in perturbation theory.
Thus we write
\[
\label{eq:realExpect}
\Re' \E(p_1 p_2 \rightarrow  p_1' &p_2' k_\eta) \delta^D(p_1 + p_2 - p_1' - p_2'-k) \equiv 
\braket{p_1' p_2'| a_\eta(k) \Re T | p_1,p_2} \,.
\]
The prime is a warning that one should treat the polarisation as real. 
(In QCD, color factors should also be taken to be real.)
Referring to equation~\eqref{eq:expecationFromT}, we define the ``imaginary'' part as
\[
\label{eq:imE}
\Im' \E(p_1 p_2 \rightarrow  p_1' &p_2' k_\eta) \delta^D(p_1 + p_2 - p_1' - p_2'-k) \\
&= \frac12 \braket{p_1' p_2'| [a_\eta(k), T^\dagger]T - T^\dagger [a_\eta(k), T]| p_1,p_2} \,.
\]
At one-loop order, the explicit $T$ matrices above may be taken as real.

Before we continue, a few words regarding the distinction between the expectation and the amplitude.
In fact, there is no distinction at the level of the real part:
\[
\Re' \mathcal{A}_{5,1} (p_1 p_2 \rightarrow p_1' p_2' k_\eta) \hdelta(p_1 + p_2 - p_1' - p_2'-k) \equiv \braket{p_1' p_2'| a_\eta(k) \Re T | p_1,p_2} \,.
\]
The real parts of the amplitude and the expectation are identical by definition (this is consistent with the situation at tree level in equation~\eqref{eq:treeExpect}).
The imaginary parts, on the other hand, are not the same.
At one loop, it is an easy exercise to see that they differ only in the sign of the second term on the right-hand-side of equation~\eqref{eq:imE}.

\subsection{Cuts and the imaginary part}
\label{sec:cutAndIm}

As usual in the study of scattering amplitudes, imaginary parts are related to cuts. 
Equation~\eqref{eq:imE} shows that the imaginary part can be decomposed in terms of two commutators, each quadratic in $T$; here, we interpret this fact in terms of cuts.

As the commutators in~\eqref{eq:imE} involve two $T$ matrices (one conjugated), and we work at one-loop order (namely $g^5$), it follows that we need one insertion of a $g^2$ tree amplitude and a $g^3$ tree amplitude.
A short list of possible amplitudes can contribute, as shown in the table:

\begin{figure}[H]
\begin{center}

 
\tikzset{
pattern size/.store in=\mcSize, 
pattern size = 5pt,
pattern thickness/.store in=\mcThickness, 
pattern thickness = 0.3pt,
pattern radius/.store in=\mcRadius, 
pattern radius = 1pt}
\makeatletter
\pgfutil@ifundefined{pgf@pattern@name@_3jlc5tfqt}{
\pgfdeclarepatternformonly[\mcThickness,\mcSize]{_3jlc5tfqt}
{\pgfqpoint{0pt}{0pt}}
{\pgfpoint{\mcSize+\mcThickness}{\mcSize+\mcThickness}}
{\pgfpoint{\mcSize}{\mcSize}}
{
\pgfsetcolor{\tikz@pattern@color}
\pgfsetlinewidth{\mcThickness}
\pgfpathmoveto{\pgfqpoint{0pt}{0pt}}
\pgfpathlineto{\pgfpoint{\mcSize+\mcThickness}{\mcSize+\mcThickness}}
\pgfusepath{stroke}
}}
\makeatother

 
\tikzset{
pattern size/.store in=\mcSize, 
pattern size = 5pt,
pattern thickness/.store in=\mcThickness, 
pattern thickness = 0.3pt,
pattern radius/.store in=\mcRadius, 
pattern radius = 1pt}
\makeatletter
\pgfutil@ifundefined{pgf@pattern@name@_ik54l71px}{
\pgfdeclarepatternformonly[\mcThickness,\mcSize]{_ik54l71px}
{\pgfqpoint{0pt}{0pt}}
{\pgfpoint{\mcSize+\mcThickness}{\mcSize+\mcThickness}}
{\pgfpoint{\mcSize}{\mcSize}}
{
\pgfsetcolor{\tikz@pattern@color}
\pgfsetlinewidth{\mcThickness}
\pgfpathmoveto{\pgfqpoint{0pt}{0pt}}
\pgfpathlineto{\pgfpoint{\mcSize+\mcThickness}{\mcSize+\mcThickness}}
\pgfusepath{stroke}
}}
\makeatother

 
\tikzset{
pattern size/.store in=\mcSize, 
pattern size = 5pt,
pattern thickness/.store in=\mcThickness, 
pattern thickness = 0.3pt,
pattern radius/.store in=\mcRadius, 
pattern radius = 1pt}
\makeatletter
\pgfutil@ifundefined{pgf@pattern@name@_esagus80x}{
\pgfdeclarepatternformonly[\mcThickness,\mcSize]{_esagus80x}
{\pgfqpoint{0pt}{0pt}}
{\pgfpoint{\mcSize+\mcThickness}{\mcSize+\mcThickness}}
{\pgfpoint{\mcSize}{\mcSize}}
{
\pgfsetcolor{\tikz@pattern@color}
\pgfsetlinewidth{\mcThickness}
\pgfpathmoveto{\pgfqpoint{0pt}{0pt}}
\pgfpathlineto{\pgfpoint{\mcSize+\mcThickness}{\mcSize+\mcThickness}}
\pgfusepath{stroke}
}}
\makeatother

 
\tikzset{
pattern size/.store in=\mcSize, 
pattern size = 5pt,
pattern thickness/.store in=\mcThickness, 
pattern thickness = 0.3pt,
pattern radius/.store in=\mcRadius, 
pattern radius = 1pt}
\makeatletter
\pgfutil@ifundefined{pgf@pattern@name@_qm4uefmv2}{
\pgfdeclarepatternformonly[\mcThickness,\mcSize]{_qm4uefmv2}
{\pgfqpoint{0pt}{0pt}}
{\pgfpoint{\mcSize+\mcThickness}{\mcSize+\mcThickness}}
{\pgfpoint{\mcSize}{\mcSize}}
{
\pgfsetcolor{\tikz@pattern@color}
\pgfsetlinewidth{\mcThickness}
\pgfpathmoveto{\pgfqpoint{0pt}{0pt}}
\pgfpathlineto{\pgfpoint{\mcSize+\mcThickness}{\mcSize+\mcThickness}}
\pgfusepath{stroke}
}}
\makeatother

 
\tikzset{
pattern size/.store in=\mcSize, 
pattern size = 5pt,
pattern thickness/.store in=\mcThickness, 
pattern thickness = 0.3pt,
pattern radius/.store in=\mcRadius, 
pattern radius = 1pt}
\makeatletter
\pgfutil@ifundefined{pgf@pattern@name@_atnmagmcb}{
\pgfdeclarepatternformonly[\mcThickness,\mcSize]{_atnmagmcb}
{\pgfqpoint{0pt}{0pt}}
{\pgfpoint{\mcSize+\mcThickness}{\mcSize+\mcThickness}}
{\pgfpoint{\mcSize}{\mcSize}}
{
\pgfsetcolor{\tikz@pattern@color}
\pgfsetlinewidth{\mcThickness}
\pgfpathmoveto{\pgfqpoint{0pt}{0pt}}
\pgfpathlineto{\pgfpoint{\mcSize+\mcThickness}{\mcSize+\mcThickness}}
\pgfusepath{stroke}
}}
\makeatother

 
\tikzset{
pattern size/.store in=\mcSize, 
pattern size = 5pt,
pattern thickness/.store in=\mcThickness, 
pattern thickness = 0.3pt,
pattern radius/.store in=\mcRadius, 
pattern radius = 1pt}
\makeatletter
\pgfutil@ifundefined{pgf@pattern@name@_osg8z11qv}{
\pgfdeclarepatternformonly[\mcThickness,\mcSize]{_osg8z11qv}
{\pgfqpoint{0pt}{0pt}}
{\pgfpoint{\mcSize+\mcThickness}{\mcSize+\mcThickness}}
{\pgfpoint{\mcSize}{\mcSize}}
{
\pgfsetcolor{\tikz@pattern@color}
\pgfsetlinewidth{\mcThickness}
\pgfpathmoveto{\pgfqpoint{0pt}{0pt}}
\pgfpathlineto{\pgfpoint{\mcSize+\mcThickness}{\mcSize+\mcThickness}}
\pgfusepath{stroke}
}}
\makeatother
\tikzset{every picture/.style={line width=0.75pt}} 


 
\end{center}
\end{figure}
\hspace{-.75cm}First a word on our diagrams.
Throughout this paper, we adopt the convention of drawing massive particle lines as solid lines.
We always indicate particle 1 with a red line and particle 2 with a blue one.
In this table, the wiggly lines indicate the appropriate messenger.

Let us first consider the second term contributing to $\Im' \E$ in equation~\eqref{eq:imE}, namely $\braket{p'_1, p'_2 | T^\dagger \, [a_\eta(k), T] | p_1, p_2}$.
The $T$ matrix here acts on the ket $\ket{p_1, p_2}$, and (because of the commutator) must involve at least one outgoing messenger. 
The only possibility in our table of possible amplitudes is the $\mathcal{O}(g^{3})$ $2 \rightarrow 3$ amplitude involving radiation of one messenger as the two scalars scatter. 
It then follows that $\bra{p'_1, p'_2} T^\dagger$ must evaluate to the four-point four-scalar amplitude as the only  $\mathcal{O}(g^2)$ amplitude with two final-state scalars. 
Thus, 
\[

\]

Next, we must understand the first term, $\braket{p'_1, p'_2 |[a_\eta(k), T^\dagger] \, T | p_1, p_2}$, on the right-hand-side of~\eqref{eq:imE}.
This is slightly more complicated because there are distinct contributions as we now discuss.
Consider the action of $T$ on the state $\ket{p_1, p_2}$. 
There are two possibilities in our table of amplitudes: (i) the $2 \rightarrow 2$ four scalar scattering amplitude, which is order $g^2$, or (ii) the $2 \rightarrow 3$ five-point amplitude involving four scalars and an outgoing messenger (order $g^3$).
In case (i), the remaining factor $\bra{p'_1, p'_2}[a_\eta(k), T^\dagger]$ must yield the $\mathcal{O}(g^3)$ five-point amplitude.
In case (ii), we must extract an $\mathcal{O}(g^2)$ amplitude from $[a_\eta(k), T^\dagger]$ to arrive at order $g^5$. 
Moreover, this amplitude must involve an outgoing messenger (because of the commutator).
Looking at the table, the only possibility is a Compton amplitude. As this Compton can attach to either of our massive lines, there are two diagrams in this case.
Thus we learn that

\[\label{eq:imcuts}

\]
%
The presence of Compton amplitudes in this cut is significant, so we refer to the relevant diagrams as ``Compton cuts''. 
More explicitly, these cuts involve cutting one massive and one massless line, exposing a product of tree amplitudes, one of which is a Compton tree.
Later in this article (see section~\ref{reactionnn}) we will argue that these Compton cuts are related to radiation-reaction effects in the classical limit\footnote{In the closed time-path (Schwinger-Keldysh) approach to computing expectation values in field theory, the $T^\dagger$ matrix arises from the part of the contour which goes ``backwards'' in time. The Schwinger-Keldysh approach  can also be used to take into account of radiation reaction \cite{Galley:2012hx,Kalin:2022hph}.}.
Because of their importance, we introduce a convenient $\CCut$ notation to take care of this class of contributions to the imaginary part of the expectation:
\[\label{ccut}

\]
In the definition above, the subscript in $\CCut_1$ indicates that the massive propagator of particle 1 is cut; $\CCut_2$ is defined analogously. 
 
\section{Technical simplifications}\label{sec:simp}

Although the full waveshape $\waveshape$, including quantum mechanical effects, can be computed using amplitudes, we are interested in the classical waveshape.
There are a wealth of classical simplifications we can take advantage of.
More specifically, we are interested in the classical limit of small angle scattering, often known as the ``post-Minkowski'' expansion in the gravitational context.
This is a relativistically covariant perturbative expansion of classical quantities --- in our case, of the classical radiation field.
A number of aspects of this expansion have been discussed in detail elsewhere, for example in references~\cite{Neill:2013wsa,Cheung:2018wkq,Kosower:2018adc,Herrmann:2021lqe,Bern:2021yeh}.
In this section, we highlight the most important tools we use to determine the classical limit of the waveshape, culminating with a summary of the strategy we propose for determining the final waveform.

\subsection{Hierarchy of momenta}\label{sec:hbar}

One simple-minded way to separate classical and quantum effects is to restore factors of $\hbar$: clearly all $\hbar$'s must disappear in classical expressions\footnote{This may involve absorbing $\hbar$ into quantities with classical meaning, such as the Land\'e $g$ factor.}, and quantum corrections will be suppressed by (dimensionless ratios involving) $\hbar$.
In this article, we follow the methods of KMOC~\cite{Kosower:2018adc} to restore $\hbar$.
For our purposes, the most important factors of $\hbar$ appear in the momenta of messengers.
Writing a generic messenger momentum as $q$ compared to a point-particle momentum $p$, we note that $q$ scales as $\hbar$,
\[
q^\mu = \hbar \bar{q}^\mu \,,
\]
while the momentum $p_i$ of particle $i$ scales as its mass:
\[
p_i^\mu = m_i u_i^\mu \,.
\]
Here we introduced a classical wavenumber $\bar q$ with dimensions of length, and the dimensionless classical proper velocity $u_i^\mu$.
The ratio of these, or rather of specific components, is of order the (reduced) Compton wavelength of the particle $\hbar / m_i$.
The classical approximation is valid only when the wavelengths of messengers, described by $\bar q$, are much larger than the Compton wavelengths of the particles.
Thus we treat messenger momenta as being very small compared to particle momenta, schematically
\[
\label{eq:hierarchyOfMomenta}
p_i \gg q \,.
\]
for each particle $i$.
Note that this can be implemented by treating the point particles as being very heavy, a point of view which is emphasised for example in references~\cite{Luna:2017dtq,Damgaard:2019lfh,  Aoude:2020onz,Haddad:2020tvs,Brandhuber:2021kpo,Brandhuber:2021eyq,Brandhuber:2021bsf}. 
Once the hierarchy~\eqref{eq:hierarchyOfMomenta} is understood, we can access the classical limit by simply Laurent expanding integrands in terms of variables suppressed by messenger momenta relative to particle momenta instead of explicitly restoring factors of $\hbar$. 

One complication with the classical limit of amplitudes is the appearance of apparently singular terms with too many inverse powers of $\hbar$. 
These terms are often colloquially called ``superclassical'' or ``hyperclassical'' terms and may lead to non-trivial Laurent expansions.
However throughout this paper this Laurent expansion will be essentially trivialised. 
The most dangerous superclassical terms are present only in the imaginary part, and (as we will see in section~\ref{vanishingcuts}) cancel directly at the level of cuts without requiring detailed computation. 
For the remainder, $\hbar$ power counting in a convenient gauge shows that all diagrams are classical at leading order in the $\hbar$ expansion.
Thus we will rarely need to make $\hbar$ explicit, and will rely instead on the inequality~\eqref{eq:hierarchyOfMomenta}.

\subsection{Classical waveshape and heavy-particle crossing}

In terms of the expectation~\eqref{eq:expectationDef} the waveshape can be written as
\begin{equation}
\label{eq:startCross}
\begin{split}
\waveshape = \int & \dd \Phi(p'_1, p'_2, p_1, p_2)\, \phi^*(p'_1, p'_2) \phi(p_1, p_2) \, e^{ib_1 \cdot (p_1 - p'_1)}e^{ib_2 \cdot (p_2 - p'_2)} 
\\ & \qquad \times i \E(p_1, p_2 \rightarrow p'_1, p'_2, k_\eta) \, \hdelta^D(p'_1 + p'_2 + k - p_1 - p_2).
\end{split}
\end{equation}
Now let's simplify this  expression in the classical limit. We will explicitly discuss the leading order approximation, because at tree level our manipulations will reveal a property of the tree amplitude in the classical limit which will be convenient below.
We will then deduce the more general case.

We will take the classical limit in two slightly different manners. 
First, write the ``outgoing'' momenta\footnote{As our observable is an expectation value, the apparent in and out states are both in states. Nevertheless it can be convenient at times to think of the primed momenta as outgoing.} as 
\[
p'_i = p_i + q_i \,.
\]
The momenta $q_i$ are messenger momenta, satisfying $k = -q_1 - q_2$.
Now, the on-shell phase space measure of the outgoing particle $i$ is
\[
\dd \Phi(p_i + q_i) = \hd^D q_i \, \Theta (p_i^0 + q_i^0) \, \hdelta(p_i^2 + 2 p_i \cdot q_i + q_i^2 - m_i^2) \,.
\]
We simplify this as follows. First, the energy $p_i^0$ is always much greater than $q_i^0$ in the classical region (since pair-production must be kinematically suppressed.) Therefore we replace the theta functions by unity. Next, we note that $p_i$ is an on-shell initial momentum so that $p_i^2 = m_i^2$. 
We further simplify the delta function noting that $q_i^2$ is suppressed relative to $p_i \cdot q_i$ because of the classical hierarchy~\eqref{eq:hierarchyOfMomenta}.
Finally, we may replace the expectation with the amplitude as they are equal at lowest order.
Thus, the leading-order waveshape becomes
\[
\waveshape = \int & \dd \Phi(p_1, p_2) \hd^D q_1 \hd^D q_2 \, \hdelta(2 p_1 \cdot q_1) \hdelta(2 p_2 \cdot q_2) \, \phi^*(p_1 + q_1, p_2 + q_1) \phi(p_1, p_2) \\
& \times e^{-ib_1 \cdot q_1}e^{-ib_2 \cdot q_2} \,
i\mathcal{A}_{5,0}(p_1, p_2 \rightarrow p_1 +q_1 , p_2 + q_2, k_\eta) \, \hdelta^D(k + q_1 + q_2)\,.
\]
Next, we simplify the wavefunctions by noting
\[
\phi(p_1 + q_1, p_2 + q_1) \simeq  \phi(p_1, p_2) \,.
\]
The origin of this fact~\cite{Kosower:2018adc} is that the messenger momenta are suppressed by a Compton wavelength relative to the particle momenta, and on this scale the wavefunctions must be rather flat for the position-space uncertainty to be negligible.
As a result, the leading-order waveshape is
\[
\label{eq:cross1}
\waveshape = \int & \dd \Phi(p_1, p_2) \hd^D q_1 \hd^D q_2 \, \hdelta(2 p_1 \cdot q_1) \hdelta(2 p_2 \cdot q_2) \, |\phi(p_1, p_2)|^2 \, e^{-ib_1 \cdot q_1}e^{-ib_2 \cdot q_2} \\
& \times i\mathcal{A}_{5,0}(p_1, p_2 \rightarrow p_1 +q_1 , p_2 + q_2, k_\eta) \, \hdelta^D(k + q_1 + q_2)\,.
\]

On the other hand, returning to equation~\eqref{eq:startCross} and instead integrating over unprimed variables via
\[
p_i = p'_i -q_i \,,
\]
we find, using the same logic,
\[
\label{eq:cross2}
\waveshape = \int & \dd \Phi(p'_1, p'_2) \hd^D q_1 \hd^D q_2 \, \hdelta(2 p'_1 \cdot q_1) \hdelta(2 p'_2 \cdot q_2) \, |\phi(p'_1, p'_2)|^2 \,
e^{-ib_1 \cdot q_1}e^{-ib_2 \cdot q_2} \\
& \times i\mathcal{A}_{5,0}(p'_1 -q_1 , p'_2 -q_2 \rightarrow p'_1, p'_2, k_\eta) \, \hdelta^D(k + q_1 + q_2)\,.
\]
There is nothing stopping us from dropping the primes in this equation, since $p'_i$ are simply variables of integration.

Comparing equations~\eqref{eq:cross1} and~\eqref{eq:cross2}, the only difference is in the details of the momentum dependence in the tree amplitude. 
The wavefunction is unspecified; we have only used properties it must have in the classical limit.
We conclude that
\[
\label{eq:classicalCross}
\mathcal{A}_{5,0}(p_1, p_2 \rightarrow p_1 +q_1 , p_2 + q_2, k_\eta) 
=
\mathcal{A}_{5,0}(p_1 - q_1, p_2 - q_2 \rightarrow p_1, p_2, k_\eta) \,.
\]
This equation only holds for the classical ``fragment'' of the amplitude, in the sense of reference~\cite{Cristofoli:2021jas}: at tree level, this fragment is defined to be the dominant term in the classical Laurent expansion.
An alternative perspective is that this crossing relation follows from the scale separation between the heavy-mass scale $m_1$ and $m_2$ in the momenta of the scalar particles, and the light scale of order $q$ in the messengers.
This decoupling is made manifest in heavy particle effective theories, which could also be used to compute these amplitudes.\

The result, then, is a kind of crossing relation valid for heavy particle effective theories. It essentially allows us to cross the messenger momentum leaving the large particle momentum untouched.
We will find this result is very useful in the next sub-section to cancel singular ``superclassical'' contributions. 
It is straightforward to check this heavy-particle crossing relation in explicit examples: 
the QED amplitude is visible in equation 5.46 of reference~\cite{Kosower:2018adc} while the gravitational five point case is written in equation 4.21 of reference~\cite{Luna:2017dtq}.
In both cases, heavy-particle crossing is achieved by eliminating the momentum $k$ in favour of $q_1 + q_2$, and then replacing $q_i \rightarrow -q_i$. This has the effect of replacing $p_i + q_i$ with the desired $p_i - q_i$ without clashing with the relation between $k$ and the $q_i$ (this relation does not pick up a sign in the crossing).

Returning to the waveshape beyond tree level, we may still simplify the phase-space measure and wavefunction exactly as discussed above. However, we must remember that the amplitude and the expectation differ beyond tree level.
So we may write 
\[\label{eq:genericWaveshape}
\waveshape = \KMOCav{\int \hd^D q_1 \hd^D q_2 \, \hdelta(2 p_1 \cdot q_1) \hdelta(2 p_2 \cdot q_2) \,& \hdelta^D(k + q_1 + q_2) \,e^{-ib_1 \cdot q_1}e^{-ib_2 \cdot q_2} 
\\ & \qquad\times i \E_5(p_1, p_2 \rightarrow p'_1, p'_2, k_\eta) },
\]
at LO and NLO, where the primed momenta are defined by $p_i' = p_i + q_i$.
The large angle brackets remind us of two points.
First, that the result must be integrated against the wavefunctions. However, once the integrand has been fully simplified in the classical limit, in particular to cancel terms involving singular powers of $\hbar$, the integrand is smooth on the scale of the wavepacket. 
We can therefore formally take the wavepacket size to zero, so that the wavepacket integral simply localises the incoming momenta $p_i$ on their classical values.
Second, the angle brackets are warning that the delta functions involving $2 p_i \cdot q_i$ can only be imposed after cancellation of singular (superclassical) terms.
An appropriate way to proceed is to use the exact on-shell conditions $2 p_i \cdot q_i \pm q_i^2 = 0$ at intermediate stages.
Once singular terms cancel, the hierarchy~\eqref{eq:hierarchyOfMomenta} allows us to drop irrelevant powers of $q_i^2$.

In essence, equation~\eqref{eq:genericWaveshape} reveals that the waveshape is an on-shell Fourier transform of the expectation. We expect this to be true to all orders of perturbation theory.
Because we will break the integrand of this Fourier transform (the expectation $\E$) into its real and imaginary parts, it is helpful to break the waveshape itself up into two corresponding parts. 
We define
\[\label{eq:genericWaveshapeRe}
\waveshape|_\textrm{Real} = \KMOCav{\int \hd^D q_1 \hd^D q_2 \, \hdelta(2 p_1 \cdot q_1) \hdelta(2 p_2 \cdot q_2) \,& \hdelta^D(k + q_1 + q_2) \,e^{-ib_1 \cdot q_1}e^{-ib_2 \cdot q_2} 
\\ & \times i \Re' \E(p_1, p_2 \rightarrow p'_1, p'_2, k_\eta) } \,,
\]
and
\[\label{eq:genericWaveshapeIm}
\waveshape|_\textrm{Im} = -\KMOCav{\int \hd^D q_1 \hd^D q_2 \, \hdelta(2 p_1 \cdot q_1) \hdelta(2 p_2 \cdot q_2) \,& \hdelta^D(k + q_1 + q_2) \,e^{-ib_1 \cdot q_1}e^{-ib_2 \cdot q_2} 
\\ & \times \Im' \E(p_1, p_2 \rightarrow p'_1, p'_2, k_\eta) } \,.
\]
Note that these are \emph{not} the real and imaginary parts of the waveshape (since the Fourier transform is complex).

\subsection{Vanishing superclassical cuts}\label{vanishingcuts}

Earlier, we advertised that our setup allows us to streamline the, possibly intricate~\cite{Kosower:2018adc}, cancellation of the cuts which have a singular $\hbar\to 0$ limit. 
We find that these are only contained in the imaginary part of the expectation $\E$, and more specifically only occur when two massive lines are cut.
That these are the only diagrams with superclassical singularities is easy to understand from the perspective of eikonal exponentiation, see particularly~\cite{Cristofoli:2021jas}.
The result of this subsection is that at the superclassical level
\[
\label{eq:twoMassiveCut}

\]
The result can be seen as a generalisation of the removal of iterated trees in an eikonal (or more general exponentiated form of the) amplitude.

To see how the cancellation works, we adjust the initial and final states under the integral signs to reach 
 \begin{figure}[H]
\begin{center}
    
 
\tikzset{
pattern size/.store in=\mcSize, 
pattern size = 5pt,
pattern thickness/.store in=\mcThickness, 
pattern thickness = 0.3pt,
pattern radius/.store in=\mcRadius, 
pattern radius = 1pt}
\makeatletter
\pgfutil@ifundefined{pgf@pattern@name@_02yvi6xbq}{
\pgfdeclarepatternformonly[\mcThickness,\mcSize]{_02yvi6xbq}
{\pgfqpoint{0pt}{0pt}}
{\pgfpoint{\mcSize+\mcThickness}{\mcSize+\mcThickness}}
{\pgfpoint{\mcSize}{\mcSize}}
{
\pgfsetcolor{\tikz@pattern@color}
\pgfsetlinewidth{\mcThickness}
\pgfpathmoveto{\pgfqpoint{0pt}{0pt}}
\pgfpathlineto{\pgfpoint{\mcSize+\mcThickness}{\mcSize+\mcThickness}}
\pgfusepath{stroke}
}}
\makeatother

 
\tikzset{
pattern size/.store in=\mcSize, 
pattern size = 5pt,
pattern thickness/.store in=\mcThickness, 
pattern thickness = 0.3pt,
pattern radius/.store in=\mcRadius, 
pattern radius = 1pt}
\makeatletter
\pgfutil@ifundefined{pgf@pattern@name@_xurj8a9h9}{
\pgfdeclarepatternformonly[\mcThickness,\mcSize]{_xurj8a9h9}
{\pgfqpoint{0pt}{0pt}}
{\pgfpoint{\mcSize+\mcThickness}{\mcSize+\mcThickness}}
{\pgfpoint{\mcSize}{\mcSize}}
{
\pgfsetcolor{\tikz@pattern@color}
\pgfsetlinewidth{\mcThickness}
\pgfpathmoveto{\pgfqpoint{0pt}{0pt}}
\pgfpathlineto{\pgfpoint{\mcSize+\mcThickness}{\mcSize+\mcThickness}}
\pgfusepath{stroke}
}}
\makeatother

 
\tikzset{
pattern size/.store in=\mcSize, 
pattern size = 5pt,
pattern thickness/.store in=\mcThickness, 
pattern thickness = 0.3pt,
pattern radius/.store in=\mcRadius, 
pattern radius = 1pt}
\makeatletter
\pgfutil@ifundefined{pgf@pattern@name@_jfwlrmva1}{
\pgfdeclarepatternformonly[\mcThickness,\mcSize]{_jfwlrmva1}
{\pgfqpoint{0pt}{0pt}}
{\pgfpoint{\mcSize+\mcThickness}{\mcSize+\mcThickness}}
{\pgfpoint{\mcSize}{\mcSize}}
{
\pgfsetcolor{\tikz@pattern@color}
\pgfsetlinewidth{\mcThickness}
\pgfpathmoveto{\pgfqpoint{0pt}{0pt}}
\pgfpathlineto{\pgfpoint{\mcSize+\mcThickness}{\mcSize+\mcThickness}}
\pgfusepath{stroke}
}}
\makeatother

 
\tikzset{
pattern size/.store in=\mcSize, 
pattern size = 5pt,
pattern thickness/.store in=\mcThickness, 
pattern thickness = 0.3pt,
pattern radius/.store in=\mcRadius, 
pattern radius = 1pt}
\makeatletter
\pgfutil@ifundefined{pgf@pattern@name@_c569v4ybj}{
\pgfdeclarepatternformonly[\mcThickness,\mcSize]{_c569v4ybj}
{\pgfqpoint{0pt}{0pt}}
{\pgfpoint{\mcSize+\mcThickness}{\mcSize+\mcThickness}}
{\pgfpoint{\mcSize}{\mcSize}}
{
\pgfsetcolor{\tikz@pattern@color}
\pgfsetlinewidth{\mcThickness}
\pgfpathmoveto{\pgfqpoint{0pt}{0pt}}
\pgfpathlineto{\pgfpoint{\mcSize+\mcThickness}{\mcSize+\mcThickness}}
\pgfusepath{stroke}
}}
\makeatother
\tikzset{every picture/.style={line width=0.75pt}} 



\end{center}
 \end{figure}
\hspace{-.75cm}
Writing out the cut at this superclassical order, we arrive at 
\[
\label{eq:zerocut}
\lexp \int & \hd^D q_1 \hd^D q_2  \, \hdelta(2 p_1 \cdot q_1) \hdelta(2 p_2 \cdot q_2) \hdelta^D(k+q_1 +q_2) \,  e^{-i b_1 \cdot q_1} e^{-i b_2 \cdot q_2} \\
& \times \int  \hd^D \ell_1 \hd^D \ell_2 \; \hdelta(2 p_1 \cdot \ell_1) \hdelta(2 p_2 \cdot \ell_2) \,  \, \hdelta^D (\ell_1 + \ell_2+k) \\
&[\mathcal{A}_{5,0} (p_1, p_2 \rightarrow p_1 + \ell_1, p_2 + \ell_2, k) \mathcal{A}_{4,0}(p_1 + \ell_1, p_2 + \ell_2 \rightarrow p_1 + q_1, p_2 + q_2) \\
&\;- \mathcal{A}_{4,0} (p_1 - q_1, p_2 -q_2 \rightarrow p_1 - \ell_1, p_2 - \ell_2) \mathcal{A}_{5,0} (p_1 - \ell_1, p_2 -\ell_2 \rightarrow p_1, p_2, k) ] \rexp \,.
\]
We recognise that the two five-point amplitudes appearing above are equal using equation~\eqref{eq:classicalCross}.
As for the four-point trees, one could make a similar use of heavy-particle crossing to show that they match.
Alternatively, it is a simple point that these tree-level four-point amplitudes, in the classical limit, only depend on the usual Mandelstam $s$ and $t$ variables which are the same in both terms.
Thus, we conclude that the cut in equation~\eqref{eq:zerocut} vanishes.
This conclusion was confirmed by a direct computation in both QED and QCD
using Mathematica and 
the results of references~\cite{Carrasco:2020ywq} and~\cite{Carrasco:2021bmu}. 
(Our Mathematica code is discussed in some more detail in section~\ref{sec:qcd}.)
Furthermore, this cancellation makes intuitive sense from the perspective of eikonal exponentiation~\cite{Cristofoli:2021jas,Britto:2021pud}.

In appendix \ref{cutextra} we discuss equation~\eqref{eq:twoMassiveCut} at the classical order.
The result is non-vanishing~\cite{Caron-Huot:2023vxl}.
As explained in detail in reference~\cite{Caron-Huot:2023vxl} these cuts arise from conservative forces (geodesic motion or the Lorentz force) and are unrelated to radiation reaction.

\subsection{Vanishing integrals}
\label{sec:vanishing}

In our one-loop computations, we will encounter topologies including pentagons, boxes, triangles etc.
Here we largely work at the level of the integrand.
Nevertheless it is very useful to simplify our integrand by dropping terms which integrate to zero.

The situation with loop integrals in the classical limit at four points and one or two loops is very well understood and is thoroughly discussed for example in references~\cite{Parra-Martinez:2020dzs,Herrmann:2021lqe}.
There are some similarities between four and five points.
For example, we note that
\[
\int \hd^D \ell \frac{(\ell - q_1)^2}{\ell^2 (\ell -q_1)^2 (p_1 \cdot \ell ) (p_2 \cdot \ell) } =
\frac{1}{m_1 m_2} \int \hd^D \ell \frac{(\ell - q_1)^2}{\ell^2 (\ell -q_1)^2 (u_1 \cdot \ell ) (u_2 \cdot \ell) } =
0\,.
\]
One viewpoint is that this occurs because the integral is scaleless in dimensional regulation (all the $q_1$ dependence goes away). 
An alternative viewpoint is that the integral is irrelevant classically with any choice of regulator. 
This is because, when integrated against the measure in equation~\eqref{eq:genericWaveshape}, one finds a factor $\delta^{D-2}(\v b)$.
This spatial contact term vanishes in the domain of validity of the classical approximation~\cite{Kosower:2018adc}.

As another example, consider the integral
\[
\int \hd^D \ell \frac{\ell^2}{\ell^2 (\ell -q_1)^2 (u_1 \cdot \ell ) (u_2 \cdot \ell) } &= \int \hd^D \ell \frac{1}{(\ell -q_1)^2 (u_1 \cdot \ell ) (u_2 \cdot \ell) } \\
&= \int \hd^D \ell \frac{1}{\ell^2 (u_1 \cdot \ell ) (u_2 \cdot (\ell + q_1)) } \,.
\]
In the second step, we simply set $\ell' = \ell - q_1$, and then dropped the prime. 
We also set $u_1 \cdot q_1 = 0$, assuming that the $\hbar$-suppressed correction term of order $q_1^2$ could be neglected; this is always the case in the main text of this paper as classical singularities are only present in two-massive particle cuts (see the previous sub-section).
This integral is \emph{not} scaleless: indeed, there is a scale $u_2 \cdot q_1 = - u_2 \cdot k$ in the integral.
Nevertheless we may still drop this integral:
\[
\int \hd^D \ell \frac{\ell^2}{\ell^2 (\ell -q_1)^2 (u_1 \cdot \ell ) (u_2 \cdot \ell) }
=
\int \hd^D \ell \frac{1}{\ell^2 (u_1 \cdot \ell ) (u_2 \cdot (\ell-k)) }
\rightarrow 0 \,.
\]
As before, the crucial point is that the integral does not depend on $q_1$ but only on $k$.
Therefore, on integration against the measure in equation~\eqref{eq:genericWaveshape}, it again leads to a contact term in $\v b$ space {since the measure and the integral are both independent of $q_1$}. Note that care must be taken for example in the context of pentagon diagrams with three massless internal propagators; pinching one of these need not necessarily lead to a vanishing contact term.

\subsection{Real parts from single cuts and principal values}\label{sec:PV}

It is useful to think of the waveshape as a Fourier transform of the expectation $\E$ as shown in equation~\eqref{eq:genericWaveshape}.
The expectation naturally has real and imaginary parts. 
Following our discussion in section~\ref{sec:cutAndIm}, the imaginary part of the expectation can be found from computing unitarity (two-particle) cuts.
In this section we will describe how to compute the real part of the expectation\footnote{As we saw earlier in section~\ref{impart}, the real part of the expectation equals the real part of the amplitude.}, in the classical limit, from single-particle cuts.

At one loop, the real part of the expectation~\eqref{eq:realExpect} is
\[\label{eq:realPart1}
2\Re' \E_{5,1}(p_1, p_2 \rightarrow & p_1', p_2', k_\eta) \\
= 
&\mathcal{A}_{(5,1)}(p_1, p_2 \rightarrow p'_1, p'_2, k_\eta) + \mathcal{A}^*_{(5,1)}(p'_1, p'_2, k_{-\eta} \rightarrow p_1, p_2) \,.
\]
For ease of discussion we introduce a short-hand notation for the two five-point amplitudes appearing here:
\[
 \mathcal{A}_{5, 1}^{(a)} &\equiv \mathcal{A}(p_1, p_2 \rightarrow p'_1, p'_2, k_\eta)\,, \\
 \mathcal{A}_{5,1}^{(b)} &\equiv \mathcal{A}(p'_1, p'_2, k_{-\eta} \rightarrow p_1, p_2) \,.
\]
Notice that the initial and final states are swapped in $\mathcal{A}_{5,1}^{(b)}$ relative to $\mathcal{A}_{5, 1}^{(a)}$.
The close relationship between the conjugated amplitude $\mathcal{A}_{5,1}^{(b)*}$ and the amplitude $\mathcal{A}_{5,1}^{(a)}$
is discussed in many quantum field theory textbooks, though the focus in typically on the imaginary part $-i (\mathcal{A}_{5,1}^{(a)} -\mathcal{A}_{5,1}^{(b)*} )/2$ because of its relevance to unitarity (see, for example,~\cite{tHooft:1973wag,Sterman:1993hfp,Srednicki:2007qs,Schwartz:2014sze} for helpful discussions in this particular context).
Because the real part $(\mathcal{A}_{5,1}^{(a)}+\mathcal{A}_{5,1}^{(b)*})/2$ is relevant to us, it is worth giving an example to see how the combination works.

We consider a one-loop diagram contributing to the amplitude $\mathcal{A}_{5,1}^{(a)}$ in Yang-Mills theory\footnote{We normalise our coupling so that the covariant derivative is $D_\mu = \partial_\mu - i g A_\mu$.}:  
\[

\]
This diagram depends on a color factor $\mathcal{C}$, a kinematic numerator $N$ and a propagator structure $P$. 
The Feynman rules lead to
\[
P^{-1} &= (\ell^2 + i \epsilon) [( q_1-\ell)^2 + i \epsilon] [(p_1 + \ell)^2 - m_1^2 + i \epsilon] [ (p_2 - \ell)^2 - m_2^2 + i \epsilon]  \\
& \hspace{220pt}\times [(p_2 - \ell - k)^2 - m_2^2 + i \epsilon] \,, \\
N &= \varepsilon^*_\eta \cdot (2 p_2 -2 \ell) (2 p_1 + \ell) \cdot (2 p_2 - \ell) (2p_1 + \ell + q_1) \cdot (2 p_2 - \ell +q_1 + 2 q_2) \,.
\]
Note the appearance of the (possibly complex) polarisation vector $\varepsilon^*_\eta$ of the outgoing messenger of momentum $k$.  
To describe the color factor, we suppose the initial color of particle $i$ is specified by a color vector $\chi_i$, while another vector $\chi_i'$ defines the final color. Let us further suppose that the outgoing gluon has adjoint color $a$. Then we have
\[
\mathcal{C} = \left(\bar \chi_1' \cdot T_1^b \cdot T_1^c \cdot \chi_1 \right) \left( \bar \chi_2' \cdot T_2^b \cdot T_2^a \cdot T_2^b \cdot \chi_2 \right) \,.
\]
The contribution of this diagram to the amplitude $\mathcal{A}_{5,1}^{(a)}$ is
\[
\label{eq:firstcontrib}
i  g^5 \mathcal{C} \int \hd^D \ell \, N P \,, 
\]
since (in our conventions) the Feynman rules evaluate to $i$ times the amplitude\footnote{This is consistent with $S = 1 + i T$, and the convention that, for example, the tree four-point amplitude in $\lambda \phi^4/4!$ theory is $\lambda$.}.

As the initial and final states are interchanged in $\mathcal{A}_{5,1}^{(b)}$, 
we instead encounter the diagram  
\begin{figure}[H]
\begin{center}

\tikzset{every picture/.style={line width=0.75pt}} 



\end{center}
\end{figure}
\hspace{-0.75cm}
The color factor, numerator and propagators are now
\[
P'^{-1} &= (\ell^2 + i \epsilon) [( q_1-\ell)^2 + i \epsilon] [(p_1 + \ell)^2 - m_1^2 + i \epsilon] [ (p_2 - \ell)^2 - m_2^2 + i \epsilon]  \\
& \hspace{220pt}\times [(p_2 - \ell - k)^2 - m_2^2 + i \epsilon] \,, \\
N' &= \varepsilon_\eta \cdot (2 p_2 -2 \ell) (2 p_1 + \ell) \cdot (2 p_2 - \ell) (2p_1 + \ell + q_1) \cdot (2 p_2 - \ell +q_1 + 2 q_2) \,, \\
\mathcal{C}' &= \left( \bar \chi_1 \cdot T_1^c \cdot T_1^b \cdot \chi'_1 \right) \left( \bar \chi_2 \cdot T_2^c \cdot T_2^a \cdot T_2^b \cdot \chi'_2 \right) \,.
\]
It is important that $N'$ is the complex conjugate of $N$, and $\mathcal{C}'$ is the complex conjugate of $\mathcal{C}$
while the propagator structures are equal: $P = P'$.
As a consequence, the contribution of this diagram to $\mathcal{A}_{5,1}^{(b)*}$ is
\[
-i  g^5 \mathcal{C} \int \hd^D \ell \, N P^* \,,
\]
which can be compared to the expression~\eqref{eq:firstcontrib}.
This is a general fact~\cite{tHooft:1973wag}: the one-loop Feynman diagrams contributing to $\mathcal{A}_{5,1}^{(b)*}$ can be obtained from the diagrams for $\mathcal{A}_{5,1}^{(a)}$ by (i) changing the overall sign, and (ii) replacing the $i \epsilon$ prescription in propagators by $-i \epsilon$.
For the two diagrams at hand, their net contribution to $\mathcal{A}_{5,1}^{(a)}+\mathcal{A}_{5,1}^{(b)*}$ is
\[ 
-g^5 \mathcal{C} \int \hd^D \ell \, N \left[ -i (P - P^*) \right]
= -2 g^5 \mathcal{C} \int \hd^D \ell \, N \Im P \,.
\]
The instruction above tells us to take the imaginary part of the propagator structure of the amplitudes. 

It is very natural to obtain the imaginary part of the propagator structure using
\[\label{plstot}
\frac{1}{p^2 - m^2 + i \epsilon} = \Principal \left(\frac{1}{p^2 - m^2}\right) - \frac{i}{2} \hdelta(p^2 - m^2) \,,
\]
where $\Principal$ is the principal value. 
The delta function here is equivalent to cutting a single particle.
By counting powers of $i$, it is clear that the imaginary part of our propagator structure is obtained by cutting an odd number of propagators. (This contrasts with the usual unitarity cuts at one loop which involve cutting two propagators.)

Our diagrams contain five propagators, so in principle there are imaginary parts when we cut one, three or five propagators. 
We drop all terms involving three point amplitudes; in Minkowski space these amplitudes (involving two massive particles and a messenger) only have support at vanishing messenger momentum and consequently we set them to zero.
Thus there is no need to consider cutting five propagators. It is also easy to see that cutting three propagators necessarily leads to one three-point amplitude. 
So, as advertised above, we see that only single cuts are relevant. 

Our one-loop diagrams involve both massive and massless propagators.
However, we find that single-cuts of massless lines do not contribute clasically.
This may seem obvious to many, as from a purely classical perspective the messengers are Fourier transforms of the field modes.
In familar situations these modes are sourced by the Coulomb field (potential modes) and therefore they cannot transport energy in the rest frame of the source. 
Thus, they cannot go on shell.
We caution that this intuition is unreliable beyond tree level as the modes of a particle's radiation field certainly go on shell.
So to demonstrate that massless single cuts do not contribute to the one-loop waveshape, we power-counted factors of $\hbar$.
We found that the single cuts of massless lines are suppressed by a power of $\hbar$ relative to the classical scaling.
(This involves choosing a specific gauge for the polarisation objects of the messengers and a remaining cancellation among Feynman diagrams.) 
We shall omit this class of cuts in the remainder of this article.
Single cuts of massive lines do contribute classically.

As a result we conclude that 
\[
\label{eq:RealViaSCut}
\Re' \E_{5,1}(p_1, p_2 \rightarrow p_1', p_2', k_\eta) = \SCut_1 &\mathcal{A}_{5,1}(p_1, p_2 \rightarrow p'_1, p'_2, k_\eta) \\
&+ \SCut_2 \mathcal{A}_{5,1}(p_1, p_2 \rightarrow p'_1, p'_2, k_\eta) \,,
\]
where the operator $\SCut_i$ instructs us to compute the single-cut on the $i$th massive line in the amplitude; all other propagators are then to be evaluated with the principal-value pole prescription~\eqref{plstot}\footnote{We emphasise that the $\Principal$ prescription appears naturally from general considerations.}.
We normalise the $\SCut$ operation specifically as
\[
\label{eq:SCutMeaning}
\SCut_i \frac{1}{2 p_i \cdot l + i \epsilon} = -\frac{1}{2} \hdelta(2 p_i \cdot \ell) \,
\]
in view of equation~\eqref{plstot}.

Finally, in the  spirit of equation \eqref{ccut}, we can characterise the real-part contributions  in a diagrammatic fashion as follows
\[

\]
Above the intermediate (red) massive propagator of particle 1 is on shell. 
The messenger lines are in principle off-shell, but as the integrals vanish if they are pinched (following the discussion in section~\ref{sec:vanishing}) they can be treated as on shell.
The situation for $\SCut_2$ is analogous.

\subsection{Overview of the Classical Waveform}

Before we begin to discuss explicit examples, it may be helpful to summarise our strategy for computing waveforms.

In equations~\eqref{eq:laterUse} and~\eqref{riemmm}, we saw that the YM field strength and Riemann curvature can be extracted from one-dimensional Fourier transforms of the waveshape, defined in equation~\eqref{eq:waveshapeDef1} to be
\[
\waveshape = \braket{\psi | S^\dagger a_\eta(k) S | \psi}\,.
\]
As is clear from its definition, the waveshape is the expectation value of a quantum-mechanical operator. 
Because we wish to extract classical waveforms, we use a state $\ket{\psi}$ which has a sensible classical limit (describing two well-separated, different, point-like classical particles in the far past.)
Since the state can be written as an appropriate integral over plane-wave states (as shown in equation~\eqref{eq:initialState}) the waveshape itself can be written as an integral over a plane-wave in-in expectation $\E$ as shown in equation~\eqref{eq:expectationDef}.
At tree level, this $\E$ is simply the five-point tree amplitude.
At NLO the situation is more involved.
The real part of the one-loop expectation is still the real part of the one-loop amplitude.
We have seen (equation~\eqref{eq:RealViaSCut}) that these real parts can be computed by single-cuts of the massive internal lines.

The imaginary part of the one-loop expectation differs from the imaginary part of the amplitude. It is given by 
\[\label{eq:imEcuts}    

\]
Once the expectation $\E$ is determined, the waveshape can be computed by integrating over the details of the state $\ket{\psi}$. 
In the classical limit, this simplifies to another Fourier transform, as shown in equation~\eqref{eq:genericWaveshape}.

So the algorithm is: first, compute the real part of $\E$ by performing single cuts. Second, determine the imaginary parts from the unitarity cuts in equation~\eqref{eq:imEcuts}.
It remains then to perform the Fourier transforms in equation~\eqref{eq:genericWaveshape}.
The time-domain waveform finally requires one more Fourier transform~\eqref{riemmm}.

\section{Radiation}\label{radiation}

In this section we turn to the computation of explicit waveforms. 
Following our algorithm, we begin with the real part $\Re' \E$ of the QED and QCD waveshape. 
The QED case is an excellent starting point: 
it is simple yet interesting, 
and closely connected to more complicated radiation fields.
It is also simple enough that we may check our work against a completely orthogonal computation in classical field theory (see section~\ref{sec:confirm}).
Thus we begin with details in QED before delving into QCD.

\subsection{QED}\label{examplecalc}

NLO waveshapes are fifth order in the coupling.
In electrodynamics we are free to give our two particles different charges $Q_1$ and $Q_2$, and correspondingly those five coupling powers
can be decomposed into four different charge sectors: $Q_1 Q_2^4$, $Q_1^2 Q_2^3$, $Q_1^3 Q_2^2$, and $Q_1^4 Q_2$. 
As is intuitively clear, there is no radiation field at order $Q_1^5$ or $Q_2^5$ since at least one photon must connect the two particles for on-shell radiation of non-zero energy to occur. 
In the language of scattering amplitudes, the one-loop five-point amplitude in QED can be decomposed into four different partial amplitudes corresponding to these four charge sectors\footnote{As the real part of the expectation is the same as the real part of the amplitude, we can discuss either. In this section we use the more familiar terminology of amplitudes.}.
There are really only two independent partial amplitudes to compute, which we can take to be the $Q_1^2 Q_2^3$ and $Q_1 Q_2^4$ amplitudes. 
The $Q_1^3 Q_2^2$ and $Q_1^4 Q_2$ partial amplitudes can be recovered by interchanging particles 1 and 2.

First, a comment. 
Throughout this section, we will omit certain cuts that could, in principle, contribute to the real part of the amplitude, but which are intuitively quantum-mechanical. 
The omitted cuts correspond to the real part of one-particle irreducible (1PI) vertex corrections and are contained in the $Q_1 Q_2^4$ and $Q_1^4 Q_2$ charge sectors. These cuts are responsible for ultraviolet divergences and must be treated by renormalisation in an appropriate scheme. In section \ref{rennn2} we will discuss the renormalisation of this class of diagrams, and demonstrate that they do not contribute to the real part of the classical waveform. 
This justifies their omission in the present section.
In QED, this entirely removes the real part of the $Q_1 Q_2^4$ partial amplitude.
We emphasise that this point does not apply to the imaginary part of the amplitude (and of the expectation) which is UV finite and is unaffected by renormalisation. 

It remains, then, to compute the real part of the $Q_1^2 Q_2^3$ partial amplitude.
By equation~\eqref{eq:RealViaSCut}, this involves computing the single cut of one-loop five-point diagrams.
As a warm-up we first look at a related computation: single cuts of one-loop four-point diagrams (involving our usual two massive particles, now connecting at one loop by massless messengers).

\subsubsection*{Single cuts at four points}

The discussion of section~\ref{sec:PV} applies essentially unchanged to four-point one-loop amplitudes, with the conclusion
\[
\Re \mathcal{A}_{4,1}(s, t) = \SCut \mathcal{A}_{4,1}(s, t) \,.
\]
As these four-point amplitudes are relevant to the impulse~\cite{Kosower:2018adc}, single cuts have a role in that observable. 
Indeed, equation 6.13 of reference~\cite{Herrmann:2021tct} shows that a particular (transverse) part of the NLO impulse is determined by the real part of the amplitude.
Thus we will be able to compare our results directly to the literature.

As in the five point case, only single-cuts of massive lines are relevant; in this context, that means we may cut either an internal propagator for line 1 or 2. 
We refer to these possibilities as $\SCut_1 \mathcal{A}_{4,1}$ (line 1 cut) or $\SCut_2 \mathcal{A}_{4,1}$ (line 2 cut).

First, we consider the result of placing the propagator for line 2 on shell.
Diagrammatically, we must then consider 
 \[\label{eq:4ptSCut}   

\]
Using equation~\eqref{eq:SCutMeaning} the explicit expression is
\[
\SCut_2 \mathcal{A}_{4,1}  = \int \hd^D \ell \frac{1}{\ell^2 (\ell-q)^2} \frac12 \hdelta ( 2 p_2 \cdot \ell)  N(\ell) \,,
\]
where $N(\ell)$ is a numerator function we must fix.
Notice the explicit photon propagators: following the logic of section~\ref{sec:vanishing} we know that if $N(\ell)$ contains any terms which cancel these propagators, the result makes a vanishing contribution to the impulse. 
Thus, for the purposes of computing $N(\ell)$ we can proceed by taking $\ell^2 = 0 = (\ell-q)^2$.
In other words we can take each of the blobs in equation~\eqref{eq:4ptSCut} to be on-shell amplitudes, so that
\[
N(\ell) = \sum_{\textrm{helicities}} \mathcal{A}_{3,0}(p_2, \ell) \mathcal{A}_{4,0}(p_1, \ell, \ell - q) \mathcal{A}_{3,0}(p_2 - \ell, \ell -q) \,. 
\]
In $D$ dimensions, the helicity sum is straightforward using formal polarisation vectors. Let us write the polarisation vector for a photon of momentum $k$ and gauge $q$ as $\varepsilon(k; q)$ (we will often suppress explicit indication of the gauge choice, writing $\varepsilon(k)$).
If we choose the gauge to be $q = p_1$, then the Compton amplitude appearing in the cut is
\[
\mathcal{A}_{4,0}(p_1, \ell, \ell - q) = 2 Q_1^2 \, \varepsilon(\ell, p_1) \cdot \varepsilon^*(\ell-q, p_1) \,.
\] 
The three-point amplitudes are trivially obtained from 
\[
\mathcal{A}_{3,0}(p_2, \ell) = 2 Q_2 \, \varepsilon^*(\ell, p_1) \cdot p_2 \,.
\]
To perform the helicity sum, we only need the completeness relation which, in case of a massive gauge vector, is
\[
\label{eq:massiveCompleteness}
\sum_{\textrm{helicities}} \varepsilon^\mu(k, q) \varepsilon^{\nu*}(k, q) = - \left( \eta^{\mu\nu} - \frac{k^\mu q^\nu + k^\nu q^\mu}{k \cdot q} + q^2 \frac{k^\mu k^\nu}{(k \cdot q)^2} \right) \,.
\]
This summation involves products of a polarisation vector and its conjugate. 
As usual in generalised unitarity, this structure naturally arises in the product of amplitudes appearing in the real part diagram above
because a photon connecting two amplitudes must be outgoing with respect to one amplitude and incoming with respect to the other.

It then follows that the numerator is
\[
\label{eq:impulseNumerator}
N(\ell) = 8 Q_1^2 Q_2^2 \left( m_2^2 + \frac{(p_1 \cdot p_2)^2}{(p_1 \cdot \ell)^2} \ell \cdot (\ell-q) \right) \,,
\]
and the single-cut is
\[
\label{eq:impulsep2}
\SCut_2 \mathcal{A}_{4,1} = 
2 Q_1^2 Q_2^2 \, m_1 m_2
\int \frac{\hd^D \ell}{\ell^2 (\ell-q)^2} \left(\frac{\hdelta(u_2 \cdot \ell)}{m_1}  \left(1 + \frac{(u_1 \cdot u_2)^2}{(u_1 \cdot \ell)^2} \ell \cdot (\ell-q) \right)  
\right) \,.
\]
Above, we used the proper velocities $u_i = p_i / m_i$. 

A few comments are in order.
First, notice that our gauge choice for the polarisation vectors has allowed us to completely bypass superclassical terms in the amplitude. 
This occurred because the condition $\varepsilon \cdot p_1 = 0$ removed the diagrams which have superclassical scaling at leading order, while the polarisation sum \eqref{eq:massiveCompleteness} recovers the subleading (ie classical) terms from these diagrams. 
Second, the expression on the right-hand-side of equation~\eqref{eq:impulsep2} can be recognised in 
the impulse given in equation 5.38 of reference~\cite{Kosower:2018adc} (as can the corresponding $\SCut_1 \mathcal{A}_{4,1}$).
Significantly more labour was required to find the same result in that reference\footnote{Terms in that equation 5.38 which involve derivatives of delta functions arise from the imaginary part of the amplitude.}.
Third, following the discussion above equation~\eqref{eq:RealViaSCut}, we did not include contributions from single-cuts of massless lines.
It is actually more obvious that massless single-cuts do not arise in the approach of~\cite{Kosower:2018adc}.
In that work, individual massive propagators were shown to combine into delta functions, performing a single cut, using a symmetry of the loop integrals which left the photon propagators invariant.
So there was never any possibility of encountering massless single cuts.
However, it did require effort to understand the emergence of the single cuts for the massive lines.
The central advantage of our current approach is that this happens automatically.

\subsubsection*{Real part, cutting line 1}

We are now prepared to compute the real part of the $Q_1^2 Q_2^3$ amplitude
by recycling much of the four-point single cut computation. 
We start by cutting line 1
leading to the diagram 
\[\label{figurina}

\]
We will soon see that this diagram gives the dominant contribution to the waveshape when the mass $m_1$ of particle 1 is large.
The main novelty relative to our discussion of the four-point single cut is the appearance of a five-point tree amplitude\footnote{Here we take the external photon to be outgoing. The direction of the internal photons is irrelevant since we ultimately sum over helicities when sewing the amplitudes.}: 

\begin{figure}[H]
\begin{center}
    
 
\tikzset{
pattern size/.store in=\mcSize, 
pattern size = 5pt,
pattern thickness/.store in=\mcThickness, 
pattern thickness = 0.3pt,
pattern radius/.store in=\mcRadius, 
pattern radius = 1pt}
\makeatletter
\pgfutil@ifundefined{pgf@pattern@name@_mcit9ylfw}{
\pgfdeclarepatternformonly[\mcThickness,\mcSize]{_mcit9ylfw}
{\pgfqpoint{0pt}{0pt}}
{\pgfpoint{\mcSize+\mcThickness}{\mcSize+\mcThickness}}
{\pgfpoint{\mcSize}{\mcSize}}
{
\pgfsetcolor{\tikz@pattern@color}
\pgfsetlinewidth{\mcThickness}
\pgfpathmoveto{\pgfqpoint{0pt}{0pt}}
\pgfpathlineto{\pgfpoint{\mcSize+\mcThickness}{\mcSize+\mcThickness}}
\pgfusepath{stroke}
}}
\makeatother

 
\tikzset{
pattern size/.store in=\mcSize, 
pattern size = 5pt,
pattern thickness/.store in=\mcThickness, 
pattern thickness = 0.3pt,
pattern radius/.store in=\mcRadius, 
pattern radius = 1pt}
\makeatletter
\pgfutil@ifundefined{pgf@pattern@name@_f95hno7q9}{
\pgfdeclarepatternformonly[\mcThickness,\mcSize]{_f95hno7q9}
{\pgfqpoint{0pt}{0pt}}
{\pgfpoint{\mcSize+\mcThickness}{\mcSize+\mcThickness}}
{\pgfpoint{\mcSize}{\mcSize}}
{
\pgfsetcolor{\tikz@pattern@color}
\pgfsetlinewidth{\mcThickness}
\pgfpathmoveto{\pgfqpoint{0pt}{0pt}}
\pgfpathlineto{\pgfpoint{\mcSize+\mcThickness}{\mcSize+\mcThickness}}
\pgfusepath{stroke}
}}
\makeatother
\tikzset{every picture/.style={line width=0.75pt}} 



\end{center}
\end{figure}

\hspace{-0.75cm}If we choose the gauge of both polarisation vectors to be $p_2$, there are only three possible Feynman diagrams leading to a compact and (for our purposes) convenient expression for the amplitude:
\[
\label{eq:5ptTree}
\mathcal{A}_{5,0} = - 4 Q_2^3 &\left[ \frac{\varepsilon^*(\ell) \cdot \varepsilon(\ell - q_1) \, \varepsilon^*(k) \cdot q_1}{2 p_2 \cdot q_1} - \frac{\varepsilon^*(\ell) \cdot \varepsilon^*(k) \, \varepsilon(\ell - q_1) \cdot q_2}{2 p_2 \cdot  (\ell - q_1)} 
\right. \\
& \hspace{150pt}\left. + \frac{\varepsilon(\ell - q_1) \cdot \varepsilon^*(k) \, \varepsilon^*(\ell) \cdot q_2}{2 p_2 \cdot \ell}   \right] \,.
\]
Notice that the second and third terms are related by swapping the momenta $\ell$ and $q_1 - \ell$.
This is a symmetry of the rest of the diagram in equation~\eqref{figurina}, so these last two terms in the five-point tree make an identical contribution in the cut.
We do not indicate the helicity of the polarisation vectors: this information washes out in the completeness relation (since each polarisation vector in the product of amplitudes is multiplied by its conjugate polarisation).

To determine this single cut, we must sum the product of the five-point tree~\eqref{eq:5ptTree} and two three-point amplitudes over helicities.
The helicity sum can be performed using the completeness relation of equation~\eqref{eq:massiveCompleteness}.
Because the last two terms in the five-point tree~\eqref{eq:5ptTree} make an identical contribution to the cut, there are only two different polarisation sums to consider.
The first term in equation~\eqref{eq:5ptTree} leads to the sum
\[
\sum_{\textrm{helicities}} p_1 \cdot \varepsilon(\ell) \, p_1 \cdot \varepsilon^*(\ell - q_1) & \frac{\varepsilon^*(\ell) \cdot \varepsilon(\ell - q_1) \, \varepsilon^*(k) \cdot q_1}{2 p_2 \cdot q_1} \\
&  = \left( m_1^2 + \frac{\ell \cdot (\ell - q_1) (p_1 \cdot p_2)^2}{p_2 \cdot \ell \, p_2 \cdot (\ell - q_1)} \right)  \frac{\varepsilon^*(k) \cdot q_1}{2 p_2 \cdot q_1} \,.
\]
Notice that this term --- specifically, the part appearing in brackets --- bears a strong structural similarity with the numerator which appeared in the impulse, equation~\eqref{eq:impulseNumerator}.
The relationship between radiation and the impulse is an example of the ``memory'' effect, encountered here at the level of the one-loop integrand.

The second class of polarisation sum to be performed is
\[
& \sum_{\textrm{helicities}} p_1 \cdot \varepsilon(\ell) \, p_1 \cdot \varepsilon^*(\ell - q_1) \frac{\varepsilon(\ell - q_1) \cdot \varepsilon^*(k) \, \varepsilon^*(\ell) \cdot q_2}{2 p_2 \cdot \ell} \\
&= \left( p_1 \cdot q_2 {-} \frac{p_1 \cdot p_2 \, \ell \cdot q_2}{p_2 \cdot \ell} \right) \left( p_1 \cdot \varepsilon^*(k) - \frac{p_1 \cdot p_2}{p_2 \cdot (\ell-q_1)} (\ell-q_1) \cdot \varepsilon^*(k) \right) \frac{1}{2 p_2 \cdot \ell} \,.
\]
Putting these together, the cut of equation~\eqref{figurina} is given by
\[
\label{eq:heavySingleCutQED}
\SCut_1 \mathcal{A}_{5, 1}^{Q_1^2 Q_2^3} = - 2 Q_1^2 Q_2^3 \int \frac{\hd^D \ell}{\ell^2 (\ell-q_1)^2} \hat{\delta}(p_1\cdot \ell) \left[ \left( m_1^2 + \frac{\ell \cdot (\ell - q_1) (p_1 \cdot p_2)^2}{p_2 \cdot \ell \, p_2 \cdot (\ell - q_1)} \right)  \frac{\varepsilon^* \cdot q_1}{p_2 \cdot q_1} \right . \\
\left. +  \left( p_1 \cdot q_2 {-} \frac{p_1 \cdot p_2 \, \ell \cdot q_2}{p_2 \cdot \ell} \right) \left( p_1 \cdot \varepsilon^* - \frac{p_1 \cdot p_2}{p_2 \cdot (\ell-q_1)} (\ell-q_1) \cdot \varepsilon^* \right) \frac{2}{p_2 \cdot \ell}
\right]   \,,
\]
where the remaining polarization vector is that of the outgoing photon (in the $p_2$ gauge).

We note that, in the large $m_1$ limit, only this cut contributes to the $Q_1^2 Q_2^3$ waveform.
Indeed, recall that equation~\eqref{eq:heavySingleCutQED} is multiplied by a factor $1/m_1 m_2$ in the waveshape because of the delta functions in equation~\eqref{eq:genericWaveshape} and therefore scales as $1/m_2^2$.

\subsection*{Real part, cutting line 2}

The remaining single-particle cut at order $Q_1^2 Q_2^3$ is  
\[\label{eq:scut2Graphs}

\]
We find it convenient to compute the cut by  choosing different gauges for the two Compton amplitudes in intermediate stages.

Both diagrams make an equal contribution to the single-cut, so 
\[
\SCut_{2}\mathcal{A}_{5, 1}^{Q_1^2 Q_2^3} = 2 \int \frac{\hd^D \ell}{\ell^2 (\ell-q_1)^2}\frac{\hat{\delta}(2p_2\cdot \ell)}{2} N(\ell) \,.
\]
The numerator $N(\ell)$ in the equation above can be found treating the blobs as on-shell amplitudes, with the result
\[\label{exkb}
N(\ell) = 8 Q_1^2 Q_2^3 \sum_{\textrm{helicities}} p_2 \cdot \varepsilon^*(\ell, p_1) \, \varepsilon(\ell, p_1) \cdot \varepsilon^*(\ell-q_1, p_1) \, \varepsilon(\ell - q_1, p_2) \cdot \varepsilon^*(k, p_2) \,.
\]
Note that we used different gauges for the polarisation vectors in different tree Compton amplitudes in the cut. 
However, it is an easy matter to change the gauge, and in particular we find it convenient to write
\[
\varepsilon_\mu(\ell-q_1, p_1) = \varepsilon_\mu(\ell - q_1, p_2) - (\ell - q_1)_\mu \frac{p_1 \cdot \varepsilon(\ell-q_1, p_2)}{p_1 \cdot \ell} \,.
\]
The helicity sum can then be performed in $D$ dimensions straightforwardly.
The contribution of the cut to the waveform is
\[\label{eq:scut2qed}
&\SCut_2 \mathcal{A}_{5,1}^{Q_1^2 Q_2^3}=
4Q_1^2 Q_2^3 
\int \hd^D \ell \frac{\hdelta(u_2 \cdot \ell)}{\ell^2 (\ell-q_1)^2}  \frac{1}{(u_1 \cdot \ell)^2} \left[ u_1 \cdot \varepsilon^* \left( u_1 \cdot \ell \, u_2 \cdot q_1 - u_1 \cdot u_2 \, \ell \cdot q_1\right) \phantom{\frac{(u_1 \cdot \ell)^2}{u_2 \cdot q_1}} \right.
        \\
&+(\ell - q_1) \cdot \varepsilon^* \left.\left( u_1 \cdot u_2 \, u_1 \cdot \ell - \frac{(u_1 \cdot u_2)^2 \ell \cdot q_1}{u_2 \cdot q_1} + \frac{(u_1 \cdot \ell)^2}{u_2 \cdot q_1}\right)  
- \ell \cdot \varepsilon^* \, u_1 \cdot u_2 \, u_1 \cdot \ell \phantom{\frac{1}{2}}\!\!\!
 \right] \, ,
\]
where the remaining polarisation vector is of the outgoing photon with momentum $k$, in the gauge $p_2 \cdot \varepsilon^* = 0$.
In this equation, we explicitly rewrote the momenta in terms of proper velocities. 
Taking the factor $1/m_1 m_2$ from the waveshape into account, we see that this contribution scales symmetrically with particle masses.

\subsubsection*{Real part, summary}

In the end the real part of the expectation $\E$ is given by the sum 
\[
\Re' \E_{5,1}^{Q_1^2 Q_2^3} = \SCut_1 \mathcal{A}_{5,1}^{Q_1^2 Q_2^3} + \SCut_2 \mathcal{A}_{5,1}^{Q_1^2 Q_2^3} \,
\]
where the single cuts are given in equations~\eqref{eq:heavySingleCutQED} and~\eqref{eq:scut2qed}.
Via equation~\eqref{eq:genericWaveshapeRe} we deduce the relevant part $\waveshape\rvert_{\text{Real}}$ of the waveshape: 
\[\label{finalwqed}
  \waveshape\rvert_{\text{Real}}= 
  \KMOCav{\int \hd^D q_1 \hd^D q_2 \, \hdelta(2 p_1 \cdot q_1) \hdelta(2 p_2 \cdot q_2) \, \hdelta^D(k + q_1 + q_2) \,e^{-ib_1 \cdot q_1}e^{-ib_2 \cdot q_2} 
\\ \times i \left( \SCut_1 \mathcal{A}_{5,1}^{Q_1^2 Q_2^3} + \SCut_2 \mathcal{A}_{5,1}^{Q_1^2 Q_2^3} + (1 \leftrightarrow 2) \right) } \,.
\]
The $1 \leftrightarrow 2$ instruction above incorporates the $Q_1^3 Q_2^2$ channels.
We will discuss the $Q_1^4 Q_2$ and $Q_1 Q_2^4$ charge sectors below in sections~\ref{rr} and~\ref{renormalix}.

We have tested these results in a number of ways. Firstly, we have compared our expressions to the one-loop five-point Yang-Mills amplitudes presented in~\cite{Carrasco:2020ywq}.
We also compared with the work of Shen~\cite{Shen:2018ebu}, who iterated the classical equations to this order. Some care has to be taken to remove divergent terms in the results of reference~\cite{Shen:2018ebu} which result from Shen's merging procedure.
Nevertheless we found agreement in this sector.
Finally, as described in section \ref{sec:confirm}, we have performed our own computation in the classical theory and find full agreement.

\subsection{QCD}
\label{sec:qcd}

Let us now move to Yang-Mills. For the purposes of our paper, the main difference between QED and Yang-Mills amplitudes is the handling of color degrees of freedom {and the appearance of new cuts involving non--abelian (pure gluon) vertices}. It is precisely these new channels  that make QCD very similar to gravity and provides us with another important motivation to study the Yang-Mills waveshape. 

As to color, we follow the setup of~\cite{delaCruz:2020bbn}.
We take our massive scalars to be in an irreducible representation of the gauge group with generators $T^a_{ij}$ and (as usual) we write the structure constants as $f^{abc}$.
When considering the classical limit of color, it is useful to introduce the color matrices $C^a = \hbar T^a$.
Although these are simple rescalings of the usual generators, it is the $C^a$ which appear in the Feynman rules when factors of $\hbar$ are restored\footnote{This choice guarantees that other factors of $\hbar$ are in the same place as in QED~\cite{delaCruz:2020bbn}.}.
We then exploit the color algebra in the form
\[\label{eq:colorAlgebra}
[C^a, C^b] &= \hbar f^{abc} \, C^c \,, \\
f^{dac}f^{cbe}-f^{dbc}&f^{cae}=f^{abc}f^{dce} \,.
\]
We further organise and expand our amplitudes (or cuts thereof) in a color basis, and  focus on each gauge invariant sector independently. 

The one-loop amplitude can be expanded in a basis of color coefficients as follows: 
\begin{equation}\label{colorbasis}
      \begin{aligned}
     \mathcal{A} _{5,1} (p_1... k) 
      &= \mathcal{C} \left(
     \begin{gathered}

      \end{gathered}
    \right) A_5
   + \cdots,
\end{aligned}      
\end{equation}
where $A_i$ is the partial amplitude corresponding to the color factor $\mathcal{C}_i$.
Classically, the five color factors shown in equation~\eqref{colorbasis} are sufficient, as we now argue.

We first note that the partial amplitudes $A_i$ may be singular in the sense that there is exactly one excess inverse power of $\hbar$ present.
We must therefore be careful to retain up to one power of $\hbar$ in color numerators when using the algebra~\eqref{eq:colorAlgebra}.
By inspection of the relevant Feynman diagrams, we see that the possible classical color structures involve five color matrices (three $C_1$ and two $C_2$ matrices or vice versa), four color matrices and one $f^{abc}$ (two $C_1$ and two $C_2$ matrices) or three color matrices and two structure constant factors (two $C_1$ and one $C_2$ matrices, or vice versa), as indicated in equation~\eqref{colorbasis}.
We refer to these structures as ``abelian'', ``pentagon'' and ``maximally non--abelian'' in what follows.
The abelian factor is distinguished by lacking any non--abelian structure constant; as we will see, it is very closely related to the QED case.

To understand the classical color structure, fix canonical abelian structures e.g. $C_1^a \cdot C_1^b \cdot C_1^d \, C_2^b \cdot C_2^d$ and $C_2^a \cdot C_2^b \cdot C_2^d \, C_1^b \cdot C_1^d$ (the $\cdot$ emphasises that these are matrix products).
Given any other structure constant of abelian type, we may use the color algebra to re-order the structure constant into the canonical form plus order $\hbar$ corrections (neglecting higher order terms which must be quantum corrections).
The correction terms can have at most one $f^{abc}$, and are therefore of pentagon color type.

Similarly, any pentagon color factor can be written in terms of one canonical pentagon, plus a $\hbar$ correction with two structure constants.
Finally, color structures with two structure constants can be written in terms of two canonical maximally non--abelian color factors and quantum corrections.
In this case we neglect contributions of gluon bubbles which are expected to be quantum.

We now take a moment to consider the partial amplitudes in \eqref{colorbasis}. As we show in Table~\ref{table_color_factors} --- where we list the topologies appearing in the \emph{classical}\footnote{The full, quantum, waveshape will involve more partial amplitudes and color sectors, but we will not study or report on these here.} partial amplitudes --- $A_1$ and $A_2$ involve only diagrams with no non--abelian vertices. We recognise these as the QED amplitude sectors computed in the previous section. The contributions from these sectors  can therefore be plugged into the QCD expression simply by dressing them with their given color factor.
In this section we therefore focus on the terms which appear for the first time in the case of QCD  -- namely $A_3$, $A_4$ and $A_5$.
As shown in Table~\ref{table_color_factors} these partial amplitudes do involve non--abelian topologies and must be calculated to find the full QCD result. 
Referring to Table~\ref{table_color_factors}, we see that $A_3$ corresponds to a pentagon partial amplitude while $A_4$ and $A_5$ correspond to the maximally non--abelian color structure.


\begin{table}
\begin{align*}
\begin{array}{c|c|c}
\text{Color type~} & A_i & \text{Topologies}
\\
\hline
&&
\\
\end{array}      
\end{align*}
\caption{Topologies contributing to the partial amplitudes of the color factors $\mathcal{C}$.}
\label{table_color_factors}
\end{table}

\subsubsection*{Pentagon}\label{pentag}

We begin by looking at the partial amplitude $A_3$, and compute this amplitude using the automated code used to generate the full color-dressed amplitude given in \eqref{colorbasis}. We write the amplitude in terms of our chosen color basis with powers of $\hbar$ restored, and consider the corresponding partial amplitudes separately. The pentagon amplitude is then given in terms of the following graphs,
\begin{equation}
\begin{aligned}
    \mathcal{A}_{5,1}^{fC_1^2C_2^2} =& \hspace{0.5cm}
    \begin{gathered}

\end{gathered}.
\end{aligned}
\end{equation}
Note that the signs of the graphs are important --- they are specifically given by the chosen color basis. As all factors of $\hbar$ have been restored to color factors and kinematic expressions, we calculate the Laurent series in $\hbar$, and consider the leading term.\footnote{There will be terms with higher powers of $\hbar$ in the denominator --- so-called super-classical terms --- but these cancel out of the final expression as described in section~\ref{vanishingcuts}.}
The resulting expression sits over massless and massive propagators, and the latter are substituted with delta functions and principal values as described in \eqref{plstot}. The one-particle cuts are then given by the sum of terms with a single delta function. The pentagon one-particle cuts are given in an ancillary notebook.

\subsubsection*{Maximally non--abelian partial amplitude}\label{a44}
 
Two of the most physically interesting gauge invariant sectors are $A_4$ and $A_5$. We will refer to them as ``maximally non--abelian" as their color factors involve two structure constants. 
Noting that these two sectors are related by particle relabelling, we will focus on $A_4$ only. The color structure corresponding to this partial amplitude is now
\begin{equation}\label{colstr}
\mathcal{C} \left(
    \begin{gathered}
\begin{tikzpicture}[scale=0.4]
\begin{feynman}
\vertex (a1m) at (-0.5, -0.5); 
\vertex (a1) at (-1.1, -0.7);
\vertex (a2m) at (-0.5, 0.5); 
\vertex (a2) at (-1.1, 0.1);
\vertex (ag) at (-1.1, 1.);
\vertex (agm) at (-0.8, 0.5);
\vertex (a3m) at (0.5, 0.5); 
\vertex (a3) at (1, 0.7);
\vertex (a4m) at (0.5, -0.5); 
\vertex (a4) at (1, -0.7);
\diagram{
(agm) -- [thick, color=red] (ag),
(a1) --  (a1m), 
(a1m) --  (a2m),
(agm)--[thick, color=red]  (a2),
(a1m) --  (a4m),
(a3m) --  (agm),
(a3) --[thick, color=blue(ncs)]  (a3m),
(a3m) --[thick, color=blue(ncs)]  (a4m) --[thick, color=blue(ncs)]  (a4)
};
\end{feynman}
\end{tikzpicture}
      \end{gathered}
    \right)=C_1^a f^{adb}f^{dAc} C_2^c \cdot C_2^b.
\end{equation}

To compute the $\SCut$ of $A_4$ we work along the same lines of \ref{examplecalc} and \ref{pentag}, so we skip explicit derivations.  Furthermore, in what follows we only detail terms which involve a $1/q_1^2$ denominator; this will also be important for radiation reaction purposes as we will later see. We find  
\begin{equation}\label{a5}
\begin{split} 
  \SCut_{2}&\mathcal{A}_{5, 1}^{f^2 C_2^2}= \frac{ 16 g^5 m_1 m_2}{q_1^2} \int \hat{\dd}^D \ell \, \hat{\delta}(u_2\cdot \ell)\,
  \varepsilon_\eta ^{*\mu}   (k)\\&
  \times  \left[ \frac{1}{\left(\ell+{q}_2\right){}^2 \left(\ell-{q}_1\right){}^2 } \left(\ell^\mu\frac{  \gamma }{  4  \, u_2\cdot q_1 }+\ell^\mu\frac{  \ell\cdot u_1  }{\ell^2  }
  -\frac{q_1^\mu}{\ell^2 }    (\ell\cdot u_1+\gamma \,u_2\cdot q_1)\right)\right. \\&\qquad
  - \frac{u_1^\mu}{\ell^2}\left(
\frac{  1}{4 (\ell+q_2)^2 } -\frac{ \left(    u_2\cdot {q}_1\right){}^2}{\left(\ell+{q}_2\right){}^2 (\ell-q_1)^2}
\right) \\&\qquad+
\frac{u_2^\mu}{\ell^2}\left( \frac{ \ell\cdot u_1  }{ 4 \left({\ell}-{q}_1\right){}^2   \, u_2\cdot q_1} \right. \left.+\frac{  \gamma }{4 \left(u_2\cdot {q}_1\right)^2} 
\right. \left.\left.  -\frac{    \gamma\,{q}_1\cdot {q}_2 -q_1\cdot u_2\,q_2\cdot u_1 }{\left(\ell+{q}_2\right){}^2 \left(\ell-{q}_1\right){}^2}\right)\right],
\end{split}
\end{equation}
from cutting line 2.  
(Here $\gamma = u_1 \cdot u_2$.)

We have checked equation~\eqref{a5} against the results of \cite{Shen:2018ebu}, by using an automated code as described above for the pentagon, and also with Feynman diagrams.  
In fact, it is a simple exercise to see that \eqref{a5} is reproduced by the classical limit of the following five Feynman diagrams:
\begin{figure}[H]
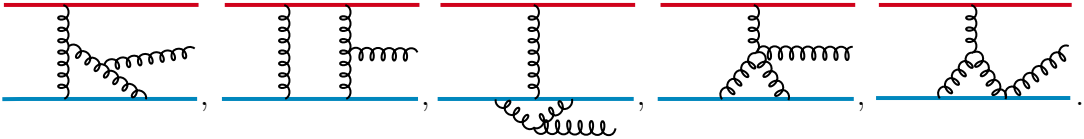

\begin{center}

\tikzset{every picture/.style={line width=0.75pt}} 



\end{center}\caption{Feynman diagrams contributing classically to $A_4.$}
\end{figure} 

The complete expression for this maximally non-Abelian single-cut can be found in our attached notebook. 

\section{Reaction}\label{reactionnn}

In the previous section, we focused on the real part $\Re' \mathcal{A}$ of the one-loop five-point amplitude. 
A complete computation of the waveshape also requires the computation of the imaginary part of the expectation $\E$, which as discussed around equation~\eqref{eq:imEcuts} is \emph{not} the same as the imaginary part of the one-loop five-point amplitude\footnote{In fact, the imaginary part of the one-loop five-point amplitude is obtained by flipping the sign of the first term on the right-hand-side of equation~\eqref{eq:imEcuts}.}.

Imaginary parts are typically associated with the physics of dissipation, and one-loop waveshapes are no exception.
In electrodynamics we will show in section~\ref{rr} that the imaginary part of the expectation, more specifically the Compton cuts of equation~\eqref{eq:imcuts} in the $Q_1 Q_2^4$ charge sector, computes the radiation emitted by a particle under the influence of its self-field, computed using the Abraham-Lorentz-Dirac (ALD) radiation-reaction force.
(See section~\ref{sec:confirm} for the classical computation of the same quantity.)

The fact that the physics of radiation reaction is relevant in one-loop computations was a surprise to us; this physics occurs instead at two loops in the impulse~\cite{Kosower:2018adc}. 
Radiation reaction is quite subtle in classical field theory because a particle's field diverges at the particle's location (see reference~\cite{Poisson:2011nh} for a detailed review of classical radiation reaction forces).
In quantum field theory, the situation seems simpler because divergences are very well understood.
The diagrams whose Compton cuts yield the ALD part of the waveshape are UV divergent. 
We discuss this topic in detail in section~\ref{rennn2}, ultimately  justifying the absence of $\SCut\mathcal{A}_{5,1}^{Q_1Q_2^4}$ from the classical waveshape.
Counterterms are real, so the imaginary part of these diagrams are physical, and are responsible for radiation reaction. 

While radiation reaction is definitely associated with Compton cuts, it is not the case that all Compton cuts are related to radiation reaction.
An example is the Compton cut in the $Q_1^2 Q_2^3$ charge sector.
This cut can be trivially deduced from the single cuts in section~\ref{examplecalc} by cutting one additional propagator.
(We present the expression in appendix~\ref{sec:fullqed}.)
The entire $Q_1^2 Q_2^3$ waveform can be classically attributed to the Lorentz force as we will see in section~\ref{sec:confirm}.

Because radiation reaction and self-force are particularly interesting topics, we further compute the contribution of the Compton cuts in QCD and gravity in sections~\ref{qcdddrrr} and~\ref{grr}.
It seems reasonable to tentatively identify these Compton cuts generally as the gauge-invariant objects associated with radiation reaction from the perspective of scattering amplitudes, though at present we can only make a sharp connection in electrodynamics.

As shown in equation~\eqref{eq:imcuts}, Compton cuts are not the only contribution to the imaginary part of the expectation.
However the remaining ``iteration'' cuts are unrelated to radiation reaction in QED and appear to be generally associated with conservative forces~\cite{Caron-Huot:2023vxl}. 
These iteration cuts are the subject of appendix~\ref{cutextra}.

We further discuss the physics of radiation reaction, cuts and the ALD force in appendix~\ref{coleman}.

\subsection{QED radiation reaction...}\label{rr}

Let us begin with electrodynamics, specifically with the Compton cut defined by equation~\eqref{ccut} in the $Q_1 Q_2^4$ charge sector. 
To ease our calculation we will employ a convenient trick. This consists of placing the photon line which connects particle one to particle two (of momentum $q_1$) on shell. 
Being rigorous, we shouldn't be allowed to do so: cutting this line isolates a three point amplitude which vanishes on the support of real Minkowski kinematics. 
Nevertheless, it turns out that we can effectively do this. 
In fact, multiplying $1/q_1^2$ by $q_1^2$ does not strictly yield zero, but  only gives a contact term which integrates to zero. 
Consequently, the Compton cuts simplify to
\[\label{figurecutqed}

\] 
On the residue of $1/q_1^2$, this cut is determined by two-particle cuts involving two tree-level Compton amplitudes.
We write the relevant Compton amplitude as
\begin{equation}\label{compt}
\mathcal{A}_{4,0} (p_1,{k}_1\to p_2,{k}_2)={2 Q^2_2} \varepsilon^{\mu}_\eta (k_1)\varepsilon^{*\nu}_{\eta'}(k_2)\mathcal{J}_{\mu\nu}(p_1,{k}_1\to p_2,{k}_2),
\end{equation}
where we are taking $p_1, \, \bk_1$ incoming and  $p_2, \, \bk_2$ outgoing. Above, we have also defined 
\begin{equation}
\mathcal{J}_{\mu\nu}(p_1,{k}_1\to p_2,{k}_2)=\frac{p_{1\mu}p_{2\nu}}{p_1\cdot {k}_1}+\frac{p_{2\mu}p_{1\nu}}{-p_1\cdot {k}_2}- \eta_{\mu\nu},\,\,\,\,\, p_1+k_1=p_2+k_2,
\end{equation}
satisfying
\begin{equation}
    \mathcal{J}_{\mu\nu}k^\mu_1 \varepsilon^\nu(k_2)=\mathcal{J}_{\mu\nu}\varepsilon^\mu(k_1)k^\nu_2 =0.
\end{equation}

Making use of  these definitions, the cut is given  explicitly by 
\[\label{earlycut}
\CCut_2 &\mathcal{A}_{5,1}^{Q_1Q_2^4}=-\frac{4Q_1Q^4_2}{q_1^2} \sum_{\eta,\eta''} p_1\cdot \varepsilon_{\eta}(q_1)\varepsilon^{*\mu}_\eta (\bq_1) \varepsilon^{*\sigma}_{\eta'}(\bk)  \int \dd \Phi(-\ell)\, \hat{\delta}(2p_2\cdot(\bl-\bq_1)) 
\\& \!\!\!\!\times
\varepsilon^{\nu}_{\eta''}(\bl)
\varepsilon^{*\rho}_{\eta''} (\bl)
\mathcal{J}_{\mu\nu}(p_2, -q_1\to p_2-q_1+\ell, -\ell)
\mathcal{J}_{\rho\sigma}(p_2 -q_1+\ell, -\ell\to p_2+q_2, k).
\]

We now proceed by using the gauge $\varepsilon\cdot p_2=0$ as in the previous sections. 
This means performing the helicity sum over $\eta''$ through \eqref{eq:massiveCompleteness} with $q=p_2$. One soon obtains (suppressing the helicity index of the outgoing photon)
\begin{equation}
\begin{split}
\CCut_2 \mathcal{A}_{5,1}^{Q_1Q_2^4}=-\frac{2Q_1Q^4_2}{m_2\,q_1^2} \sum_{\eta} p_1\cdot \varepsilon_{\eta}(q_1) & \int \dd \Phi(-\ell)\, \hat{\delta}(u_2\cdot(\bl-\bq_1))  
\\& \!\!\!\!\times\left( \frac{\varepsilon^*_\eta(\bq_1)\cdot\bl\,\varepsilon^*(\bk)\cdot  \bl}{( u_2\cdot \bq_1)^2}+\varepsilon^*_{\eta }(\bq_1) \cdot\varepsilon^* (\bk)\right).
\end{split}
\end{equation}

The loop integrals are easy to do here. The scalar one was first evaluated in \cite{Kosower:2018adc}. Taking $D=4$, we find  
\begin{equation}
\int  \dd \Phi(-\ell) \, \hat{\delta}(u_2\cdot(\bl-\bq_1)) =\frac{u_2\cdot k}{2\pi}\Theta(u_2\cdot k),
\end{equation}
and the tensor one follows   by reduction, we find  
\begin{equation}
\int  \dd \Phi(-\ell) \, \hat{\delta}(u_2\cdot(\bl-\bq_1)) \bl^\mu \bl^\nu=-\frac{(u_2\cdot k)^3}{6\pi}\left(
\eta^{\mu\nu}-4\, u_2^\mu u_2^\nu
\right)\Theta(u_2\cdot k).
\end{equation} 
In the end, after summing over the remaining helicity states, we arrive at the following expression for the cut
\begin{equation}\label{finea}
\begin{split}
\CCut_2 \mathcal{A}_{5,1}^{Q_1Q_2^4}&=\frac{4 Q_1 Q_2^4}{m_2^2   }  \frac{p_2\cdot \bk }{6\pi}\frac{1}{\bq_1^2} \left(
p_1\cdot \varepsilon^*(\bk) +
\frac{p_1\cdot p_2\, \varepsilon^*(\bk)\cdot \bq_1}{p_2\cdot \bk} 
\right).
\end{split}
\end{equation} 

This result is quite remarkable and simple: the cut isolating two Compton amplitudes is essentially $p_2\cdot k$ times the  tree-level five-point amplitude, all multiplied by a geometric factor $1/6\pi$ coming from the loop angular integration.
Above we have also set $\Theta(u_2\cdot k)=1$, which holds on the support of the $\dd\Phi (k)$ integral of \eqref{eq:genFieldStrength}.
So what is the physical origin of this beautiful and simple loop correction? 
The answer is that it is a leading-order consequence of the ALD force.
This is most easily seen by a direct classical computation with the ALD force law, which we discuss in section~\ref{sec:confirm}.

The imaginary part of the expectation is given in general by Compton cuts and iteration cuts. 
However in the $Q_1 Q_2^4$ charge sector the iteration cuts vanish, so we conclude that 
\[\label{qedimpartfinalll}
\Im' \E_{5,1}^{Q_1Q_2^4}&=\frac{4 Q_1 Q_2^4}{m_2^2   }  \frac{p_2\cdot \bk }{6\pi}\frac{1}{\bq_1^2} \left(
p_1\cdot \varepsilon^*(\bk) +
\frac{p_1\cdot p_2\, \varepsilon^*(\bk)\cdot \bq_1}{p_2\cdot \bk} 
\right) \,.
\]
The contribution to the waveshape follows from~\eqref{eq:genericWaveshapeIm}.
We can of course deduce the contribution from the $Q_1^4 Q_2$ charge sector by permuting particle labels.

\subsection{...QCD radiation reaction...} \label{qcdddrrr}

Radiation reaction in non-abelian theories is far less well understood than in electrodynamics or in gravity.
It therefore seems interesting to begin an analysis of this topic from the perspective of amplitudes.
Our idea here is that Compton cuts captured radiation reaction in electrodynamics, and it seems reasonable to expect this to continue in the more complicated cases of Yang-Mills and gravity.
So let us analyse the Compton cuts in QCD. As it happens, the preparatory work of section \ref{a44} will be very convenient for us.

Radiation reaction in QED was accompanied by a $1/q_1^2$ photon pole.
The remainder of the diagram involved a photon interacting with the particle which produced it: a self-field interaction.
Thus in this section we also focus on the coefficient of the $1/q_1^2$ pole.
We will soon find that the non-Abelian radiation reaction channels, defined as the $1/q_1^2$ part of the Compton cut on line 2, are precisely those characterised by $A_4$ (and $A_5$) which we studied in section \ref{a44}. 

\begin{figure}[H]
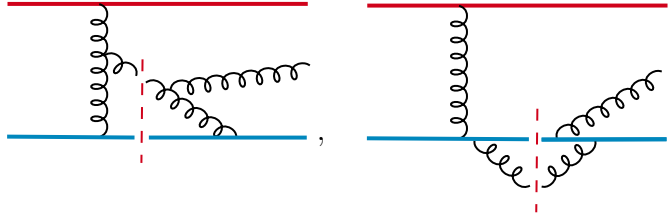

\begin{center}
   
\tikzset{every picture/.style={line width=0.75pt}} 



 \caption{Two cuts contributing to non-Abelian radiation reaction. We have also effectively cut the single gluon line as explained above.}
\end{center}
\end{figure}

Let us then compute these ``radiation reaction'' diagrams of QCD. As before, we consider the following cut diagram
\begin{figure}[H]
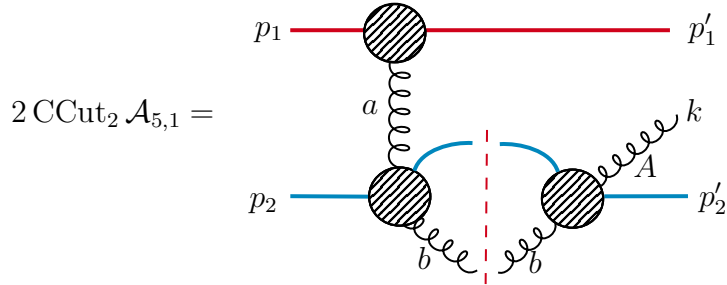

\begin{center}

 
\tikzset{
pattern size/.store in=\mcSize, 
pattern size = 5pt,
pattern thickness/.store in=\mcThickness, 
pattern thickness = 0.3pt,
pattern radius/.store in=\mcRadius, 
pattern radius = 1pt}
\makeatletter
\pgfutil@ifundefined{pgf@pattern@name@_u0zpcl5r7}{
\pgfdeclarepatternformonly[\mcThickness,\mcSize]{_u0zpcl5r7}
{\pgfqpoint{0pt}{0pt}}
{\pgfpoint{\mcSize+\mcThickness}{\mcSize+\mcThickness}}
{\pgfpoint{\mcSize}{\mcSize}}
{
\pgfsetcolor{\tikz@pattern@color}
\pgfsetlinewidth{\mcThickness}
\pgfpathmoveto{\pgfqpoint{0pt}{0pt}}
\pgfpathlineto{\pgfpoint{\mcSize+\mcThickness}{\mcSize+\mcThickness}}
\pgfusepath{stroke}
}}
\makeatother

 
\tikzset{
pattern size/.store in=\mcSize, 
pattern size = 5pt,
pattern thickness/.store in=\mcThickness, 
pattern thickness = 0.3pt,
pattern radius/.store in=\mcRadius, 
pattern radius = 1pt}
\makeatletter
\pgfutil@ifundefined{pgf@pattern@name@_mcwhbxxmr}{
\pgfdeclarepatternformonly[\mcThickness,\mcSize]{_mcwhbxxmr}
{\pgfqpoint{0pt}{0pt}}
{\pgfpoint{\mcSize+\mcThickness}{\mcSize+\mcThickness}}
{\pgfpoint{\mcSize}{\mcSize}}
{
\pgfsetcolor{\tikz@pattern@color}
\pgfsetlinewidth{\mcThickness}
\pgfpathmoveto{\pgfqpoint{0pt}{0pt}}
\pgfpathlineto{\pgfpoint{\mcSize+\mcThickness}{\mcSize+\mcThickness}}
\pgfusepath{stroke}
}}
\makeatother

 
\tikzset{
pattern size/.store in=\mcSize, 
pattern size = 5pt,
pattern thickness/.store in=\mcThickness, 
pattern thickness = 0.3pt,
pattern radius/.store in=\mcRadius, 
pattern radius = 1pt}
\makeatletter
\pgfutil@ifundefined{pgf@pattern@name@_fn01dsdhn}{
\pgfdeclarepatternformonly[\mcThickness,\mcSize]{_fn01dsdhn}
{\pgfqpoint{0pt}{0pt}}
{\pgfpoint{\mcSize+\mcThickness}{\mcSize+\mcThickness}}
{\pgfpoint{\mcSize}{\mcSize}}
{
\pgfsetcolor{\tikz@pattern@color}
\pgfsetlinewidth{\mcThickness}
\pgfpathmoveto{\pgfqpoint{0pt}{0pt}}
\pgfpathlineto{\pgfpoint{\mcSize+\mcThickness}{\mcSize+\mcThickness}}
\pgfusepath{stroke}
}}
\makeatother

 
\tikzset{
pattern size/.store in=\mcSize, 
pattern size = 5pt,
pattern thickness/.store in=\mcThickness, 
pattern thickness = 0.3pt,
pattern radius/.store in=\mcRadius, 
pattern radius = 1pt}
\makeatletter
\pgfutil@ifundefined{pgf@pattern@name@_v05jckv0c}{
\pgfdeclarepatternformonly[\mcThickness,\mcSize]{_v05jckv0c}
{\pgfqpoint{0pt}{0pt}}
{\pgfpoint{\mcSize+\mcThickness}{\mcSize+\mcThickness}}
{\pgfpoint{\mcSize}{\mcSize}}
{
\pgfsetcolor{\tikz@pattern@color}
\pgfsetlinewidth{\mcThickness}
\pgfpathmoveto{\pgfqpoint{0pt}{0pt}}
\pgfpathlineto{\pgfpoint{\mcSize+\mcThickness}{\mcSize+\mcThickness}}
\pgfusepath{stroke}
}}
\makeatother

 
\tikzset{
pattern size/.store in=\mcSize, 
pattern size = 5pt,
pattern thickness/.store in=\mcThickness, 
pattern thickness = 0.3pt,
pattern radius/.store in=\mcRadius, 
pattern radius = 1pt}
\makeatletter
\pgfutil@ifundefined{pgf@pattern@name@_hymsk8kde}{
\pgfdeclarepatternformonly[\mcThickness,\mcSize]{_hymsk8kde}
{\pgfqpoint{0pt}{0pt}}
{\pgfpoint{\mcSize+\mcThickness}{\mcSize+\mcThickness}}
{\pgfpoint{\mcSize}{\mcSize}}
{
\pgfsetcolor{\tikz@pattern@color}
\pgfsetlinewidth{\mcThickness}
\pgfpathmoveto{\pgfqpoint{0pt}{0pt}}
\pgfpathlineto{\pgfpoint{\mcSize+\mcThickness}{\mcSize+\mcThickness}}
\pgfusepath{stroke}
}}
\makeatother

 
\tikzset{
pattern size/.store in=\mcSize, 
pattern size = 5pt,
pattern thickness/.store in=\mcThickness, 
pattern thickness = 0.3pt,
pattern radius/.store in=\mcRadius, 
pattern radius = 1pt}
\makeatletter
\pgfutil@ifundefined{pgf@pattern@name@_o11butkr5}{
\pgfdeclarepatternformonly[\mcThickness,\mcSize]{_o11butkr5}
{\pgfqpoint{0pt}{0pt}}
{\pgfpoint{\mcSize+\mcThickness}{\mcSize+\mcThickness}}
{\pgfpoint{\mcSize}{\mcSize}}
{
\pgfsetcolor{\tikz@pattern@color}
\pgfsetlinewidth{\mcThickness}
\pgfpathmoveto{\pgfqpoint{0pt}{0pt}}
\pgfpathlineto{\pgfpoint{\mcSize+\mcThickness}{\mcSize+\mcThickness}}
\pgfusepath{stroke}
}}
\makeatother
\tikzset{every picture/.style={line width=0.75pt}} 

\caption{Cut relevant to radiation reaction in QCD. For clarity we have just indicated color indices, momentum routing is the same as in equation~\eqref{figurecutqed}. }
\end{center}
\end{figure}
  
Indicating the left/right YM tree-level Compton amplitudes entering the cut with $\mathcal{A}^{ab}_{L, R}$ the diagram reads
\begin{equation} \label{c2qcd}
\begin{split}
\CCut_2{\mathcal{A}_{5,1}}= \frac{1}{q_1^2}&\sum_{\text{helicities}}{2p_1\cdot \varepsilon}  \int \dd\Phi (-\ell)\, \hat{\delta}(2p_2\cdot(\bl-\bq_1)) 
\mathcal{A}^{ab}_{L} \cdot\mathcal{A}^{bA}_{R}.
\end{split}
\end{equation}
At this point it is useful to note again that the non-Abelian Compton amplitude $\mathcal{A}^{ab}\equiv \mathcal{A}^{ab}_{4,0}$ is
\[
\mathcal{A}^{ab}=2g^2C^a_i \cdot C^b_i \varepsilon_1\cdot\varepsilon_2+2g^2f^{abc}C_i^c \frac{2p_i\cdot k_1}{(k_1+k_2)^2}\varepsilon_1\cdot\varepsilon_2,
\]
so that it can be written more schematically as
\begin{equation}
\mathcal{A}^{ab}=C^a\cdot C^b {A} +f^{abc}C^c A',
\end{equation}
where $A$ and $A'$ are both proportional to abelian Compton amplitudes, differing only by a factor.
Using this knowledge we expand the integrand in the following manner
\begin{equation}\label{qcdcut}
\begin{split}
\mathcal{A}^{ab}_L\cdot \mathcal{A}^{bA}_R&=\left( C_2 ^a\cdot C_2^b A_L +f^{abd}C_2^d A'_L
\right)\cdot\left( C_2 ^b\cdot C_2^A A_R +f^{bAe}C_2^e A'_R
    \right)\\&
    \approx C_2 ^a\cdot C_2^b \cdot C_2 ^b\cdot C_2^A  A_L A_R+ f^{abd} f^{bAe} C_2^d\cdot C_2^e A'_L A'_R \,.
    \end{split}
\end{equation}
We ignored the cross terms which are quantum due to the algebra in equation~\eqref{eq:colorAlgebra}.

Now, the simple relation above makes it very easy to interpret the structure of the cut in non-Abelian gauge theories.
The first term in the last line of equation \eqref{qcdcut}, the one proportional to $C_2 ^a\cdot C_2^b \cdot C_2 ^b\cdot C_2^A $, is exactly the one already encountered in QED! 
In other words, what we had computed in QED was also part of the QCD story,  only now multiplied by a constant color structure:
\[
\CCut_2\mathcal{A}_{5,1}^{C^4_2}=C_2 ^a\cdot C_2^b \cdot C_2 ^b\cdot C_2^A \times \frac{g^5}{Q_1Q_2^4}\CCut_2\mathcal{A}_{5,1}^{Q_1 Q_2^4}.
\]

The last term in equation \eqref{qcdcut} is obviously non-abelian given its color factor.
Moreover its color structure is precisely that of equation~\eqref{colstr} --- namely of the partial amplitude $A_{4}$. 
Thus we can compute the relevant $1/q_1^2$ pole of the Compton cut from 
equation~\eqref{a5} by simply placing an additional messenger on shell.
From \eqref{a5}, the $\CCut_2$ of the color-stripped amplitude is immediately obtained and it reads 
\begin{equation}
\CCut_2{A}_4^{f^2C_2^2}=\frac{ 8 g^5 m_1 m_2}{q_1^2} \varepsilon^*_\mu  \!\!\int\!\! \dd\Phi(-\ell) \hat{\delta}(u_2\cdot(\ell-q_1))  \frac{   (u_1\cdot \ell+\gamma q_1\cdot u_2 ) q_1^\mu 
 -(u_2\cdot q_1)^2   u_1^\mu
  }{(q_1-\ell)^2(\ell+k)^2},
\end{equation}
in a gauge where $\varepsilon\cdot p_2=0$.

\subsection{...GR radiation reaction}\label{grr}

Finally for this section we tackle the gravitational case. 
Following the examples of QED and QCD we again focus on the two-particle Compton cut, just involving a graviton line now. 
One could proceed as done in sections \ref{rr} and \ref{qcdddrrr}, that is computing this cut by sitting on the $1/q_1^2$ propagator pole.
However in this case the tree amplitudes contributing to the cut are all available in a convenient form, so we
compute the full Compton cut.

The cut is diagrammatically given by
\[\label{cutgravv}

\]
and its explicit expression is
\begin{equation}\label{c2gr}
\begin{split}
\CCut_2&\mathcal{M}_{5,1}= 
\frac{1}{2}\sum_{\text{helicities}} \int  \dd\Phi (-\ell)\, \hat{\delta}(2p_2\cdot(\bl-\bq_1))\\&\times\mathcal{M}_{5,0} (p_1,p_2,\to p_1+q_1,p_2-q_1+\ell, -\ell)\mathcal{M}_{4,0}(p_2 -q_1+\ell, -\ell\to p_2+q_2, k).
\end{split}
\end{equation}
At this point we need the classical five-point tree amplitude, which can be written as~\cite{Luna:2017dtq} 
\[
\mathcal{M}_{5,0}(p_1,p_2\to p_1+q_1, p_2-q_1+\ell, -\ell)=-\frac{\kappa^3}{4}m_1^2m_2^2 \varepsilon^{*\mu}_{\eta}(\ell)\varepsilon^{*\nu}_{\eta}(\ell)\mathcal{T}_{\mu\nu}(q_1, \ell),
\]
where $\mathcal{T}^{\mu\nu}$ reads
\[\label{tensorT}
\mathcal{T}^{\mu\nu}=\frac{4 P_{12}^\mu P_{12}^\nu }{q_1^2 (q_1-\ell)^2}+\frac{2\gamma Q_{12}^{(\mu}P_{12}^{\nu)}}{q_1^2 (q_1-\ell)^2}+\left(\gamma^2-\frac{1}{D-2}\right)\left(
\frac{Q_{12}^\mu Q_{12}^\nu }{q_1^2(q_1-\ell)^2}-\frac{P_{12}^\mu P_{12}^\nu}{(\ell\cdot u_1 \ell\cdot u_2)^2}
\right) \,.
\]
Following \cite{Luna:2017dtq}, we defined 
\[
P_{12}^\mu=\ell\cdot u_2 u_1^\mu-\ell\cdot u_1 u_2^\mu  , \qquad Q_{12}^\mu=(\ell-2q_1)^\mu+\frac{q_1^2}{\ell\cdot u_1}u_1^\mu-\frac{(q_1-\ell)^2}{\ell\cdot u_2}u_2^\mu,
\]
such that $P_{12}\cdot \ell=0=Q_{12}\cdot \ell$, ensuring the gauge-invariance of the amplitude.

The quantity in equation~\eqref{cutgravv} can  be greatly simplified by choosing, as we did above, a gauge in which graviton polarisations are orthogonal to $p_2$: $\varepsilon_\mu \varepsilon_\nu p^\mu_2=\varepsilon_\mu \varepsilon_\nu p^\nu_2=0$. This gauge kills all the terms proportional to $u_2^\mu$ in \eqref{tensorT} and drastically simplifies the gravitational tree-level Compton down to  a single contraction  \cite{Bern:2002kj} 
\begin{equation}
 \mathcal{M}_{4,0}(p_2, k, \ell)=-\frac{\kappa^2m_2^2}{2} \frac{(u_2\cdot k)^2}{(\bl+k)^2  }(\varepsilon(\ell)\cdot \varepsilon^* (k))^2.
\end{equation}

Next, we have to evaluate the sum over physical states. Note that we haven't been explicit about helicity assignments since these always come with opposite signs inside the loop. We  have  
\begin{equation}
\sum_{\text{helicities}} \mathcal{T}_{\mu\nu}(q_1, \ell) \varepsilon^{*\mu} (\ell)\varepsilon^{*\nu} (\ell)(\varepsilon(\ell)\cdot \varepsilon^* (k))^2 = \mathcal{T}^{\mu\nu}(q_1, \ell)\varepsilon^{*\rho}(k)\varepsilon^{*\sigma}(k) P_{\mu\nu\rho\sigma}(\ell),
\end{equation}
where the projector over physical states reads, for the case
of a massive gauge vector~$p_2$,
\[
P_{\mu\nu\rho\sigma}(\ell)&=\sum_{\text{helicities}}\varepsilon_\mu (\ell)\varepsilon_\nu (\ell)\varepsilon_\rho^*(\ell)\varepsilon_\sigma^* (\ell) \\
&=\left[\frac{1}{2}P_{\mu(\rho}(\ell,u_2)P_{\sigma)\nu}(\ell,u_2)
-\frac{1 }{D-2}P_{\mu\nu}(\ell,u_2)P_{\rho\sigma}(\ell,u_2)\right].
\]
Above we defined 
\begin{equation}
    P^{\mu\nu}(\ell,u_2)=- \left( \eta^{\mu\nu} - \frac{\ell_{\phantom{2}}^{(\mu}u_2^{\nu)}  }{\ell \cdot u_2} +   \frac{\ell^\mu \ell^\nu}{(\ell \cdot u_2)^2} \right),
\end{equation}
which is the projector we used in the electromagnetic case, in the gauge $u_2$.
The contraction that we have to evaluate is then straightforward and yields
\[
  \mathcal{T}^{\mu\nu}(q_1, \ell)\varepsilon^{*\rho}(k)\varepsilon^{*\sigma}(k)P_{\mu\nu\rho\sigma}(\ell)=\frac{N}{(D-2)^2 q_1^2\left(2\left(\ell \cdot q_1\right)-q_1^2\right)\left(\ell \cdot u_1\right)^2\left(q_1 \cdot u_2\right)^4} ,
\]
where
\[
N =-2\left(( D - 2 ) \left(2(D-2)\left( \varepsilon^* (k) \cdot u_1\right)^2\left(\ell \cdot u_1\right)^2\left(q_1 \cdot u_2\right)^2+2(D-2) \gamma\left(\varepsilon^*(k) \cdot u_1\right)\right.\right. \\
\left(\ell \cdot u_1\right)\left(\left(\varepsilon^*(k) \cdot u_1\right) q_1^2-2\left(\varepsilon^*(k) \cdot q_1\right)\left(\ell \cdot u_1\right)\right)\left(q_1 \cdot u_2\right)+\left((D-2) \gamma^2-1\right) \\
\left.\left(2\left(\varepsilon^*(k) \cdot q_1\right)^2\left(\ell \cdot u_1\right)^2+\left(\varepsilon^*(k) \cdot u_1\right)\left(\left(\varepsilon^*(k) \cdot u_1\right)\left(\ell \cdot q_1\right)-2\left(\varepsilon^*(k) \cdot q_1\right)\left(\ell \cdot u_1\right)\right) q_1^2\right)\right) \\
\left(q_1 \cdot u_2\right)^4+2(D-2)(\varepsilon^*(k) \cdot l)\left(( \varepsilon^*(k) \cdot u _ { 1 } ) ( q _ { 1 } \cdot u _ { 2 } ) \left(-2(D-2) \gamma\left(\ell \cdot q_1\right)\left(\ell \cdot u_1\right)^2+\right.\right. \\
\left(2(D-2)\left(\ell \cdot u_1\right)^2-\left((D-2) \gamma^2+1\right) q_1^2\right)\left(q_1 \cdot u_2\right)\left(\ell \cdot u_1\right)+ \\
\left.\gamma\left(2(D-2)\left(\ell \cdot u_1\right)^2+\left(1-(D-2) \gamma^2\right)\left(\ell \cdot q_1\right)\right) q_1^2\right)+\left(\varepsilon^*(k) \cdot q_1\right)\left(\ell \cdot u_1\right) \\
\left(2\left(\ell \cdot u_1\right)\left(q_1 \cdot u_2\right)^2+\gamma\left((D-2) \gamma^2-1\right) q_1^2-2(D-2)\left(\ell \cdot u_1\right)^2\right)\left(q_1 \cdot u_2\right)+ \\
\left.\left.\left((D-2) \gamma^2-1\right)\left(\ell \cdot u_1\right)\left(2\left(\ell \cdot q_1\right)-q_1^2\right)\right)\right)\left(q_1 \cdot u_2\right)^2+(\varepsilon^*(k) \cdot l)^2 \\
\left(-4(D-2)\left(\ell \cdot u_1\right)^2\left(q_1 \cdot u_2\right)^4+\left(2(D-3)\left((D-2)\left(\ell \cdot u_1\right)^2+2\left(\ell \cdot q_1\right)\right)\left(\ell \cdot u_1\right)^2+\right.\right. \\
\left.\left(\left((D-2) \gamma^2-1\right)^2\left(\ell \cdot q_1\right)-2\left((D-2)^2 \gamma^2+D-4\right)\left(\ell \cdot u_1\right)^2\right) q_1^2\right)\left(q_1 \cdot u_2\right)^2- \\
2(D-3)(D-2) \gamma\left(\ell \cdot u_1\right)^3\left(2\left(\ell \cdot q_1\right)-q_1^2\right)\left(q_1 \cdot u_2\right)+ \\
\left.\left.(D-3)\left((D-2) \gamma^2-1\right)\left(\ell \cdot q_1\right)\left(\ell \cdot u_1\right)^2\left(2\left(\ell \cdot q_1\right)-q_1^2\right)\right)\right) \,.
\]
Combining this information, we finally have
\begin{equation}\label{imgr}
\begin{split}
\CCut_2\mathcal{M}_{5,1}=&\frac{\kappa^5 m_1^2 m_2^4}{16} \int  \dd\Phi (-\ell)\, \hat{\delta}(2p_2\cdot(\bl-\bq_1))\\&\times \frac{1}{(D-2)^2 q_1^2 \left(k \cdot u_2\right)^2}\frac{N}{ (\bl+k)^2 \left(2\left(\ell \cdot q_1\right)-q_1^2\right)\left(\ell \cdot u_1\right)^2} \,.
\end{split}
\end{equation}

Let us make a few comments before turning to a new topic.
Firstly, we note that the integral in equation~\eqref{imgr} is IR divergent. We will discuss this issue in more detail in the next section.
Secondly, it seems that the $1/q_1^2$ pole of equation~\eqref{imgr} may be most relevant to radiation reaction: after all, it is the radiation pole which should lead to reaction.
It is easy to extract this pole if needed.
Thirdly, it would be interesting to replicate \eqref{imgr} using purely classical methods, as we will do for QED in section~\ref{sec:confirm}.
One way to do this could be to use the ``MiSaTaQuWa" equations of \cite{Mino:1996nk, Quinn:1996am}, which are known to describe linear self-force in gravity. 
Obviously confidence in our identification of Compton cuts (perhaps with $1/q_1^2$ poles) as radiation reaction would be improved if one could see how they emerge from other methods.
We leave this for future work.

\section{Renormalisation}\label{rennn2}

We have just seen that diagrams like those in equation~\eqref{figurecutqed} are essential for understanding the physics of radiation reaction.
So why did we omit these diagrams while discussing real parts in section~\ref{radiation}?
For example, in QED, we could compute single-cuts in the $Q_1 Q_2^4$ charge sector --- yet this charge sector was (almost) entirely ignored in that section.
Our next goal is to explain that the real part of this class of diagram makes no contribution to the classical waveshape in electrodynamics.
The basic physics point is that the graphs are UV divergent, and once they are renormalised the result is quantum.

The fact that Feynman diagrams responsible for radiation reaction are UV divergent is easy to understand. 
The self-field of a particle is infinite at the location of the particle --- but this does not mean that the self-field is unphysical.
In quantum field theory the divergence is renormalised using textbook methods as we describe below.
In classical field theory the divergence must also be renormalised; see, for example, the comprehensive review~\cite{Poisson:2011nh}.
As divergences are far more familiar in quantum field theory, we believe that the quantum treatment has an advantage in this discussion.
Potentially confusing issues regarding the systematic nature of renormalisation (including scheme dependence) are better understood in quantum field theory.
Moreover our one-loop discussion shows how the class-room story of divergences in scalar QED resolves the glaring issue of the divergent self-field of a particle in classical electrodynamics, while elegantly incorporating the physics of radiation reaction.

While we are on the topic of divergences, we take the opportunity to study infrared divergences of the waveshape, closely following Weinberg's peerless treatment~\cite{Weinberg:1965nx}. 
One additional ingredient is that we perform a $\hbar$ expansion of the IR divergent diagrams to extract their classical contributions to the waveform.
The classical waveshape is in fact infrared divergent.

We start by discussing infrared divergences before turning to the renormalisation of UV divergences.

\subsection{Infrared divergences}

\newcommand*\bell{\ensuremath{\boldsymbol\ell}}

We first discuss infrared divergences arising from soft virtual photons in loop amplitudes. These virtual IR divergences arise from diagrams where a virtual soft photon is attached to on-shell external legs in the manner illustrated in figure~\ref{fig:ir}.
\begin{figure}[h!]

 
\tikzset{
pattern size/.store in=\mcSize, 
pattern size = 5pt,
pattern thickness/.store in=\mcThickness, 
pattern thickness = 0.3pt,
pattern radius/.store in=\mcRadius, 
pattern radius = 1pt}
\makeatletter
\pgfutil@ifundefined{pgf@pattern@name@_gnpnpbb0w}{
\pgfdeclarepatternformonly[\mcThickness,\mcSize]{_gnpnpbb0w}
{\pgfqpoint{0pt}{0pt}}
{\pgfpoint{\mcSize+\mcThickness}{\mcSize+\mcThickness}}
{\pgfpoint{\mcSize}{\mcSize}}
{
\pgfsetcolor{\tikz@pattern@color}
\pgfsetlinewidth{\mcThickness}
\pgfpathmoveto{\pgfqpoint{0pt}{0pt}}
\pgfpathlineto{\pgfpoint{\mcSize+\mcThickness}{\mcSize+\mcThickness}}
\pgfusepath{stroke}
}}
\makeatother

 
\tikzset{
pattern size/.store in=\mcSize, 
pattern size = 5pt,
pattern thickness/.store in=\mcThickness, 
pattern thickness = 0.3pt,
pattern radius/.store in=\mcRadius, 
pattern radius = 1pt}
\makeatletter
\pgfutil@ifundefined{pgf@pattern@name@_xr3yc0wi5}{
\pgfdeclarepatternformonly[\mcThickness,\mcSize]{_xr3yc0wi5}
{\pgfqpoint{0pt}{0pt}}
{\pgfpoint{\mcSize+\mcThickness}{\mcSize+\mcThickness}}
{\pgfpoint{\mcSize}{\mcSize}}
{
\pgfsetcolor{\tikz@pattern@color}
\pgfsetlinewidth{\mcThickness}
\pgfpathmoveto{\pgfqpoint{0pt}{0pt}}
\pgfpathlineto{\pgfpoint{\mcSize+\mcThickness}{\mcSize+\mcThickness}}
\pgfusepath{stroke}
}}
\makeatother

 
\tikzset{
pattern size/.store in=\mcSize, 
pattern size = 5pt,
pattern thickness/.store in=\mcThickness, 
pattern thickness = 0.3pt,
pattern radius/.store in=\mcRadius, 
pattern radius = 1pt}
\makeatletter
\pgfutil@ifundefined{pgf@pattern@name@_iucl9rgf0}{
\pgfdeclarepatternformonly[\mcThickness,\mcSize]{_iucl9rgf0}
{\pgfqpoint{0pt}{0pt}}
{\pgfpoint{\mcSize+\mcThickness}{\mcSize+\mcThickness}}
{\pgfpoint{\mcSize}{\mcSize}}
{
\pgfsetcolor{\tikz@pattern@color}
\pgfsetlinewidth{\mcThickness}
\pgfpathmoveto{\pgfqpoint{0pt}{0pt}}
\pgfpathlineto{\pgfpoint{\mcSize+\mcThickness}{\mcSize+\mcThickness}}
\pgfusepath{stroke}
}}
\makeatother
\tikzset{every picture/.style={line width=0.75pt}} 



\caption{Infrared divergent diagrams in the in the 1-loop amplitude. Each diagram corresponds to a particular value of $(n,m)$ indicating the external particles to which the soft photon is attached.}\label{fig:ir}
\end{figure}
We choose $\Lambda$ to be the scale that defines the soft photons, and take $\mu$ as an IR cutoff in our loop integrals\footnote{As we use a cutoff regulator in this discussion, we take $D=4$.}.
Following Weinberg, we perform the $\ell^0$ integral using residues.
The remaining spatial momenta satisfy $\mu \leq|\bell| \leq \Lambda$.
IR divergences arise from the region where the virtual loop momentum $|\bell|$ is much smaller than the momenta of external particles $p_i$. 
In this region, it is possible to show that the IR divergent amplitudes take the form \cite{Weinberg:1965nx}: 
\begin{equation}\label{eq:IRexp}
    \mathcal{A}^{IR} =  \left(\frac{1}{2} \sum_{n m}  Q_n Q_m \eta_n \eta_m J_{n m} \right)  \times \mathcal{A}^{\text{Hard}},
\end{equation}
where $\mathcal{A}^{\text{Hard}}$ is what is left of the amplitude after removing the virtual photons lines. 
The divergent factor $J_{nm}$ is given by
 \begin{equation}
J_{n m} \equiv-i\left(p_n \cdot p_m\right) \int_{\mu \leq|\bell| \leq \Lambda} \frac{\hat{\dd}^4 \ell}{\left[\ell^2+i \epsilon\right]\left[p_n \cdot \ell+i \eta_n \epsilon\right]\left[-p_m \cdot \ell+i \eta_m \epsilon\right]},
\end{equation}
where $\eta_n = \pm 1$ for outgoing and incoming particles respectively. 

We evaluate the integral $J_{nm}$ by residues in the complex $\ell_0$ plane. The remaining integral over the spatial momenta $\bell$ yields the following contributions to the real and imaginary parts of $J_{nm}$:
\[\label{RealDivergenece}
&\Re  J_{nm} = \frac{1}{8\pi^2} \; \beta^{-1}_{n m} \; \ln \left(\frac{1+\beta_{n m}}{1-\beta_{n m}}\right) \ln \left(\frac{\Lambda}{\mu}\right), \\
&\Im  J_{nm} = -\frac{ \delta_{\eta_n \eta_m,1} }{4 \pi } \; \beta^{-1}_{n m}\ln \left(\frac{\Lambda}{\mu}\right),
\]
where $\beta_{nm}$ is the relative velocity
\begin{equation}
\beta_{n m} \equiv \sqrt{1-\frac{m_n^2 m_m^2}{\left(p_n \cdot p_m\right)^2}} .
\end{equation}
Notice that the factor $\delta_{\eta_n \eta_m,1}$ in \eqref{RealDivergenece} restricts the imaginary part to diagrams where particles $n$ and $m$ are either both outgoing or both incoming. In other words, diagrams of the crossed box topology do not contribute to the imaginary IR divergences at one loop. 
Let us also mention that equation \eqref{RealDivergenece} can be derived by cutting massive propagators in the spirit of section~\ref{sec:PV} before performing any integration. 

Noting that the one-loop amplitude involves an additional factor $Q^2 / \hbar$ relative  to the tree amplitude, we see that infrared divergences could introduce classically singular ``superclassical'' terms, in addition to classical IR divergences at one subleading order $\hbar$.
Quantum parts of $J_{nm}$ are those of order at least $\hbar^2$, and will be neglected here.
We denote the incoming and outgoing massive momenta by $p_1,p_2$ and $p_1' , p_2'$ respectively such that $\eta_1 = \eta_2 = -1$ and $\eta_1' = \eta_2' =+1$ and  $Q_1 = Q_1'$ , $Q_2 = Q_2'$.  
Using momentum conservation, we write
\begin{equation}
\begin{aligned}
      &p_1' = p_1 + q, \\
      & p_2' = p_2 - q - k .
\end{aligned}
\end{equation}
Recall from section~\ref{sec:hbar} that $q$ and $k$ are of order $\hbar$. Furthermore, from the on-shell conditions we have
\begin{equation}\label{Kinematics}
    \begin{aligned}
        &p_1 \cdot q  = \mathcal{O}(\hbar^2), \\
        &p_2 \cdot q =  - p_2 \cdot k + \mathcal{O}(\hbar^2).
    \end{aligned}
\end{equation}
Using this information, it is straightforward to expand $J_{nm}$ in powers of $\hbar$; we will find it convenient to express our results in terms of the Lorentz factor $\gamma \equiv \gamma_{12}$ and $\beta_{12} \equiv \beta$. 

\subsubsection*{Real divergence}

We now examine the $\hbar$ expansion of the real divergences in \eqref{RealDivergenece}, which are IR divergences of the real part of the amplitude.
(Recall that the real parts of the amplitude and expectation are equal.)
In particular, we examine the $\hbar$ expansion of the sum
\begin{equation}\label{sum}
\sum_{n m} Q_n Q_m \eta_n \eta_m J_{n m} 
\end{equation}
which controls the total IR divergence of equation~\eqref{eq:IRexp}.

To ensure that these divergences are  quantum, we must show that the $\mathcal{O}(\hbar^0)$ and $\mathcal{O}(\hbar)$ terms vanish in the sum over $(n,m)$ in the expression~\eqref{sum}. 
Considering the  $\mathcal{O}(\hbar^0)$ terms first, it is easy to check the sum of terms cancels exactly. For example the terms proportional to $Q_1^2$ are
\begin{equation}
     Q_1^2 \Re \; (\eta_1\eta_1 J_{1 1} +  2 \eta_1 \eta_1' J_{1 1'} +  \eta'_1\eta'_1 J_{1' 1'}).
\end{equation}
To this order in $\hbar$, we know $p_1 \cdot p_1 = p_1 \cdot p_1' = p_1' \cdot p_1' = m_1^2 $ so that $J_{11} = J_{11'} = J_{1'1'}$. 
The sum vanishes due to the sign differences arising from the $\eta$ factors. 
It is easy to verify that a similar cancellation occurs for the terms proportional to $Q_2^2$ and $Q_1 Q_2$. We conclude that the $O(\hbar^0)$ terms do not contribute to the real IR divergences. 

Turning to the $O(\hbar)$ terms, we start by noting that the terms in the sum proportional to $Q_i^2$ for $i=1,2$ still cancel in the same manner as before. This is because the equality $p_i \cdot p_i = p_i \cdot p_i' = p_i' \cdot p_i'$ still holds to this order in $\hbar$. We therefore only need to look at the terms proportional to $Q_1 Q_2$. Noting that the terms $J_{12} = J_{21}$ do not contribute powers of $\hbar$, we are left with: 
\begin{equation}
    Q_1 Q_2  \text{Re} \; (\eta_1\eta'_2 J_{1 2'} +  \eta'_1 \eta_2 J_{1' 2} +  \eta'_1\eta'_2 J_{1' 2'}).
\end{equation}
Expanding each term to linear order in $\hbar$ using \eqref{RealDivergenece} and the kinematics \eqref{Kinematics} we find
\begin{equation}
\begin{split}
 &\text{Re} \; \eta_1\eta'_2 J_{1 2'} = \frac{  u_1 \cdot q_2\left(\beta \ln (\frac{1+\beta}{1-\beta}) - 2\gamma^2 +2 \right)}{8 \pi^2 \, m_2 \, \beta^4 \gamma^3}\ln \left(\frac{\Lambda}{\mu}\right),
  \\&
   \text{Re} \; \eta'_1\eta_2 J_{1' 2} = \frac{  u_2 \cdot q_1\left(\beta \ln (\frac{1+\beta}{1-\beta}) - 2\gamma^2 +2 \right)}{8 \pi^2 \, m_1 \, \beta^4 \gamma^3}\ln \left(\frac{\Lambda}{\mu}\right),
  \\& \text{Re} \; \eta'_1\eta'_2 J_{1' 2'} =  - \text{Re} \; \eta_1\eta'_2 J_{1 2'}-\text{Re} \; \eta'_1\eta_2 J_{1' 2}.
    \end{split}
\end{equation}
Thus, these terms once again cancel in the sum. Having established this, we conclude that the real infrared divergences do not contribute classically. 

\subsubsection*{Imaginary part}

While studying imaginary parts of divergences, we do need to distinguish between the amplitude and the expectation.
Otherwise the $\hbar$ expansion of the imaginary IR divergences proceeds in a similar manner as in the previous subsection, now expanding $\Im  J_{nm}$ in equation~\eqref{RealDivergenece}.
In the imaginary case, the sum does not run over all pairs $(n,m)$ but only those for which $\eta_n = \eta_m$. 
This restriction on the sum prevents both the superclassical and classical contributions from cancelling in the one-loop amplitude. However, the cancellation of superclassical terms occurs at the level of the expectation $\E(p_1 p_2 \rightarrow  p_1' p_2' k_\eta)$, where the superclassical IR divergence of the one-loop amplitude is compensated by similar IR divergences originating from the unitarity cut. On the other hand, the classical IR divergence of the amplitude is 
\[\label{ImIRQED}
\Im \; J_{1'2'} =-\frac{1}{4 \pi} \frac{p_1 \cdot k + p_2 \cdot k}{m_1 m_2 (\gamma^2 -1)^{3/2}} \ln \left(\frac{\Lambda}{\mu}\right) \,.
\]
Unlike the superclassical part, this IR divergence does not cancel at the level of the expectation, but is in fact amplified by a factor of two~\cite{Caron-Huot:2023vxl}. Nevertheless, these imaginary IR divergences exponentiate into an overall phase in the waveform.  

\subsubsection*{QCD and gravity}
The analysis of IR divergences in QCD and gravity proceeds broadly in the same manner, albeit with two major differences. 
First, in both theories, soft divergences also arise in diagrams where a soft messenger connects a massive to a \emph{massless} line. Imaginary IR divergences in such diagrams also have classical implications, and are discussed in references~\cite{Goldberger:2009qd,Porto:2012as,wave1,wave2}.
Second, in QCD, collinear divergences arise at the level of the amplitude. It would be interesting to explore the classical implications of these collinear divergences in future.

\subsection{Ultraviolet divergences}\label{renormalix}

\begin{figure}[t]
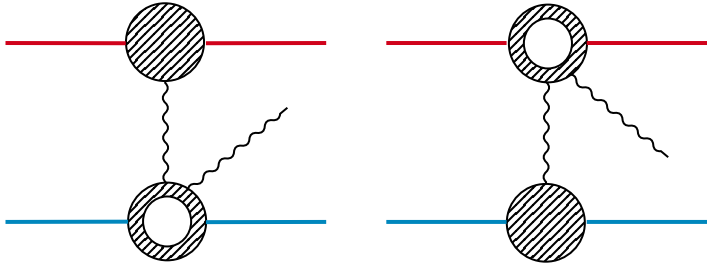

    \centering

 
\tikzset{
pattern size/.store in=\mcSize, 
pattern size = 5pt,
pattern thickness/.store in=\mcThickness, 
pattern thickness = 0.3pt,
pattern radius/.store in=\mcRadius, 
pattern radius = 1pt}
\makeatletter
\pgfutil@ifundefined{pgf@pattern@name@_83g5r5roy}{
\pgfdeclarepatternformonly[\mcThickness,\mcSize]{_83g5r5roy}
{\pgfqpoint{0pt}{0pt}}
{\pgfpoint{\mcSize+\mcThickness}{\mcSize+\mcThickness}}
{\pgfpoint{\mcSize}{\mcSize}}
{
\pgfsetcolor{\tikz@pattern@color}
\pgfsetlinewidth{\mcThickness}
\pgfpathmoveto{\pgfqpoint{0pt}{0pt}}
\pgfpathlineto{\pgfpoint{\mcSize+\mcThickness}{\mcSize+\mcThickness}}
\pgfusepath{stroke}
}}
\makeatother

 
\tikzset{
pattern size/.store in=\mcSize, 
pattern size = 5pt,
pattern thickness/.store in=\mcThickness, 
pattern thickness = 0.3pt,
pattern radius/.store in=\mcRadius, 
pattern radius = 1pt}
\makeatletter
\pgfutil@ifundefined{pgf@pattern@name@_mn6djfd9c}{
\pgfdeclarepatternformonly[\mcThickness,\mcSize]{_mn6djfd9c}
{\pgfqpoint{0pt}{0pt}}
{\pgfpoint{\mcSize+\mcThickness}{\mcSize+\mcThickness}}
{\pgfpoint{\mcSize}{\mcSize}}
{
\pgfsetcolor{\tikz@pattern@color}
\pgfsetlinewidth{\mcThickness}
\pgfpathmoveto{\pgfqpoint{0pt}{0pt}}
\pgfpathlineto{\pgfpoint{\mcSize+\mcThickness}{\mcSize+\mcThickness}}
\pgfusepath{stroke}
}}
\makeatother

 
\tikzset{
pattern size/.store in=\mcSize, 
pattern size = 5pt,
pattern thickness/.store in=\mcThickness, 
pattern thickness = 0.3pt,
pattern radius/.store in=\mcRadius, 
pattern radius = 1pt}
\makeatletter
\pgfutil@ifundefined{pgf@pattern@name@_g16io79yu}{
\pgfdeclarepatternformonly[\mcThickness,\mcSize]{_g16io79yu}
{\pgfqpoint{0pt}{0pt}}
{\pgfpoint{\mcSize+\mcThickness}{\mcSize+\mcThickness}}
{\pgfpoint{\mcSize}{\mcSize}}
{
\pgfsetcolor{\tikz@pattern@color}
\pgfsetlinewidth{\mcThickness}
\pgfpathmoveto{\pgfqpoint{0pt}{0pt}}
\pgfpathlineto{\pgfpoint{\mcSize+\mcThickness}{\mcSize+\mcThickness}}
\pgfusepath{stroke}
}}
\makeatother

 
\tikzset{
pattern size/.store in=\mcSize, 
pattern size = 5pt,
pattern thickness/.store in=\mcThickness, 
pattern thickness = 0.3pt,
pattern radius/.store in=\mcRadius, 
pattern radius = 1pt}
\makeatletter
\pgfutil@ifundefined{pgf@pattern@name@_yh6y5sqjk}{
\pgfdeclarepatternformonly[\mcThickness,\mcSize]{_yh6y5sqjk}
{\pgfqpoint{0pt}{0pt}}
{\pgfpoint{\mcSize+\mcThickness}{\mcSize+\mcThickness}}
{\pgfpoint{\mcSize}{\mcSize}}
{
\pgfsetcolor{\tikz@pattern@color}
\pgfsetlinewidth{\mcThickness}
\pgfpathmoveto{\pgfqpoint{0pt}{0pt}}
\pgfpathlineto{\pgfpoint{\mcSize+\mcThickness}{\mcSize+\mcThickness}}
\pgfusepath{stroke}
}}
\makeatother
\tikzset{every picture/.style={line width=0.75pt}} 


\caption{Diagrams contributing to the $Q_1Q_2^4$ and $Q_1^4Q_2$ charge sectors. }
\label{fig:ren1}
\end{figure}

We now turn to ultraviolet divergences arising in the waveshape. 
More specifically, we study the real part of the waveshape in the $Q_1 Q_2^4$ and $Q_1^4 Q_2$ charge sectors, which was omitted in section~\ref{examplecalc}.
This charge sector is ultraviolet divergent\footnote{The imaginary part of the waveshape is UV finite. At one loop, all UV divergences must be absorbed by the real counterterms.}.
In this section, we carry out the renormalisation of the relevant diagrams in the on-shell scheme. 
We then examine the $\hbar$ scaling of the remaining finite contributions and show that they do not contribute to the classical waveshape. 

The diagrams in this charge sector are constructed by sewing a tree-level three-point amplitude with a one-loop Compton amplitude as shown in figure \ref{fig:ren1}, with the latter containing all UV divergences. It is therefore sufficient for our purposes to consider the renormalisation of the one-loop Compton amplitude. Specifically, we consider loop corrections to the scalar propagator, the three-point vertex and four-point vertex. We will not consider corrections to the photon propagator since these can be easily seen to be purely quantum corrections as shown in \cite{Kosower:2018adc}. 

To start, we rewrite the bare Lagrangian for scalar QED in terms of the renormalised fields by defining
\begin{equation}
\begin{aligned}
      \phi = \sqrt{Z_2}\phi_0 \,,  \\
      A^\mu = \sqrt{Z_3} A_0^\mu \,, \\
      Q = \sqrt{Z_Q}Q_0 \,,
\end{aligned} 
\end{equation}
where the bare fields and charges are distinguished by a subscript. Using this definition we can write the bare Lagrangian in terms of renormalised fields so that
\begin{equation}
\begin{aligned}
\mathcal{L}_{\text {bare }}
= & -\frac{1}{4} Z_3 F_{\mu \nu} F^{\mu \nu}+Z_2 \partial^\mu \phi^* \partial_\mu \phi-Z_2 m_0^2 \phi^* \phi \\
& +i Q Z_1 A^\mu\left(\phi^* \partial_\mu \phi-\phi \partial_\mu \phi^*\right)+Q^2 Z_4 A^\mu A_\mu \phi^* \phi,
\end{aligned}
\end{equation}
where we have further defined $ Z_1 = \sqrt{Z_3 Z_Q}Z_2 $ and $Z_4 = Z_3  Z_2 Z_Q$. 
Expanding the bare Lagrangian using $Z_i = 1 + \delta_i$ and $Z_2 (m_0^2 -  m^2 ) = \delta_m$ leads to counterterms in the usual fashion.
\begin{figure}
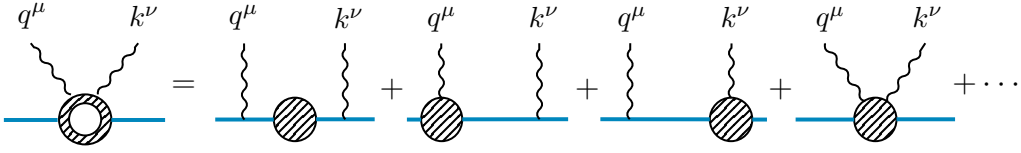

    \centering

 
\tikzset{
pattern size/.store in=\mcSize, 
pattern size = 5pt,
pattern thickness/.store in=\mcThickness, 
pattern thickness = 0.3pt,
pattern radius/.store in=\mcRadius, 
pattern radius = 1pt}
\makeatletter
\pgfutil@ifundefined{pgf@pattern@name@_6szu5rn01}{
\pgfdeclarepatternformonly[\mcThickness,\mcSize]{_6szu5rn01}
{\pgfqpoint{0pt}{0pt}}
{\pgfpoint{\mcSize+\mcThickness}{\mcSize+\mcThickness}}
{\pgfpoint{\mcSize}{\mcSize}}
{
\pgfsetcolor{\tikz@pattern@color}
\pgfsetlinewidth{\mcThickness}
\pgfpathmoveto{\pgfqpoint{0pt}{0pt}}
\pgfpathlineto{\pgfpoint{\mcSize+\mcThickness}{\mcSize+\mcThickness}}
\pgfusepath{stroke}
}}
\makeatother

 
\tikzset{
pattern size/.store in=\mcSize, 
pattern size = 5pt,
pattern thickness/.store in=\mcThickness, 
pattern thickness = 0.3pt,
pattern radius/.store in=\mcRadius, 
pattern radius = 1pt}
\makeatletter
\pgfutil@ifundefined{pgf@pattern@name@_u0x3renxm}{
\pgfdeclarepatternformonly[\mcThickness,\mcSize]{_u0x3renxm}
{\pgfqpoint{0pt}{0pt}}
{\pgfpoint{\mcSize+\mcThickness}{\mcSize+\mcThickness}}
{\pgfpoint{\mcSize}{\mcSize}}
{
\pgfsetcolor{\tikz@pattern@color}
\pgfsetlinewidth{\mcThickness}
\pgfpathmoveto{\pgfqpoint{0pt}{0pt}}
\pgfpathlineto{\pgfpoint{\mcSize+\mcThickness}{\mcSize+\mcThickness}}
\pgfusepath{stroke}
}}
\makeatother

 
\tikzset{
pattern size/.store in=\mcSize, 
pattern size = 5pt,
pattern thickness/.store in=\mcThickness, 
pattern thickness = 0.3pt,
pattern radius/.store in=\mcRadius, 
pattern radius = 1pt}
\makeatletter
\pgfutil@ifundefined{pgf@pattern@name@_auycuhz17}{
\pgfdeclarepatternformonly[\mcThickness,\mcSize]{_auycuhz17}
{\pgfqpoint{0pt}{0pt}}
{\pgfpoint{\mcSize+\mcThickness}{\mcSize+\mcThickness}}
{\pgfpoint{\mcSize}{\mcSize}}
{
\pgfsetcolor{\tikz@pattern@color}
\pgfsetlinewidth{\mcThickness}
\pgfpathmoveto{\pgfqpoint{0pt}{0pt}}
\pgfpathlineto{\pgfpoint{\mcSize+\mcThickness}{\mcSize+\mcThickness}}
\pgfusepath{stroke}
}}
\makeatother

 
\tikzset{
pattern size/.store in=\mcSize, 
pattern size = 5pt,
pattern thickness/.store in=\mcThickness, 
pattern thickness = 0.3pt,
pattern radius/.store in=\mcRadius, 
pattern radius = 1pt}
\makeatletter
\pgfutil@ifundefined{pgf@pattern@name@_q0o8xjq9q}{
\pgfdeclarepatternformonly[\mcThickness,\mcSize]{_q0o8xjq9q}
{\pgfqpoint{0pt}{0pt}}
{\pgfpoint{\mcSize+\mcThickness}{\mcSize+\mcThickness}}
{\pgfpoint{\mcSize}{\mcSize}}
{
\pgfsetcolor{\tikz@pattern@color}
\pgfsetlinewidth{\mcThickness}
\pgfpathmoveto{\pgfqpoint{0pt}{0pt}}
\pgfpathlineto{\pgfpoint{\mcSize+\mcThickness}{\mcSize+\mcThickness}}
\pgfusepath{stroke}
}}
\makeatother

 
\tikzset{
pattern size/.store in=\mcSize, 
pattern size = 5pt,
pattern thickness/.store in=\mcThickness, 
pattern thickness = 0.3pt,
pattern radius/.store in=\mcRadius, 
pattern radius = 1pt}
\makeatletter
\pgfutil@ifundefined{pgf@pattern@name@_z33qrmdpf}{
\pgfdeclarepatternformonly[\mcThickness,\mcSize]{_z33qrmdpf}
{\pgfqpoint{0pt}{0pt}}
{\pgfpoint{\mcSize+\mcThickness}{\mcSize+\mcThickness}}
{\pgfpoint{\mcSize}{\mcSize}}
{
\pgfsetcolor{\tikz@pattern@color}
\pgfsetlinewidth{\mcThickness}
\pgfpathmoveto{\pgfqpoint{0pt}{0pt}}
\pgfpathlineto{\pgfpoint{\mcSize+\mcThickness}{\mcSize+\mcThickness}}
\pgfusepath{stroke}
}}
\makeatother
\tikzset{every picture/.style={line width=0.75pt}} 


\caption{Diagrams contributing to the Compton amplitude at one loop. The shaded blobs denote 1PI corrections to the propagator, cubic vertices, and quartic vertex respectively. The ellipses contain the crossed ($q^\mu \leftrightarrow k^\nu$) counterparts of the first three diagrams.}
\label{fig:ren2}
\end{figure}
The counterterms $\delta_1, \delta_2, \delta_m$ and $\delta_4$ are chosen to cancel the divergences of the one-loop diagrams. There remains, however, the freedom to choose the finite part of these counterterms --- that is, the choice of scheme.
We find it convenient to work in the on-shell scheme defined below. 

In what follows, we will consider the three types of vertex corrections indicated in figure~\ref{fig:ren2}. Starting with the self energy corrections, we define
\begin{equation}
    \Sigma(p^2) =  \Sigma_{\text{loop}}(p^2) + \Sigma_{\text{ct}}(p^2) ,
\end{equation}
where the first and second terms refer to the one-loop  and counterterm contributions respectively. The 1PI cubic and quartic vertices are defined in a similar manner
\begin{equation}
    \begin{aligned}
         &\Gamma^{{\mu}}(p,q) =  \Gamma^{{\mu}}_{\text{loop}}(p,q) + \Gamma^{{\mu}}_{\text{ct}}(p,q), \\
         &\Gamma^{\mu \nu}(p,q , k) =  \Gamma^{\mu \nu}_{\text{loop}}(p,q , k) + \Gamma^{\mu \nu}_{\text{ct}}(p,q ,k ),
    \end{aligned}
\end{equation}
where $\Gamma^{{\mu}}(p,q)$ and $\Gamma^{\mu \nu}(p,q , k)$ refer to cubic and quartic vertices with incoming scalar momentum $p$ and outgoing photon momenta $q$ and $k$ (see figure~\ref{fig:ren2}). 
Note that we do not include the tree-level vertices in our definition of $\Gamma^{{\mu}}$ and $\Gamma^{\mu \nu}$. 

We now define our renormalisation conditions in the on-shell scheme. 
For the self-energy terms the values of $\delta_2$ and $\delta_m$ are chosen such that, in the limit $p^2 \rightarrow m^2$, the following renormalization conditions are satisfied: 
\begin{equation}\label{OnShellSelf}
    \begin{aligned}
        &\;\Sigma( m^2) = 0, \\
        &\frac{\dd}{\dd p^2} \Sigma(p^2 \rightarrow m^2 )  =  0 ,
    \end{aligned}
\end{equation}
where it is understood that the on-shell condition is applied after taking the required derivative. These conditions ensure that the renormalised propagator coincides with the free propagator near the pole $p^2 = m^2$. 
This allows us to neglect self energy corrections in external matter lines since the on-shell renormalised propagator is still truncated by the inverse (free) propagator factors in the LSZ formula\footnote{Notice that infrared divergences are no obstruction to the on-shell scheme in this case. In particular there are no real, classical, IR divergences.}. 

Turning to interaction vertices, we will soon see that only the renormalisation of the quartic vertex will be of direct relevance to us\footnote{Of course the Ward identity relates the counterterms $\delta_4$, $\delta_1$ and $\delta_2$.}.
The renormalisation conditions for the  vertices are defined in the limit of zero photon momenta. 
In this limit the quartic vertex is of the form:
\begin{equation}\label{Form}
\begin{aligned}
    & i\Gamma^{\mu \nu}(p,q \rightarrow 0 , k \rightarrow 0) = Q^2 F_1(p^2) \eta^{\mu \nu} + Q^2 F_2(p^2) p^\mu p^\nu. 
\end{aligned}
\end{equation}
It is easy to verify that the coefficient $F_2(p^2)$ is UV finite, as is expected from simple power counting. The divergence in $F_1(p^2)$ is absorbed by the counterterm $\delta_4$ according to the renormalisation condition
\begin{equation}\label{ren}
    F_1( m^2) = 0 \,.
\end{equation}

It remains to determine whether the renormalised vertex corrections survive in the classical limit. 
To do this we consider the $\hbar$ expansion of the renormalised one-loop Compton amplitude. 
The one-loop amplitude is obtained from the tree-level one by replacing (one of) the tree-level vertices by their loop-corrected counterparts,
\[
&\Gamma_{\text{tree}}^{\mu }(p,q) \rightarrow \Gamma^{\mu }(p,q), \\
&\Gamma_{\text{tree}}^{\mu \nu}(p,q) \rightarrow \Gamma^{\mu \nu}(p,q),
\]
or by replacing the free propagator by 
\[
\frac{i}{p^2 - m^2 + i \epsilon} \rightarrow G(p) =  \frac{i}{p^2 - m^2 + i \epsilon}\left[i\Sigma(p^2)\right] \frac{i}{p^2 - m^2 + i \epsilon}. \\
\]
Applying these replacements to the tree-level amplitude generates three types of diagrams: the self-energy diagrams 
\begin{equation}\label{DSelf}
    \begin{aligned}
        \mathcal{D}^{\mu\nu}_{\text{self}}(p,k,q) =  (iQ)^2 (2p - q)^\mu G(p-q) (2p-2 q-k)^\nu  \\
        +  (iQ)^2 (2p - k)^\nu G(p-k) (2p-q-2k)^\mu \,;
    \end{aligned}
\end{equation}
the cubic vertex corrections 
\begin{equation}\label{Cubic}
\begin{aligned}
    \mathcal{D}^{\mu \nu}_{\text{Cubic}}(p,k,q) = (iQ) \frac{ i\Gamma^\mu(p,q) (2p-2q-k)^\nu + (2p-q)^\mu  i\Gamma^\nu(p-q,k) }{(p-q)^2 - m^2 + i \epsilon}  \\
    + (iQ) \frac{ i\Gamma^\nu(p,k) (2p-2k-q)^\mu + (2p-k)^\nu  i\Gamma^\mu(p-k,q) }{(p-k)^2 - m^2 + i \epsilon}; 
\end{aligned}
\end{equation}
and finally,  the four-point vertex correction
\begin{equation}
    \mathcal{D}^{\mu \nu}_{\text{Quartic}}(p,k,q) = i \Gamma^{\mu \nu}(p,q,k) .
\end{equation}
At this point we can exploit gauge invariance of the Compton amplitude. 
It is very convenient to choose polarisation vectors such that
\[
\varepsilon^*(k) \cdot p = \varepsilon^*(q) \cdot p =  0.
\]
Now, considering the gauge invariant amplitude
\[
\mathcal{M}(p,q,k) \equiv \varepsilon^*_\mu(q) \varepsilon^*_\nu(k) \left( \mathcal{D}^{\mu\nu}_{\text{self}} + \mathcal{D}^{\mu \nu}_{\text{Cubic}}+ \mathcal{D}^{\mu\nu}_{\text{Quartic}} \right),
\]
where each external line is on shell, it is easy to verify that
\[
\varepsilon^*_\mu(q) \varepsilon^*_\nu(k) \,    \mathcal{D}^{\mu\nu}_{\text{self}}(p,k,q) = 0.  
\]
The contribution $\mathcal{D}^{\mu \nu}_{\text{Cubic}}(p,k,q)$ also vanishes since we may expand
\[
\Gamma^\mu(p,q) = f_1(p,q) \, p^\mu + f_2(p,q) \, q^\mu
\]
as a consequence of Lorentz invariance.
The tensor structure is annihilated by the polarisation vectors so we do not need to study the form factors $f_1(p,q)$ and $f_2(p,q)$.
The quartic vertex, being a symmetric function of $p$ and $p'\equiv p-q-k$, takes the generic form:
\[
i \Gamma^{\mu \nu}(p,p') = Q^2 F_1(p,p') \eta^{\mu \nu} + Q^2 \left[F_{2,a}(p,p')  (p^{\mu}p^{\nu} 
+ p'^\mu p'^\nu) + F_{2,b}(p,p')  p^{(\mu}p'^{\nu)} \right],
\]
where the functions $F_1(p,p')$, $F_{2,a}(p,p')$ and $F_{2,b}(p,p')$ reduce appropriately to $F_1(p^2)$ and $F_2(p^2)$ in the $q,k \rightarrow 0$ limit. 
Contracting this with the polarisation vectors, we find that the only nonvanishing terms are
\[\label{EpsGamma}
i \varepsilon^*_\mu(q) \varepsilon^*_\nu(k) \,  \Gamma^{\mu \nu}(p,p') = Q^2 \left( F_1(p,p') \, \varepsilon^*(q) \cdot \varepsilon^*(k) 
+ F_{2,a}(p,p') \, \varepsilon^*(k) \cdot q \, \varepsilon^*(q) \cdot k \right) .
\]
It remains to check whether this contribution vanishes in the classical limit. 
To do this, let us note that the tree-level Compton amplitude carries classical scaling in this gauge. 
Using the Feynman rules, it is easy to check that $F_i(p,p')$ involve one power of $1/\hbar$ relative to the tree.
This implies that the second term in \eqref{EpsGamma} is $\mathcal{O}(\hbar)$ since both $q$ and $k$ are proportional to $\hbar$. 
In fact, the first term also contributes at $\mathcal{O}(\hbar)$. 
To see this, we expand $F_1(p,p')$ as 
\[
F_1(p,p-q-k) = F_1(p^2)  - 2 p \cdot ( q +k) F_1'(p^2) + \cdots .
\]
As $p^2 = m^2$ in the amplitude, we see that first term above vanishes by the renormalisation condition \eqref{ren}, while the second term is $\mathcal{O}(\hbar)$ since $2p \cdot (q+k) = (q+k)^2 \sim \mathcal{O}(\hbar^2)$ on shell. Overall, we find that the one-loop Compton amplitude is suppressed by a power of $\hbar$ relative to the tree-level amplitude. This establishes that the one-loop vertex corrections do not survive in the classical limit, and justifies omitting the $Q_1 Q_2^4$ charge sector from our discussion in section~\ref{examplecalc}.

\section{Classical confirmation}\label{sec:confirm}

This part of the paper is devoted to confirming our work by integrating the classical equations of motion.
We begin by focusing on the conservative part. Since the methods are standard, we will be brief.

The first order of business is to decide what classical object should be compared to the waveshape. 
In Lorenz gauge, the momentum-space Maxwell equation relate the gauge potential to a source according to
\[\label{eq:MEj}
-k^2 \tilde A_\mu(k) = \tilde J_\mu(k) \,.
\]
We have written $\tilde A_\mu(k)$ for the Fourier components of the gauge field $A_\mu(x)$, and similarly $\tilde J_\mu(k)$ are the Fourier components of the source.
The position-space field strength is 
\[
F_{\mu\nu}(x) = i \int \hd^D k \, \frac{e^{-ik \cdot x}}{k^2} k_{[\mu} \, \tilde J_{\nu]}(k) \,,
\]
where the integral over $k$ is defined with retarded boundary conditions.
Standard manipulation of the $k$ integral at large distances $x$ from the scattering event leads to the field strength in the form
\[
F_{\mu\nu}(x) = \frac{1}{4\pi |\v x|} 2 \Re \int_0^\infty \hd \omega \, e^{-i \omega u} \sum_\eta k_{[\mu} \varepsilon^\eta_{\nu]}(k) \left( i \varepsilon^{\eta*}(k) \cdot \tilde J(k) \right) \,,
\]
evaluated in the on-shell limit $k^2 \rightarrow 0$ with $k = (\omega, \omega {\v x}/ |\v x|)$. The retarded time appearing here is $u = x^0 - |\v x|$. Comparing with equation~\eqref{eq:laterUse}, we identify the classical counterpart of the waveshape as
\[
\label{eq:classicalrelation}
\waveshape &= \braket{ \psi | S^\dagger a_\eta(k) S | \psi }  \\
&= -i \varepsilon^{\eta*}(k) \cdot \tilde J(k) \,.
\]

We first discuss this classical waveshape in the large $m_1$ limit, which simplifies the problem since we can presume that particle 1 is stationary and therefore does not radiate. 
Leaving aside the self-field of particle 2 for now (we will include it below using the ALD force) we need only consider the motion of particle 2 in the Coulomb field of particle 1. The radiation field is entirely due to this accelerated motion, so the problem is to determine the field of particle 2 in perturbation theory. 
The Maxwell equation to be solved is then explicitly
\[
\label{eq:MEft}
-k^2 \tilde A^\mu(k) = \tilde J^\mu(k) = Q_2 \int \dd \tau \, v_2^\mu(\tau) \, e^{i k \cdot x_2(\tau)} \,,
\]
where $x_2(\tau)$ is the position of particle 2 and $v_2(\tau)$ is its proper velocity. The position and velocity can be written as a perturbative series
\[
x_2(\tau) = b_2 + u_2 \tau + \Delta^{(1)}x_2(\tau) + \Delta^{(2)} x_2(\tau) + \cdots\,,
\]
around the straight-line trajectory $b_2 + u_2 \tau$. 
The objects $\Delta^{(n)}x_2(\tau)$ are corrections to the trajectory at a given perturbative order.
Assuming that the perturbations of the motion of particle 2 are entirely due to the Lorentz force of particle 1, it is an easy power-counting exercise to show that
$\Delta^{(n)}x_2(\tau)$ is of order $Q_1^n Q_2^n$. Similarly, the velocity can be written as
\[\label{velpert}
v_2 (\tau) = u_2 + \Delta^{(1)}v_2(\tau) + \Delta^{(2)} v_2(\tau) + \cdots\,.
\]
In this notation, the order $Q_1^2 Q_2^3$ part of the acceleration field is obtained by expanding equation~\eqref{eq:MEft} with the result
\[\label{ClassicalRadiation}
i k^2 \Delta^{(2)} \tilde A^\mu(k)  = -iQ_2 \int \dd \tau \,& e^{i k \cdot (b_2 + u_2 \tau)} \left[ \Delta^{(2)} v_2^\mu(\tau) + i k \cdot \Delta^{(1)}x_2(\tau) \Delta^{(1)}v_2^\mu(\tau) \phantom{\frac12} \right. \\
&\quad\left. + u_2^\mu \left( i k \cdot \Delta^{(2)} x_2(\tau) - \frac12 (k \cdot \Delta^{(1)} x_2(\tau))^2 \right) \right]  \,.
\]
In view of equations~\eqref{eq:MEj} and~\eqref{eq:classicalrelation} we are only interested in the projection of this radiation field onto a polarisation vector in the on-shell limit $k^2 \rightarrow 0$. We are then free to choose the gauge of the polarisation vector to suit us. It is clear that the choice $\varepsilon^{\eta*}(k) \cdot u_2 = 0$ simplifies the calculation (and indeed this is the choice we made in section~\ref{examplecalc}). 

It remains to compute the quantities $\Delta^{(2)} v_2(\tau)$, $\Delta^{(1)}v_2(\tau)$ and $\Delta^{(1)}x_2(\tau)$. Of these, the first order perturbations $\Delta^{(1)}v_2(\tau)$ and $\Delta^{(1)}x_2(\tau)$ can be found in section 6 of reference~\cite{Kosower:2018adc}. The second-order term $\Delta^{(2)} v_2(\tau)$ can be computed by iterating the Lorentz force at one further order than performed in that reference, using precisely the same method. The result is
\[
\Delta^{(2)} v_2^\mu(\tau) = \frac{i Q_1^2 Q_2^2}{m_2^2} \int \hd^D \ell_1 \hd^D \ell_2 \frac{\hdelta(\ell_1 \cdot u_1) \hdelta(\ell_2 \cdot u_2)}{\ell_1^2 \ell_2^2} \frac{e^{- i (\ell_1 + \ell_2) \cdot (b_2 + u_2\tau)}}{(\ell_1 \cdot u_2 + i \epsilon)^2 ((\ell_1 + \ell_2) \cdot u_2 + i \epsilon)^2} \\
 \times \ell_2^{[\mu} u_1^{\nu]} \left( \ell_1 \cdot u_2 \ell_{1[\nu} u_{1\rho]} u_2^\rho + u_{2 \nu} \ell_{2\rho} \ell_1^{[\rho} u_1^{\sigma]} u_{2\sigma} \right) \,.
\]

We must now combine the results to extract the waveshape. 
After some relabelling of the variables of integration, we find 
\[\label{classico}
\hspace{-5pt}
-i \varepsilon^{\eta*} \cdot \tilde J(k) = \int \hd^D q_1 \hd^D q_2 \, \hdelta(2 p_1 \cdot q_1)\hdelta(2 p_2 \cdot q_2) & \hdelta^D (k+ q_1 + q_2) e^{-i b_1 \cdot q_1} e^{-i b_2 \cdot q_2} \\
& \hspace{70pt} \times 4i m_1 m_2\E^\textrm{class} ,
\]
where
\[\label{classico2}
\E^\textrm{class} &\equiv \frac{Q_1^2 Q_2^3}{m_2^2} \int \hd^D \ell\frac{ \hdelta (\ell \cdot u_1)}{\ell^2 (\ell - q_1)^2}  \left[ \varepsilon^* \cdot u_1 \frac{u_1 \cdot u_2 \, k \cdot \ell -k\cdot u_1 \, \ell \cdot u_2 - \ell \cdot (\ell -q_1) u_1 \cdot u_2}{(\ell \cdot u_2 + i \epsilon)^2}  \right.\\&
\left. \hspace{2cm} + \varepsilon^* \cdot (\ell - q_1) \left( \frac{1}{k \cdot u_2} + \frac{1}{k \cdot u_2} \frac{(u_1 \cdot u_2)^2 \ell \cdot (\ell-q_1)}{(\ell \cdot u_2 + i \epsilon)^2}  \right.\right.\\&\left.\left. \hspace{5cm} +u_1 \cdot u_2 \frac{k \cdot u_1 \, \ell \cdot u_2 - k \cdot \ell \, u_1 \cdot u_2}{(\ell \cdot u_2 + i \epsilon)^2 ((\ell - q_1) \cdot u_2 + i \epsilon)}\right) \right] \,.
\]
The polarisation vector here is a shorthand: $\varepsilon^* = \varepsilon^{\eta*}(k)$.
We finally see the emergence of the generic structure of a waveshape at any order, as shown in equation~\eqref{eq:genericWaveshape}.
This is one advantage of the approach from amplitudes: the general form of the observable is clear at the outset.

To see how $\E^\textrm{class}$ relates to the QED waveshape in section~\ref{examplecalc}, we note it is of order $Q_1^2 Q_2^3$ and evaluated in the heavy $m_1$ limit.
Thus we compare to the single-cut in equation~\eqref{eq:heavySingleCutQED}. 
However, that single-cut is the real part of the five-point amplitude while $\E^\textrm{class}$ is complex on account of the $i \epsilon$'s.
Taking the real part of $\E^\textrm{class}$ converts these $i \epsilon$'s into PV pole prescriptions as used in section~\ref{examplecalc}.
The imaginary part of $\E^\textrm{class}$ is non-vanishing; we discuss it further in appendix~\ref{cutextra}.

To see that the real parts do indeed match, it is necessary to take advantage of properties of the integrals involved. For example we may neglect $\ell^2$ in the numerator of the loop integral: as this cancels one of the photon propagators the result will be a contact term. It is also useful to note, for example, that
\[
\int \hd^D \ell\frac{ \hdelta (\ell \cdot u_1)}{\ell^2 (\ell - q_1)^2}   \varepsilon^* \cdot (\ell - q_1) & \frac{\ell \cdot (\ell - q_1)}{\ell \cdot u_2 \, (\ell-q_1) \cdot u_2} 
\\ = - &\frac{1}{2}\int \hd^D \ell\frac{ \hdelta (\ell \cdot u_1)}{\ell^2 (\ell - q_1)^2}  \varepsilon^* \cdot  q_1 \frac{\ell \cdot (\ell - q_1)}{\ell \cdot u_2 \, (\ell-q_1) \cdot u_2} \,.
\]
When the dust settles, we do indeed find a complete match of $\Re \E^\textrm{class}$ with the single cut in equation~\eqref{eq:heavySingleCutQED}. 
The simplest point of comparison is the $\varepsilon^* \cdot u_1 \, u_1 \cdot u_2$ coefficient in equation~\eqref{classico2}.
Using conservation of momentum, and throwing away vanishing integrals, the term is
\[
-\frac{Q_1^2 Q_2^3}{m_1 m_2} \int \hd^D \ell\frac{ \hdelta (\ell \cdot p_1)}{\ell^2 (\ell - q_1)^2}  \varepsilon^* \cdot p_1 \frac{p_1 \cdot p_2 \, \ell \cdot q_2 }{(\ell \cdot p_2)^2}  \,.
\]
Taking the factor $4 m_1 m_2$ in equation~\eqref{classico} into account, this matches the same term in equation~\eqref{eq:heavySingleCutQED}.

Beyond the large $m_1$ limit, additional terms contribute with a symmetric mass dependence.
Returning to equation \eqref{ClassicalRadiation}, and recalling that the second line vanishes in $\varepsilon^{\eta*}(k) \cdot u_2 = 0$ gauge, we see that the symmetric-mass terms must originate from $\Delta^{(2)} v_2^\mu(\tau)$. This is because $\Delta^{(1)}x_2(\tau)$ and $  \Delta^{(1)}v_2^\mu(\tau)$ are proportional to $1/m_2$. We determine the correction $\Delta^{(2)} v_2^\mu(\tau)$ using the Lorentz force as before, this time extracting the terms proportional to $1/m_1m_2$ to obtain
\[
\Delta^{(2)} v_2^\mu(\tau) = \frac{Q_1^2 Q_2^2}{m_1 m_2 } \int  \hd^D \ell \; \hd^D q_1  &\; e^{i  (\ell-q_1) \cdot b} e^{- i \ell \cdot u_2 \tau} \frac{\hdelta(u_1 \cdot ( \ell -q_1)) \hdelta(u_2 \cdot q_1) }{q_1^2 \ell^2 (u_2 \cdot \ell - i \varepsilon)} \\
&\left[ \frac{- q_1^\mu \; \ell \cdot u_2 \; u_1 \cdot u_2 - \ell^{[\mu} u_2^{\nu]} \; q_1 \cdot u_1 u_{2\nu}}{q_1 \cdot u_1 + i \epsilon} \right.\\
& \left. - \frac{(\ell \cdot q_1 \; u_1 \cdot u_2 - \ell \cdot u_2  \; q_1 \cdot u_1)(\ell^{[\mu}u_1^{\nu]}u_{2\nu})}{( q_1 \cdot u_1+ i \epsilon)^2} \right ].
\]
Inserting this into \eqref{ClassicalRadiation} and changing variables we obtain the symmetric-mass part of the classical waveshape:
\[\label{ClassSymm}
\E^\textrm{class}_{m_1 m_2}= \frac{Q_1^2 Q_2^3}{m_1 m_2} \int & \frac{\hd^D \ell}{\ell^2 [(\ell-q_1)^2 + i \epsilon]} \; \frac{\hdelta(u_2 \cdot \ell)}{(u_2 \cdot q_1)} \varepsilon_\mu^{\eta*}(k) \left[ -\frac{\ell^\mu \; u_2 \cdot q_1 \; u_1 \cdot u_2- (\ell-q_1)^\mu u_1 \cdot \ell }{u_1 \cdot \ell + i \epsilon} \right.\\
& \left. - \frac{(\ell \cdot q_1\; u_1 \cdot u_2 - u_2\cdot q_1 \; u_1 \cdot \ell)((\ell-q_1)^\mu u_1 \cdot u_2 + u_1^\mu \; u_2 \cdot q_1 ) }{(u_1 \cdot \ell + i \epsilon)^2}\right]. 
\]
The real part of this quantity now reproduces the single-cut of equation~\eqref{eq:scut2qed} as discussed above.
Notice that we included a pole prescription for the $\ell-q_1$ Green's function. 
This photon can go on-shell, and is one classical source of Compton cuts. 
The retarded pole prescription for the photon yields the same result because the cut conditions force $\ell^0 - q_1^0 >0$.

Let us now turn to radiation reaction.
In electrodynamics, we are lucky that radiation reaction can be systematically computed through the ALD  force \cite{lorentz1892theorie, abraham, Dirac:1938nz}. This non-conservative force prescribes the following momentum kick on particle 2:
\begin{equation}\label{ald}
\frac{\dd p^\mu_2  }{\dd \tau}=\frac{Q_2^2}{6\pi m_2}\left(
\frac{\dd^2 p^\mu_2 }{\dd \tau^2}+\frac{p_2^\mu}{m_2^2} \frac{\dd p_2}{\dd\tau}\cdot \frac{\dd p_2}{\dd\tau}
\right),
\end{equation}
which supplements the lower order deflection due to the Lorentz force.

To see how \eqref{ald} feeds in $F^{\mu\nu}(x)$ at order $Q_1 Q_2^4$ we start from the LO velocity correction  updated through the Lorentz force only. If we expand perturbatively the  four velocity  of particle 2 just as in \eqref{velpert},
 then the first correction to the constant term $u_2^\mu$ is \cite{Kosower:2018adc}
\begin{equation}
\begin{split}
& \Delta^{(1)} v^\mu_2(\tau)=-\frac{Q_1 Q_2}{m_2} \int \hat{\dd}^D \bq \,\hat{\delta}(\bq\cdot u_1)e^{i \bq\cdot (b_{12}- u_2 \tau)} \frac{u_{2\nu}  \bq^{[\mu}u_1^{\nu]}    }{\bq^2 \,\bq\cdot u_2},
\end{split}
\end{equation}
with $b_{12}=b_1-b_2$.
Now we  feed this into  \eqref{ald} noting that, at the lowest order, we only need to consider the first term $\propto {\dd^2 p^\mu}/{\dd \tau^2}$. 
Integrating, we obtain corrections to the  velocity and  particle trajectory given by
\begin{equation}\label{alddata}
\begin{split}
&\Delta^{(2)} v^\mu_{2,\,\text{ALD}}(\tau)   =\frac{Q_2^2}{6\pi m_2} \int_{-\infty}^\tau \dd\tau \frac{\dd^2  \Delta^{(1)} v^\mu_2 }{\dd \tau^2}=
i\frac{Q_1 Q_2^3}{ 6\pi m_2^2} \int \hat{\dd}^D \bq \,\hat{\delta}(\bq\cdot u_1)e^{i \bq\cdot (b_{12}- u_2 \tau)} \frac{ u_{2\nu}  \bq^{[\mu}u_1^{\nu]}}{\bq^2},
\\& \Delta^{(2)} x^\mu_{2,\,\text{ALD}}(\tau)= \int_{-\infty}^\tau \dd\tau\, \Delta^{(2)} v^\mu_{2,\,\text{ALD}}   =-\frac{Q_1 Q_2^3}{ 6\pi m_2^2} \int \hat{\dd}^D \bq \,\hat{\delta}(\bq\cdot u_1)e^{i \bq\cdot (b_{12}- u_2 \tau)} \frac{ u_{2\nu}  \bq^{[\mu}u_1^{\nu]}}{\bq^2\, \bq\cdot u_2}.
\end{split}
\end{equation}

The crucial step is now to use the ALD perturbations \eqref{alddata} to solve the Maxwell equations.  
After going to momentum space, a short calculation leads to
\begin{equation}
\begin{split}
-\bk^2 \tilde{A}^\mu(\bk) \rvert_{\text{ALD} }
& =  Q_2 \int \dd \tau  \, e^{i\bk\cdot (b_2+u_2\tau)}\left(
\Delta^{(2)} v^\mu_{2, \text{ALD}}  (\tau)  +i\bk \cdot \Delta^{(2)} x_{2,\text{ALD}}(\tau) u_2^\mu
\right) \\&=
i\frac{Q_1 Q_2^4}{ 6\pi m_2^2}  \int\hat{\dd}^D \bq_1 \hat{\dd}^D \bq_2\, e^{i\bq_1\cdot b_1}e^{i\bq_2\cdot b_2}\hat{\delta}(\bq_1\cdot u_1)\hat{\delta}(\bq_2 \cdot u_2)\hat{\delta}^D(\bk+\bq_1+\bq_2)\\&
\hspace{2cm}\times\frac{\bk\cdot u_2}{\bq_1^2}\left(
u_1^\mu+\bq_1^\mu \frac{ u_1\cdot u_2}{\bk\cdot u_2} -u_2^\mu\frac{ \bk\cdot u_1}{\bk\cdot u_2}-u_2^\mu \frac{ u_1\cdot u_2  \,\bk\cdot \bq_1}{(\bk \cdot u_2)^2}
\right),
\end{split}
\end{equation}
at the relevant $Q_1 Q_2^4$ order. One can already see that things start to look familiar: the integrand above is highly reminiscent of  the $\CCut$ found earlier in equation \eqref{finea}.  To make the equivalence of the two calculation even more manifest we simply have to contract with a polarisation vector according to \eqref{eq:classicalrelation}; in doing so we pick a gauge choice where  $\varepsilon\cdot u_2$=0 as done
in section~\ref{rr}. 
We have
\begin{equation}\label{classicalald}
\begin{split}
-i  \varepsilon^* (k) \cdot  \tilde{A}(\bk)\rvert_{\text{ALD} } \,  \bk^2
 =  - \int\hat{\dd}^D \bq_1 \hat{\dd}^D \bq_2\,& e^{i\bq_1\cdot b_1}e^{i\bq_2\cdot b_2}\hat{\delta}(\bq_1\cdot u_1)\hat{\delta}(\bq_2 \cdot u_2)\hat{\delta}^D(\bk+\bq_1+\bq_2)\\
&\hspace{-9pt} \times\frac{Q_1 Q_2^4}{ 6\pi m_2^2}\frac{\bk\cdot u_2}{\bq_1^2} \left(\varepsilon^*(k)\cdot
u_1+\varepsilon^*(k)\cdot\bq_1 \frac{ u_1\cdot u_2}{\bk\cdot u_2}  
\right) \,.
\end{split}
\end{equation}
It is a trivial exercise to see that this agrees with equation~\eqref{qedimpartfinalll} plugged into the general expression for the imaginary part of the waveform, equation~\eqref{eq:genericWaveshapeIm}. 
Notice also that there is no way to generate additional terms in the $Q_1 Q_2^4$ sector which could give rise to a real part, in agreement with the discussion in section~\ref{renormalix}.
 
Let us remark that the term in the ALD acceleration  $\propto \dd^2 p^\mu/\dd\tau^2$ which gave rise to the relevant correction  \eqref{alddata} is known in the literature as a ``Schott term" \cite{schott1912electromagnetic}.
This contribution is enhanced in the coupling relative to the term $\propto p^\mu (\dd p /\dd \tau)^2$ but, importantly, is a total time derivative.
The fact that Schott terms average to zero in periodic situations is often used in ``derivations'' of the ALD force (see, for example, the discussion in Jackson's textbook~\cite{Jackson:1998nia}).
Even in scattering scenarios, Schott terms can often be neglected.
For example, defining the total ALD impulse\footnote{The impulse is defined as the total change in momentum of a particle during the scattering event.} to be the total time integral of the force, the Schott term contributes a term proportional to
\[
\int_{-\infty}^\infty \dd \tau \, \frac{\dd^2 p^\mu}{\dd\tau^2} = \left. \frac{\dd p^\mu}{\dd\tau}\right|_{-\infty}^\infty \,.
\]
If the momentum of the particle is asymptotically constant, again the Schott term does not contribute.
Nevertheless the Schott term is relevant for understanding the time-dependent motion of the particle, and it is through this mechanism that the ALD force contributes to the radiation field of the particle at one-loop order.

It may be worth commenting in this context on an interesting connection between radiation-reaction effects in the one-loop waveform and the impulse at two loops.
In reference~\cite{Kosower:2018adc}, the impulse was related to four-point amplitudes.
Radiation reaction affects the impulse as mechanical momentum is radiated away.
Now one can build a subset of the two-loop diagrams contributing to the impulse by attaching the outgoing photon line (of momentum $k$) back to particle 1 in the diagram in equation~\eqref{figurecutqed}.
So in this sense one can recycle computations concerning radiation reaction and the one-loop waveform into information about radiation reaction at two loops in the impulse.
However, the (classical) physics captured by this recycling is different in the waveform and the impulse.
In the case of the waveform, the acceleration of the particle, arising from a Schott term, is transient.
However in the impulse the particle's momentum has been irretrievably lost.
Presumably the fact that the same diagrams compute these two superficially different aspects of radiation reaction is an amplitudes-based way of understanding the tight link between the two terms on the RHS of the ALD force law equation~\eqref{ald}. 
Classically the two terms are related by consistency with $p \cdot \dd p / \dd \tau = 0$.

Finally, we stress that, even in electrodynamics, it is significantly easier to determine the waveshape using amplitudes once the relevant cuts are known: the QFT-based approach separates structural aspects of the waveshape, e.g. equation~\eqref{classico}, from the dynamics at the outset. 
Dynamical information is then elegantly captured in the cuts, avoiding rather lengthy algebra.
 
\section{Conclusions}
\label{sec:concl}

In this paper we investigated next-to-leading-order radiation fields, and described how they can be computed elegantly using the techniques of modern scattering amplitudes.  

Building upon the KMOC formalism of \cite{Kosower:2018adc}, we characterised NLO radiation fields in terms of the real and imaginary parts of a waveshape. 
The real part is extracted by cutting one massive line of a five-point one-loop amplitude, whereas the imaginary part is obtained by a double (unitarity) cut of this amplitude. 
With this arrangement, all remaining propagators are defined through a principal-value prescription. 
This propagator structure emerges directly from the KMOC setup, the Feynman $i \epsilon$ prescription, and the split into real / imaginary parts, with no further intervention by hand.

Our organisation of the observable provides two key benefits.
First, it improves computational efficiency: the cancellation of apparently singular inverse powers of $\hbar$ (the ``superclassical'' terms) can be trivialised.
Second, this organisation clarifies the underlying physics.
Both real and imaginary parts have separate, gauge invariant, physical meaning.

In QED, the one-loop real part describes the radiation emitted by a body moving under the influence of essentially conservative forces: for example, a charge accelerated by the Lorentz force in the field of a different charge.
The imaginary part in addition captures intrinsically dissipative effects: radiation generated under the influence of the particle's self field.
In electrodynamics, this can be understood as the portion of radiation generated by the action of the Abraham-Lorentz-Dirac force on the charge.
As we discussed, the acceleration of the charge at this order originates in a Schott term; the total impulse on the charge vanishes, but the time-dependent acceleration nevertheless leaves an imprint on the radiation field of the particle.
In our description, this aspect of radiation reaction at one loop order is directly related to a simple unitarity cut involving a product of two Compton amplitudes in electrodynamics, Yang-Mills theory and gravity.

Radiation reaction is a consequence of self-force. For point-like objects, this inevitably entails some kind of regulation and renormalisation of singularities.
Our approach is directly rooted in traditional quantum field theory, so we took advantage of the opportunity to explain how the usual procedure of renormalisation in quantum field theory removes the part of diagrams whose imaginary part is associated with radiation reaction.
Using the on-shell renormalisation scheme, we showed that this class of diagram indeed cancels from classical computations.
As the counterterms are real, they do not affect the imaginary parts of these graphs which capture the effects of radiation reaction, and are perfectly classical.
As the waveform is observable we anticipate that any scheme dependence cancels in the final result.

We tested our QFT-based computations in electrodynamics by comparing to a fully classical computation of the complete radiation field at order $Q^5$, finding detailed agreement.
The classical computation, although not especially arduous, is nevertheless more involved than the elegant approach based on generalised unitarity, once the relevant cuts are understood.
Although electrodynamics is a comparatively simple theory, nevertheless it is rich enough to provide a very stimulating laboratory for understanding many aspects of the dialogue between amplitudes and classical physics, especially since the electrodynamic case is often a useful step towards a computation in Yang-Mills theory~\cite{delaCruz:2020bbn,delaCruz:2021gjp,Bern:2021xze}.

Turning to future directions, it would be very interesting to understand the physical meaning of the real and imaginary parts of the waveshape beyond one loop.
Our initial motivatation to study these real and imaginary parts arose from studying references~\cite{Herrmann:2021lqe,Herrmann:2021tct}, where the authors simplified other observables (the impulse and radiated momentum) after splitting into real and imaginary parts.
The authors of references~\cite{Herrmann:2021lqe,Herrmann:2021tct} found this approach useful at both one and two loops.
A first topic, then, would be to study the waveshape at two loops.
Is it possible to link a well-defined part of the observable to radiation reaction at this order?
What pole prescription emerges for the various propagators?

A related motivation for studying real and imaginary parts of the waveshape emerges from eikonal and related approaches to amplitudes in the classical limit, especially~\cite{Bern:2021dqo,DiVecchia:2021bdo,DiVecchia:2022nna,Cristofoli:2021jas,DiVecchia:2022piu}.
The essence of these approaches is that classical physics arises from a stationary-phase approximation at the level of the path integral. 
Amplitudes in the classical limit are essentially perturbative expansions of this phase.
Because the phase has the structure $\exp(i S/\hbar)$, the expansion introduces inverse powers of $\hbar$; these cancel in observables.
The relevant product in the expansion of the phase has the momentum-space interpretation of a convolution, emerging from a cut (see, for example~\cite{Parra-Martinez:2020dzs} for more on this link).
Thus we should expect an interplay between real and imaginary parts and the cancellation of ``superclassical'' terms in amplitudes at \emph{all} orders.
Indeed, in our work, we found that superclassical terms cancelled completely at the level of cuts after splitting into real and imaginary parts.
This greatly simplifies the computation of the relevant cuts.
Consequently we think it is very likely that this kind of organisation will be particularly useful at higher orders.

There is also still a great deal of interesting physics to be understood without facing the extremely challenging situation at two loops.
Capturing the physics of black hole spin in the tree-level waveform already requires an understanding of the Compton amplitude with spin, including relevant contact terms.
These have been recently studied by different groups, for example~\cite{Aoude:2022trd, Bjerrum-Bohr:2023jau, Cangemi:2022bew,Bern:2022kto, Bautista:2022wjf}. 
At one loop, we also need the five-point analogue of the tree Compton amplitude which appears in the relevant one-loop cuts.

Throughout this article, our focus was on the waveshape generated during a scattering experiment.
It would obviously be very exciting if our computations could be analytically continued using some appropriate algorithm to the bound state case, 
perhaps along the lines of references~\cite{Kalin:2019rwq,Kalin:2019inp,Adamo:2022ooq}.

In section~\ref{reactionnn}, we saw that certain Compton cuts capture radiation reaction effects in the EM waveform. We then tentatively identified Compton cuts with radiation reaction in Yang-Mills theory and in gravity.
It would be very interesting to explore this further.
For example perhaps it could be possible to recover the gravitational Compton cut from a computation with the MiSaTaQuWa force.

We believe that our work shows once again that the perturbative structure of classical interactions is clarified when generalised unitarity and the double copy are exploited.
Methods based on scattering amplitudes successfully factorise observables into a general kinematic structure (eg an integral over an in-in expectation) and a dynamical object (the expectation) which must be determined. 
Generalised unitarity determines the dynamical content (the expectation) from foreknowledge of its general analytic structure.
The double copy allows us to bypass much of the complexity of gravity.
Although our methods at first sight seem foreign to classical physics, 
our work demonstrates that they indeed have a new home in the classical domain. 

\acknowledgments

We particularly thank John Joseph Carrasco for collaborating with us in the early stages of this work, when he contributed significant ideas which enhanced our understanding of the physics.
We also thank Andreas Brandhuber, Graham Brown, Gang Chen, Stefano De Angelis, Joshua Gowdy, Aidan Herderschee, Radu Roiban, Fei Teng, and Gabriele Travaglini for cooperating with us with the submission of this manuscript.
Our work has also benefited from useful discussions with  Andrea Cristofoli, Leonardo de la Cruz, Kays Haddad, Hofie Hannesd\'ottir, Franz Herzog, Anton Ilderton, Sebastian Mizera, Alasdair Ross, Justin Vines, Pablo Vives Matasan and Mao Zeng.

This research was supported in part by the National Science Foundation under Grant No. NSF PHY-1748958.
AE is sponsored by a Higgs Fellowship.
DOC is supported by the U.K. Science and Technology Facility Council (STFC) grant ST/P000630/1. 
MS is supported by a Principal's Career Development Scholarship from the University of Edinburgh and the School of Physics and Astronomy.
IVH is supported by the Knut and Alice Wallenberg Foundation under grants KAW 2018.0116 (From Scattering Amplitudes to Gravitational Waves) and KAW 2018.0162.
For the purpose of open access, the author has applied a Creative Commons Attribution (CC BY) licence to any Author Accepted Manuscript version arising from this submission.

\appendix

\section{Iteration cuts}\label{cutextra}

In section~\ref{sec:cutAndIm} we explained the structure of the imaginary part of the expectation $\E$. 
The result was summarised succintly in equation~\eqref{eq:imEcuts}. 
It can be expressed simply in terms of the Compton cuts (which we defined in equation~\eqref{ccut}) and a new kind of iteration cut, $\ICut$, which we now define as
\[\label{eq:iCutDef}    

\]
We can therefore write the imaginary part of the expectation as
\[\label{eq:niceimcuts}
\Im' \E_{5,1}(p_1, p_2 \rightarrow p_1', p_2', k_\eta) = \sum_{i=1,2}& \CCut_i \mathcal{A}_{5,1} (p_1, p_2 \rightarrow p_1', p_2', k_\eta)
\\&+
\ICut \mathcal{A}_{5,1}(p_1, p_2 \rightarrow p_1', p_2', k_\eta) \,.
\]
In a preprint version of this article, we omitted the $\ICut$, which in fact does not vanish.
What is true is that superclassical terms cancel in the combination of equation~\eqref{eq:iCutDef} as discussed in section~\ref{vanishingcuts} (see equation \eqref{eq:zerocut}.)
A finite classical remainder remains~\cite{Caron-Huot:2023vxl}.

For completeness, we include a brief discussion of the iteration cut in electrodynamics.
To compute the $\ICut$, it is enough to expand the cuts appearing in equation~\eqref{eq:iCutDef} in powers of $\hbar$.
As usual, we write the primed momenta in terms of $q_i$ as $p'_i = p_i + q_i$.
Then, following the discussion in section~\ref{sec:hbar}, the $\hbar$ expansion is equivalent to expanding the amplitudes to one sub-leading order in $q_i$ and $k$.
In doing so, we may still take advantage of the simplifications described in section~\eqref{sec:vanishing}, in particular ignoring terms which contribute powers of messenger propagators in the numerator.
Nevertheless it is important to take care about expanding the delta functions which constrain the loop momenta in equation~\eqref{eq:iCutDef}.
 
It is convenient to write the total cut in terms of two parts,
\[
\ICut\mathcal{A}_{5,1}(p_1, p_2 \rightarrow p_1', p_2', k_\eta) = 
\ICut^{\text{H}}&\mathcal{A}_{5,1}(p_1, p_2 \rightarrow p_1', p_2', k_\eta) 
\\
&+ \ICut^{\text{Sym}} \mathcal{A}_{5,1}(p_1, p_2 \rightarrow p_1', p_2', k_\eta) \,.
\]
We choose these so that $\ICut^{\text{H}} \mathcal{A}_{5,1}$ is proportional to $m_1 / m_2$ while $\ICut^{\text{Sym}} \mathcal{A}_{5,1}$ is symmetric in its mass dependence.
This choice ensures that $\ICut^{\text{H}}$ and $\ICut^{\text{Sym}}$ can be compared to the classical quantities given in equations~\eqref{classico2} and~\eqref{ClassSymm}, respectively.
After some straightforward algebra, taking care to use the appropriate on-shell conditions for each cut, we find the following results: 
\[\label{eq:HeavyItCut}
\ICut^{\text{H}} \mathcal{A}_{5,1}&(p_1, p_2 \rightarrow p_1', p_2', k_\eta) = \frac{2Q_1^2Q_2^3 m_1}{m_2} 
\int \hd^D \ell \;\frac{\hdelta(u_1 \cdot \ell)}{\ell^2 (\ell -q_1)^2} \\
\times & \left[ \hdelta(u_2 \cdot \ell)  \left(-\varepsilon^*(k) \cdot u_1\, q_2 \cdot u_1 + \varepsilon^*(k) \cdot q_1 \left(\frac{u_1 \cdot u_2}{u_2\cdot q_1}\right)^2 (\ell\cdot k + \frac{q_1 \cdot u_2 \, q_2 \cdot u_1}{u_1\cdot u_2}) \right)  \right. \\
 & \left. -\ell \cdot q_2 \hdelta'(u_2 \cdot \ell) \left( \varepsilon^*(k)\cdot u_1 (u_1 \cdot u_2)+\varepsilon^*(k) \cdot(\ell -q_1)\frac{(u_1 \cdot u_2)^2}{u_2 \cdot q_1} \right)  \right.\\
  &\left. + \hdelta(u_2 \cdot \ell) \varepsilon^*(k)\cdot \ell \left(\frac{u_1 \cdot u_2}{u_2\cdot q_1}\right)^2(q_2^2/2 - q_1^2/2 - 2 \ell \cdot k) \right] \,,
\]
and
\[
&\ICut^{\text{Sym}}  \mathcal{A}_{5,1}(p_1, p_2 \rightarrow p_1', p_2', k_\eta) = -2 Q_1^2Q_2^3   \int \hd^D \ell \;\frac{ 1}{\ell^2 (\ell -q_1)^2} \\
&\quad\times\left[\hdelta(u_1 \cdot \ell)\hdelta(u_2 \cdot \ell)  \left (\varepsilon^*(k) \cdot u_1 - \varepsilon^*(k) \cdot  q_1 \frac{u_1 \cdot u_2}{u_2\cdot q_1}\right) q_1 \cdot u_2
 \right. \\
&\qquad \left. +  \ell \cdot q_1 \hdelta'(u_1 \cdot \ell)\hdelta(u_2 \cdot \ell)\left(\varepsilon^*(k) \cdot u_1 (u_1 \cdot u_2)+\varepsilon^*(k)\cdot(\ell -q_1)\frac{(u_1 \cdot u_2)^2}{u_2 \cdot q_1} \right)  \right].
 \]
As elsewhere in our discussion of electrodynamics, these results are valid in the gauge $\varepsilon(k) \cdot p_2 = 0$.

Classically, the same contributions arise from the $i\epsilon$ prescription in the matter propagators in equations \eqref{classico2} and \eqref{ClassSymm}, respectively, using equation~\eqref{plstot} and the related identity
\begin{equation}
\left(\frac{1}{x+i\epsilon}\right)^2=  \text{PV}\left(\frac{1}{x^2} \right)+\frac{i}{2}\hat{\delta}'(x)
\end{equation}
bearing in mind the factor $4 m_1 m_2$ which appeared in equation~\eqref{classico}.

Iteration cuts in YM and GR results can be obtained in the same way from the tree-level amplitudes of \cite{delaCruz:2020bbn,Luna:2017dtq} respectively.

\section{A derivation of the ALD force with cuts}\label{coleman}

In this appendix we propose a classical derivation of radiation reaction which further supports the theory  developed in \ref{impart}. There, we related non conservative effects to Compton-amplitude cuts of the five-point 
 one-loop amplitude.  Specifically, here we will demonstrate how the Schott term in \eqref{ald} arises from a double cut integral which highly resembles the ones in section \ref{reactionnn}. We are inspired by Coleman's lectures on relativistic radiation \cite{RM-2820-PR}.   
 
Let us start by considering the electromagnetic field strength tensor $F^{\mu\nu}$ and compute it in a close neighborhood of the particle. We will work in an arbitrary number of dimensions $D$ to consistently drop scaleless contributions, and only at the end take $D=4$.
The field's Fourier transform with retarded boundary conditions reads
\begin{equation}\label{fcol}
    F_{\mu\nu}(0)=i Q \int \hd^D k\int_{-\infty}^0 \dd \tau\, e^{ik\cdot r(\tau)}\frac{k_{[\mu}u_{\nu]}(\tau)}{k^2},
\end{equation}
with $k^2\equiv (k^0+i\epsilon)^2-\v{k}^2$. Note that we set for simplicity $x=0$ in the field's argument.  We now sit on top of the particle and expand all space-time dependent quantities  in a series of small proper time such that $\tau\sim k^{-1}\ll 1 $. Guided by the fact that \ref{ald} involves quantities with three $\tau$-derivatives, we expand the position $r^\mu(\tau)$ up to terms which involve the particle's acceleration change $\dot{a}^\mu$. Thus we have
\begin{equation}
\begin{split}
r^\mu(\tau)\approx  u^\mu \tau + \frac{1}{2}  a^\mu \tau^2+\frac{1}{6}\tau^3 \dot{a}^\mu,  \,\,\,\,\,\, u^\mu(\tau)\approx  u^\mu  +  a^\mu \tau+\frac{1}{2}\tau^2 \dot{a}^\mu, 
\end{split}
\end{equation}
which we can substitute inside \eqref{fcol} to obtain 
\begin{equation} \label{expanded}
\begin{split}
    F_{\mu\nu}(0)=i Q \int& \hd^D k \int_{-\infty}^0 \dd \tau\, e^{ik\cdot u \,\tau}\frac{k_{[\mu}}{k^2}\left(
u_{\nu]}+\tau a_{\nu]}+\frac{1}{2}i \tau^2 k\cdot a \,u_{\nu]}
    \right.\\& \left. 
+\frac{1}{2}i \tau^3 k\cdot a \, a_{\nu]}+\frac{1}{2}\tau^2\dot{a}_{\nu]} -\frac{1}{8}\tau^4 (k\cdot a)^2 u_{\nu]}+\frac{1}{6}i \tau^3 k\cdot \dot{a}\, u_{\nu ]}
    \right)+\cdots
    \end{split}
\end{equation}
Note again  that in this expansion  $\tau k\sim 1$ so above we have kept terms up to order $\tau^3/k \sim \tau^2/k^2\sim \tau^4 $ in the second line.

Next, we begin to simplify our expression using standard symmetry arguments and dimensional analysis. In fact, we will exploit the result that scaleless integrals vanish in dimreg: $\int \dd^D k \, k^{-p}=0$. Let us look at the zero-th order term. Here, we will often find it useful to use the decomposition   $k^\mu = k\cdot u \, u^\mu + k_\perp^\mu$. Then, the first piece of   \ref{expanded} is 
\begin{equation}
    \int \hd^D k \int_{-\infty}^0 \dd \tau\, e^{ik\cdot u \,\tau}\frac{k^{[\mu}u^{\nu]}}{k^2} =  \int \hd^D k\frac{i}{k\cdot u -i \epsilon}\frac{k^{[\mu}_\perp u^{\nu]}}{k^2}=0,
\end{equation}
 because of the anti-symmetric $k^\mu _\perp$ integrand\footnote{In a frame where $u^\mu=(1, \v{0})$, the  $k^\mu_\perp$ directions are simply the spatial ones.}. 
Regarding to the second term in \eqref{expanded}, we find it to be zero  since it is scaleless. Indeed, using similar steps, one brings this integral to the form
\begin{equation}
    \int \hd^D k \int_{-\infty}^0 \dd \tau\, e^{ik\cdot u \,\tau}\frac{k_{[\mu}a_{\nu]}}{k^2} \tau = i u_{[\mu}a_{\nu]}\int \hat{\dd}^{D-1}|\v{k}|\frac{1}{|\v{k}|^2}=0,
\end{equation}
 in dimensional regularization. We won't discuss each term singularly, but similar arguments can be applied to the other pieces in \eqref{expanded}. Some of them involve a tensor numerator that yields a vanishing result once the integral is reduced, and  on the support of $k^2=0$ and $a\cdot u=0$.

In the end, we find that only two terms survive, after integrating out $\tau$ the field strength  looks like 
\begin{equation}
\begin{split}
F_{\mu\nu}(0)=Q \int \hd^D k  \frac{k_{[\mu}}{k^2}\left(
\frac{\dot{a}_{\nu]}}{(k\cdot u-i\epsilon)^3}-\frac{k\cdot \dot{a}\,u_{\nu]}}{(k\cdot u-i\epsilon)^4 }
\right),
\end{split}
\end{equation}
For the first term we exploit symmetry along the $k_\perp$ integral to write ${k_{[\mu }\dot{a}_{\nu]}= k\cdot u \,u_{[\mu }\dot{a}_{\nu]}}$. Then, a simple tensor reduction of the $k$ integral of the second one yields
\begin{equation}
    \int \hd^D k  \frac{k_{[\mu}u_{\nu]}}{k^2} \frac{k\cdot \dot{a}}{(k\cdot u-i\epsilon)^4 } = \frac{ \dot{a}_{[\mu}\,u_{\nu]}}{D-1}  \int \hd^D k  \frac{1}{k^2} \frac{1}{(k\cdot u-i\epsilon)^2 } ,
\end{equation}
in the end we obtain the following 
\begin{equation}
\begin{split}
F_{\mu\nu}(0)=-Q\frac{D}{D-1} \dot{a}_{[\mu}\,u_{\nu]} \int \hd^D k  \frac{1}{k^2} \frac{ 1}{(k\cdot u-i\epsilon)^2 }.
\end{split}
\end{equation}

Our final task is to perform the $k$ integration. Therefore,  we first write  
\begin{equation}
   \frac{1}{(k\cdot u -i\epsilon)^2}=-\frac{\dd}{\dd(k\cdot u)}\frac{1}{k\cdot u - i\epsilon} 
\end{equation}
and then use the Sokhotski–Plemelj formula  \eqref{plstot} for both denominators. At this point, dimreg instructs once more to drop all principal value parts of the propagators. Essentially, we are taking the imaginary part of the integral now. We finally remain with
\begin{equation}\label{finalfmn}
\begin{split}
F_{\mu\nu}(0)=\frac{Q}{3} \dot{a}_{[\mu}\,u_{\nu]} \int \hd^4 k \, \hat{\delta}(k^2)\hat{\delta}'(k\cdot u)=\frac{Q}{6\pi} \dot{a}_{[\mu}\,u_{\nu]},
\end{split}
\end{equation}
having taken  $D=4$. 

Nicely, this last expression \eqref{finalfmn} matches the Schott term in \eqref{ald}. What is interesting here is to see how the double cut structure of \ref{impart} reverberates in our final steps of the proof. To this end one can interpret the fact the differentiated delta function $\hat{\delta}'(k\cdot u)$ in \eqref{finalfmn}   as a small $q$-expansion of  $\hat{\delta}(u\cdot (\ell -q))$ in \ref{reactionnn}.

\section{Electrodynamic waveform}
\label{sec:fullqed}

In this appendix we gather results for the complete integrand of the classical waveform in electrodynamics from the main text of the paper. We also provide explicit expressions for some cuts which were not described in detail above.

\subsection*{Overall structure}

The radiation field due to a two-particle scattering process is given in terms of the waveshape $\waveshape$ by equation~\eqref{eq:laterUse}.
Specialised to QED, the field is
\[
\label{eq:fieldAppendix}
F_{\mu\nu}(x) = \frac{-1}{4\pi |\v x|} 2 \Re \int_0^\infty \hd \omega \, e^{-i \omega u} \sum_\eta k_{[\mu} \varepsilon^\eta_{\nu]}(k) \,\waveshape  \, .
\]
We found it convenient to express the waveshape in terms of a more primitive object, the expectation $\E$ which is formally defined in equation~\eqref{eq:expectationDef}. We will present detailed expressions for the expectation below.
Using $\E$, the waveshape can be written as a sum of two parts (equations~\eqref{eq:genericWaveshapeRe} and~\eqref{eq:genericWaveshapeIm}:
\[\label{eq:apReWS}
\waveshape|_\textrm{Real} = \KMOCav{\int \hd^D q_1 \hd^D q_2 \, \hdelta(2 p_1 \cdot q_1) \hdelta(2 p_2 \cdot q_2) \,& \hdelta^D(k + q_1 + q_2) \,e^{-ib_1 \cdot q_1}e^{-ib_2 \cdot q_2} 
\\ & \times i \Re' \E(p_1, p_2 \rightarrow p'_1, p'_2, k_\eta) } \,,
\]
and
\[\label{eq:apImWS}
\waveshape|_\textrm{Im} = -\KMOCav{\int \hd^D q_1 \hd^D q_2 \, \hdelta(2 p_1 \cdot q_1) \hdelta(2 p_2 \cdot q_2) \,& \hdelta^D(k + q_1 + q_2) \,e^{-ib_1 \cdot q_1}e^{-ib_2 \cdot q_2} 
\\ & \times \Im' \E(p_1, p_2 \rightarrow p'_1, p'_2, k_\eta) } \,.
\]
The prime on the real and imaginary part symbols above indicate that polarisation vectors are to be treated as real quantities.

Thus given knowledge of $\E$, one has to compute the integrals in equations~\eqref{eq:apReWS} and~\eqref{eq:apImWS} to get the waveshape as a function; this yields direct knowledge of the fieldstrength in frequency space via equation~\eqref{eq:fieldAppendix}.

In the rest of this appendix, we present our results for the expectation at one-loop order, which is five powers of the charge.
As all of the amplitudes we discuss in this appendix are five-point one-loop amplitudes, we write $\mathcal{A}$ for $\mathcal{A}_{5,1}$ (and similarly for $\E$).
We also work throughout in the gauge $\varepsilon(k) \cdot p_2$ = 0. 
If necessary, it is straightforward to switch to polarisation vectors in a generic gauge, $\varepsilon'(k)$, using
\begin{equation}
\varepsilon_\mu(k)= \varepsilon'_\mu(k)-\frac{\varepsilon'(k)\cdot p_2}{p_2\cdot k }k_\mu .
\end{equation}

\subsection*{Order $Q_1 Q_2^4$}

At this order, the real part of the expectation is zero. The imaginary part is given by
\[
\Im' \E_{5,1}^{Q_1Q_2^4}&=\frac{4 Q_1 Q_2^4}{m_2^2   }  \frac{p_2\cdot \bk }{6\pi}\frac{1}{\bq_1^2} \left(
p_1\cdot \varepsilon^*(\bk)  +
\frac{p_1\cdot p_2\, \varepsilon^*(\bk)\cdot \bq_1}{p_2\cdot \bk} 
\right) \,.
\] 
For clarity, $\E^{Q_1Q_2^4}$ is a short-hand notation for $\E^{Q_1Q_2^4}(p_1,p_2 \rightarrow p_1', p_2', k_\eta)$. 
The expression for the $Q_1^4 Q_2$ charge sector can be deduced by simply relabelling the particle label $1 \rightarrow 2$.

\subsection*{Order $Q_1^2 Q_2^3$}

This charge sector is significantly more complicated and has both real and imaginary contributions.

We deduced the real part as a sum of two single-cuts,
\[
\Re' \E_{5,1}^{Q_1^2 Q_2^3} = \SCut_1 \mathcal{A}_{5,1}^{Q_1^2 Q_2^3} + \SCut_2 \mathcal{A}_{5,1}^{Q_1^2 Q_2^3} \,.
\]
The cuts themselves are (equations~\eqref{eq:heavySingleCutQED} and~\eqref{eq:scut2qed}):
\[\label{eq:appSingleCut1}
\SCut_1 \mathcal{A}^{Q_1^2 Q_2^3} = - 2 Q_1^2 Q_2^3 \int \frac{\hd^D \ell}{\ell^2 (\ell-q_1)^2} \hat{\delta}(p_1\cdot \ell) \left[ \left( m_1^2 + \frac{\ell \cdot (\ell - q_1) (p_1 \cdot p_2)^2}{p_2 \cdot \ell \, p_2 \cdot (\ell - q_1)} \right)  \frac{\varepsilon^* \cdot q_1}{p_2 \cdot q_1} \right . \\
\left. +  \left( p_1 \cdot q_2 {-} \frac{p_1 \cdot p_2 \, \ell \cdot q_2}{p_2 \cdot \ell} \right) \left( p_1 \cdot \varepsilon^* - \frac{p_1 \cdot p_2}{p_2 \cdot (\ell-q_1)} (\ell-q_1) \cdot \varepsilon^* \right) \frac{2}{p_2 \cdot \ell}
\right]   \,,
\]
and
\[\label{eq:appSingleCut2}
&\SCut_2 \mathcal{A}^{Q_1^2 Q_2^3}=
4Q_1^2 Q_2^3 
\int \frac{\hd^D \ell}{\ell^2 (\ell-q_1)^2}  \frac{\hdelta(u_2 \cdot \ell)}{(u_1 \cdot \ell)^2} \left[ u_1 \cdot \varepsilon^* \left( u_1 \cdot \ell \, u_2 \cdot q_1 - u_1 \cdot u_2 \, \ell \cdot q_1\right) \phantom{\frac{(u_1 \cdot \ell)^2}{u_2 \cdot q_1}} \right.
        \\
&+(\ell - q_1) \cdot \varepsilon^* \left.\left( u_1 \cdot u_2 \, u_1 \cdot \ell - \frac{(u_1 \cdot u_2)^2 \ell \cdot q_1}{u_2 \cdot q_1} + \frac{(u_1 \cdot \ell)^2}{u_2 \cdot q_1}\right)  
- \ell \cdot \varepsilon^* \, u_1 \cdot u_2 \, u_1 \cdot \ell \phantom{\frac{1}{2}}\!\!\!
 \right] \, ,
\]
In these equations the polarisation vector is associated with the outgoing photon of momentum $k$. 

There are two types of contribution to the imaginary part of the expectation (diagrammatically shown in equation~\eqref{eq:imEcuts}): Compton cuts (one for each massive particle) and an iteration cut.
We therefore write the imaginary part of the waveshape as (see equation~\eqref{eq:niceimcuts})
\[
\Im' \E^{Q_1^2 Q_2^3} = \CCut_1 \mathcal{A}^{Q_1^2 Q_2^3} + \CCut_2 \mathcal{A}^{Q_1^2 Q_2^3} + \ICut \mathcal{A}^{Q_1^2 Q_2^3} \,.
\]
The Compton cut for a given massive line can be easily deduced from the associated single cut by placing a photon on shell.
A brief examination of the cut conditions shows that there is no Compton cut associated with~\eqref{eq:appSingleCut1} (that is, with line 1 in this charge channel) essentially because the additional cut constraint isolates three-point amplitudes\footnote{We do not consider contributions from zero-energy photons.}.
There is a Compton cut for line 2 associated with equation~\eqref{eq:appSingleCut2}.
To extract the value of the cut, note that only the first diagram in equation~\eqref{eq:scut2Graphs} contributes. 
The cut can be obtained from the numerator in equation~\eqref{exkb}, integrated over the appropriate phase space. 
The result is\footnote{In comparing this cut to the real part in equation~\eqref{eq:scut2Graphs}, it can be useful to note that the step function in the phase space measure is always unity. This is very similar to the situation discussed in section~\ref{rr}.}
\[
&\CCut_2 \E^{Q_1^2 Q_2^3}=
-2Q_1^2 Q_2^3 
\int \frac{\dd \Phi(\ell-q_1)}{\ell^2} \frac{\hdelta(u_2 \cdot \ell)}{(u_1 \cdot \ell)^2} \left[ u_1 \cdot \varepsilon^* \left( u_1 \cdot \ell \, u_2 \cdot q_1 - u_1 \cdot u_2 \, \ell \cdot q_1\right) \phantom{\frac{(u_1 \cdot \ell)^2}{u_2 \cdot q_1}} \right.
        \\
&+(\ell - q_1) \cdot \varepsilon^* \left.\left( u_1 \cdot u_2 \, u_1 \cdot \ell - \frac{(u_1 \cdot u_2)^2 \ell \cdot q_1}{u_2 \cdot q_1} + \frac{(u_1 \cdot \ell)^2}{u_2 \cdot q_1}\right)  
- \ell \cdot \varepsilon^* \, u_1 \cdot u_2 \, u_1 \cdot \ell \phantom{\frac{1}{2}}\!\!\!
 \right] \, .
\]
(This Compton cut arises classically from equation~\eqref{ClassSymm}, extracting the imaginary part of the photon propagator using equation~\eqref{plstot}.)
Finally, as discussed in appendix~\ref{cutextra} iteration cuts also make a contribution given by
\[\label{eq:HeavyICut}
\ICut^{\text{H}} & \mathcal{A}_{5,1}(p_1, p_2 \rightarrow p_1', p_2', k_\eta) = \frac{2Q_1^2Q_2^3 m_1}{m_2} 
\int \hd^D \ell \;\frac{\hdelta(u_1 \cdot \ell)}{\ell^2 (\ell -q_1)^2} \\
\times & \left[ \hdelta(u_2 \cdot \ell)  \left(-\varepsilon^*(k) \cdot u_1\, q_2 \cdot u_1 + \varepsilon^*(k) \cdot q_1 \left(\frac{u_1 \cdot u_2}{u_2\cdot q_1}\right)^2 (\ell\cdot k + \frac{q_1 \cdot u_2 \, q_2 \cdot u_1}{u_1\cdot u_2}) \right)  \right. \\
 & \left. -\ell \cdot q_2 \hdelta'(u_2 \cdot \ell) \left( \varepsilon^*(k)\cdot u_1 (u_1 \cdot u_2)+\varepsilon^*(k) \cdot(\ell -q_1)\frac{(u_1 \cdot u_2)^2}{u_2 \cdot q_1} \right)  \right.\\
  &\left. + \hdelta(u_2 \cdot \ell) \varepsilon^*(k)\cdot \ell \left(\frac{u_1 \cdot u_2}{u_2\cdot q_1}\right)^2(q_2^2/2 - q_1^2/2 - 2 \ell \cdot k) \right] \,,
\]
and
\[
&\ICut^{\text{Sym}}  \mathcal{A}_{5,1}(p_1, p_2 \rightarrow p_1', p_2', k_\eta) = -2 Q_1^2Q_2^3   \int \hd^D \ell \;\frac{ 1}{\ell^2 (\ell -q_1)^2} \\
&\quad\times\left[\hdelta(u_1 \cdot \ell)\hdelta(u_2 \cdot \ell)  \left (\varepsilon^*(k) \cdot u_1 - \varepsilon^*(k) \cdot  q_1 \frac{u_1 \cdot u_2}{u_2\cdot q_1}\right) q_1 \cdot u_2
 \right. \\
&\qquad \left. +  \ell \cdot q_1 \hdelta'(u_1 \cdot \ell)\hdelta(u_2 \cdot \ell)\left(\varepsilon^*(k) \cdot u_1 (u_1 \cdot u_2)+\varepsilon^*(k)\cdot(\ell -q_1)\frac{(u_1 \cdot u_2)^2}{u_2 \cdot q_1} \right)  \right].
 \]

\bibliographystyle{JHEP}

\providecommand{\href}[2]{#2}\begingroup\raggedright\endgroup

\end{document}